\documentclass[12pt,preprint]{aastex}

\newcommand{\CIVdblt}{{C}\kern 0.1em{\sc iv}~$\lambda\lambda 1548, 1550$}
\newcommand{\MgIIdblt}{{Mg}\kern 0.1em{\sc ii}~$\lambda\lambda 2796, 2803$}
\newcommand{\SiIVdblt}{{Si}\kern 0.1em{\sc iv}~$\lambda\lambda 1393, 1402$}
\newcommand{\NVdblt}{\hbox{{N}\kern 0.1em{\sc v}~$\lambda\lambda 1239,1243$}}
\newcommand{\OVIdblt}{{O}\kern 0.1em{\sc vi}~$\lambda\lambda 1032, 1038$}
\newcommand{\AlIIIdblt}{{Al}\kern 0.1em{\sc iii}~$\lambda\lambda 1855, 1863$}
\newcommand{\CII}{\hbox{{C}\kern 0.1em{\sc ii}}}
\newcommand{\CIII}{\hbox{{C}\kern 0.1em{\sc iii}}}
\newcommand{\CIV}{\hbox{{C}\kern 0.1em{\sc iv}}}
\newcommand{\HI}{\hbox{{H}\kern 0.1em{\sc i}}}
\newcommand{\NaI}{\hbox{{Na}\kern 0.1em{\sc i}}}
\newcommand{\HII}{\hbox{{H}\kern 0.1em{\sc ii}}}
\newcommand{\HeI}{\hbox{{He}\kern 0.1em{\sc i}}}
\newcommand{\HeII}{\hbox{{He}\kern 0.1em{\sc ii}}}
\newcommand{\AlII}{\hbox{{Al}\kern 0.1em{\sc ii}}}
\newcommand{\AlIII}{\hbox{{Al}\kern 0.1em{\sc iii}}}
\newcommand{\NII}{\hbox{{N}\kern 0.1em{\sc ii}}}
\newcommand{\Lya}{\hbox{{Ly}\kern 0.1em$\alpha$}}
\newcommand{\Lyb}{\hbox{{Ly}\kern 0.1em$\beta$}}
\newcommand{\Lyg}{\hbox{{Ly}\kern 0.1em$\gamma$}}
\newcommand{\Lyd}{\hbox{{Ly}\kern 0.1em$\delta$}}
\newcommand{\Lye}{\hbox{{Ly}\kern 0.1em$\epsilon$}}
\newcommand{\FeII}{\hbox{{Fe}\kern 0.1em{\sc ii}}}
\newcommand{\MgI}{\hbox{{Mg}\kern 0.1em{\sc i}}}
\newcommand{\MgII}{\hbox{{Mg}\kern 0.1em{\sc ii}}}
\newcommand{\OVI}{\hbox{{O}\kern 0.1em{\sc vi}}}
\newcommand{\OVII}{\hbox{{O}\kern 0.1em{\sc vii}}}
\newcommand{\OVIII}{\hbox{{O}\kern 0.1em{\sc viii}}}
\newcommand{\NV}{\hbox{{N}\kern 0.1em{\sc v}}}
\newcommand{\SiII}{\hbox{{Si}\kern 0.1em{\sc ii}}}
\newcommand{\SiIII}{\hbox{{Si}\kern 0.1em{\sc iii}}}
\newcommand{\SiIV}{\hbox{{Si}\kern 0.1em{\sc iv}}}
\newcommand{\kms}{\ensuremath{\mathrm{km~s^{-1}}}}
\newcommand{\cmsq}{\ensuremath{\mathrm{cm^{-2}}}}
\newcommand{\cc}{\ensuremath{\mathrm{cm^{-3}}}}

\usepackage{epsfig,rotating}

\begin{document}

\title{The Kinematic Evolution of Strong {\MgII} Absorbers\footnotemark[1]}

\footnotetext[1]{Based on public data obtained from the ESO archive of observations from the UVES spectrograph at the VLT, Paranal, Chile.}

\author{Andrew C. Mshar\altaffilmark{2}, Jane C. Charlton\altaffilmark{2},  Ryan S. Lynch\altaffilmark{2,3},Chris Churchill\altaffilmark{4}, and Tae-Sun Kim\altaffilmark{5,6}}

\altaffiltext{2}{Department of Astronomy and Astrophysics, The Pennsylvania State University, University Park, PA 16802, \textit{amshar, charlton@astro.psu.edu}}
\altaffiltext{3}{Astronomy Department, University of Virginia, P. O. Box 3818, Charlottesville, VA 22903, \textit{rsl4v@mail.astro.virginia.edu}}
\altaffiltext{4}{Department of Astronomy, New Mexico State University, Las Cruces, NM 88003, \textit{cwc@nmsu.edu}}
\altaffiltext{5}{Institute of Astronomy, Madingley Road, Cambridge CB3 0HA, UK}
\altaffiltext{6}{Astrophysikalisches Institut Potsdam, An der Sternwarte 16, 14482 Potsdam, Germany, \textit{tkim@aip.de}}

\begin{abstract}

We consider the evolution of strong ($W_r(2796) > 0.3$~{\AA}) {\MgII}
absorbers, most of which are closely related to luminous galaxies.
Using 20 high resolution quasar spectra from the VLT/UVES public
archive, we examine 33 strong {\MgII} absorbers in the redshift range
$0.3 < z < 2.5$.  We compare and supplement this sample with 23 strong
{\MgII} absorbers at $0.4 < z < 1.4$ observed previously with
HIRES/Keck.  We find that neither equivalent width nor kinematic
spread (the optical depth weighted second moment of velocity) of
{\MgII}~$\lambda$2796 evolve.  However, the kinematic spread is
sensitive to the highest velocity component, and therefore not as
sensitive to additional weak components at intermediate velocities
relative to the profile center.  The fraction of absorbing pixels
within the full velocity range of the system does show a trend of
decreasing with decreasing redshift.  Most high redshift systems
(14/20) exhibit absorption over the entire system velocity range,
which differs from the result for low redshift systems (18/36) at the
95\% level.  This leads to a smaller number of separate subsystems for
high redshift systems because weak absorping components tend to
connect the stronger regions of absorption.  We hypothesize that low
redshift {\MgII} profiles are more likely to represent well formed
galaxies, many of which have kinematics consistent with a disk/halo
structure.  High redshift {\MgII} profiles are more likely to show
evidence of complex protogalactic structures, with multiple accretion
or outflow events.  Although these results are derived from
measurements of gas kinematics, they are consistent with hierarchical
galaxy formation evidenced by deep galaxy surveys.

\end{abstract}


\section{Introduction}
\label{sec:intro}

We study absorption line spectra in order to gain perspective on the
dynamics and evolution of astrophysical systems, such as galaxies.
The strength, relative abundance, and kinematics of absorption in
specific ions can constrain many properties of a given system.  For
example, metallicity, density, and abundance pattern can be determined
for various clouds that represent different parts of a galaxy and its
surroundings.  It is also possible to suggest, based on kinematics,
whether it is statistically likely that a specific absorption system
is produced primarily in the disk of a galaxy or in a halo
\citep{cc98}.  For a line of sight through a spiral galaxy, some
of the absorption components are likely to arise from the disk and
others from the halo.  \citet{cc98} suggest that the
dominant region of absorption in a strong {\MgII} absorber is often produced in
the disk of the galaxy, and that the outlying components are related to the
halo.

The {\MgIIdblt} doublet is of particular interest because of its
relative strength, ease of detection, and association with star
formation.  The close proximity of the two members of the doublet
(separated by only 7~{\AA} in the rest frame) makes it more convenient
to locate than multiplet metals such as iron.  Once a {\MgII}
absorption system is detected, other metal transitions can be located
and analyzed.  The ratio of {\FeII} to {\MgII} provides information
about the ionization parameter/density of the gas, and about the star
formation history of the system.  Type II supernovae enrich the ISM
with a large magnesium to iron ratio within the first billion years of the
formation of a stellar population.  Iron is mostly generated in Type
Ia supernovae, indicating that a system must have a longer history in
order to develop a relatively large {\FeII} to {\MgII} ratio.  The
{\MgI}~$\lambda$2853 transition is also of interest for understanding
the physical conditions of strong {\MgII} absorbers.  \citet{vpfits}
found that many clouds seem to require two separate phases of gas to
produce both the observed {\MgI} and {\MgII} absorption.  Comparisons
between {\MgII}, {\FeII}, and {\MgI} can give insight into the phase
structure and ionization state of the absorber.

Strong {\MgII} absorbers are defined by $W_r(2796)~{\ge}~0.3$~{\AA},
and have historically been interesting because of a direct
correspondence to luminous galaxies \citep{Berg91, Berg92, LeBrun93,
Steid94, Steid95}.  Although the {\MgII} absorbing gas is patchier
than once thought, it is still clear that the majority of strong
{\MgII} absorbers arise within $\sim 40 h^{-1}$~kpc of an $L > 0.05
L^*$ galaxy \citep{Church05}.  For the subset of strong {\MgII}
absorbers with $W_r(2796)>0.8$~{\AA}, \citet{zibetti07} find based on
stacking of images that half the light surrounding an absorber is
found within 50~kpc and the other half between 50 and 100~kpc.  The
ensemble of {\MgII} profiles for lines of sight through 0.4 $<$ z $<$
1.4 galaxies have kinematics consistent with expectations for a
representative mixture of disk and halo kinematics \citep{cc98}.
Spiral galaxies are expected to typically give rise to a blend of
several strong components spanning 30-100~{\kms}, and one or more
``satellite'' clouds beyond this velocity range.  \citet{Steid02}
measured rotation curves and found that 4/5 {\MgII} absorbers had
kinematics consistent with rotation in the same sense as the disk,
though in some cases a thick disk or a halo rotating in the same sense
was required.  Elliptical galaxies will tend to have components more
uniformly spread over a typical velocity range of 100-300~{\kms}.  The
kinematics and overall {\MgII} absorption strength are also influenced
by asymmetries in the gas caused by interactions and mergers
\citep{Kacp05}.

Our goal is to determine the evolution of the kinematics of strong
{\MgII} absorbers over the redshift range 0.3 $<$ z $<$ 2.5.  This
range covers about 7.7 Gyrs of cosmic history, from 2.7~Gyrs to
10.4~Gyrs after the Big Bang, assuming $H_0=73$~{\kms}{\rm Mpc}$^{-1}$,
$\Omega_{matter} = 0.26$, and $\Omega_{\lambda}=0.74$.
We divide this period into two halves, primarily based on the cutoff
in the study by \citet{CV01} (hereafter CV01).  0.3 $<$ z $<$ 1.2 is
referred to as our low redshift range, and 1.2 $<$ z $<$ 2.5 as our
high redshift range.

These redshift ranges are of particular interest for galaxy evolution
because several influential factors are known to change over this time
period.  First, galaxy morphologies tend to be more irregular at high
redshift.  \citet{Con04} found that large diffuse star-forming
objects, suggested to be the predecessors of spiral disks are found
primarily at 1 $<$ z $<$ 2, while asymmetric star-forming objects
(perhaps mergers that produce ellipticals) peak in abundance at $z \sim
1$.  Similarly, \citet{Elm05} found a predominance of ``chain'' and
``clump-cluster'' galaxies among the fainter magnitudes in the Hubble
Ultra Deep Field while traditional spirals and ellipticals dominated
among the brightest galaxies at low redshift.  The second (and
related) factor is the evolution in the galaxy merger rate, which
dramatically increases with increasing redshift \citep{Lef00,Pat02}.

The third factor, evolution in the extragalactic background radiation
(EBR), can dramatically influence absorption line systems because it
leads to a shift in the ionization balance of the various metal-line
transitions.  The EBR has been modeled as the cumulative radiation
from quasars and star-forming galaxies, modulated by the {\Lya} forest
\citep{Haardt96,Haardt01}.  The amplitude of the EBR is expected to be
relatively constant, with a number density of ionizing photons
$\log n_{\gamma} \sim -4.75$~{\cc}, over most
of our high redshift range, down to $z\sim1.5$, but then decreases by
nearly an order of magnitude to $\log n_{\gamma} \sim -5.64$~{\cc} from
$z=1.5$ to $z=0.3$.

The last factor is the evolution of the global star formation rate in
galaxies.  This rate is relatively constant from $z=4$ to $z=1$, then
decreases significantly to $z=0$ \citep{Gab04}.  The peak epoch of
star formation occurs earlier for giant galaxies than for dwarfs
\citep{Bauer05}.  We expect $\alpha$-enhanced metal build-up for the
first billion years past the birth of a stellar population, and an
increase in the ratio of iron to magnesium subsequently due to
contributions from Type Ia supernovae.

CV01 studies the kinematics of 23 strong {\MgII} systems in the
redshift range 0.4 $< z <$ 1.2.  The spectra were obtained with the
HIRES spectrograph with a resolution of $\sim 6.7$~{\kms}.  When
possible, {\FeII} and {\MgI} for these systems are compared to the
{\MgII}.  The authors found that strong {\MgII} absorbers are
typically not characterized by multiple subsystems of comparable
equivalent width or kinematic spread, but instead have a dominant
subsystem, often with subsystems of significantly smaller equivalent
width.  It is important to note, however, that there are a wide
variety of kinematic profiles within the CV01 sample.  Among systems
with the same equivalent width, they found some cases with weak
components widely spread in velocity and others with a single,
saturated component.  They also noted the interesting trend for
systems with multiple subsystems to have the intermediate and high
($>$40~{\kms}) velocity subsystems located either all redward or all
blueward of the dominant subsystem.  They interpreted this to mean
that the dominant subsystem is related to a rotating disk that is
systematically offset in velocity from the halo material that produces
weaker subsystems.  Within the redshift range, 0.4 $<z <$ 1.2, there
was no significant evolution in system or subsystem properties.

CV01 compares the smaller subsystems at intermediate and high
velocities to both single cloud, weak {\MgII} absorbers, and Galactic
high velocity clouds (HVCs). In order to make a quantitative
comparison, a slope was fit to the observed equivalent width
distributions of these ``satellite clouds'' and the single cloud, weak
{\MgII} absorbers \citep{cwc99b}.  Because of the large errors in the
fit for the subsystems of the strong absorbers (see Figure~8a from
CV01), they cannot distinguish between the slopes for the two samples.
However, \citet{cwc99b} does not find a turnoff in the equivalent
width distribution of weak {\MgII} absorbers, complete down to
$W_r(2796)$ = 0.02~{\AA}.  The equivalent width distribution of the
weak subsystems in CV01 have a turnoff at $W_r(2796)$ = 0.08~{\AA},
well above the drop in completeness, indicating a fundamental
difference between the two samples.

Galactic HVCs refer to clouds with velocities $v \ge 90$~{\kms}
relative to material in the Milky Way disk along the same line of
sight \citep{wakker97}.  Located in the Galaxy and its surroundings,
they are likely to have a variety of origins, ranging from a
``galactic fountain", to accretion of dwarf galaxies and tidal debris,
to infalling filaments and sheets from large-scale structure (see
\citet{Sem06} and references therein).  The satellites of strong
{\MgII} absorbers have {\HI} column densities less than the detection
threshold of 21-cm surveys for HVCs, which led CV01 to conclude they
are not analogous.  However, {\OVI} HVCs cover a larger fraction of
the sky ($\sim 60$-$85$\%; \citet{Sem03}) than 21-cm {\HI} HVCs ($\sim
37$\%; \citet{Lockman02}).  This implies that some of the {\OVI} HVCs
have lower {\HI} column densities.  Also, low ionization stages are
detected in a separate phase at the same velocities with most of the
Milky Way {\OVI} HVCs \citep{Col05}.  Thus, it now seems plausible that
the satellite clouds of some {\MgII} absorbers are analogs to some
types of Milky Way HVCs.

The strongest {\MgII} absorbers, those with $W_r(2796) > 1$~{\AA} may
be produced by different physical processes than the typical strong
{\MgII} absorber.  Using Sloan Digital Sky Survey (SDSS) and high resolution
Keck spectra, \citet{prochter06} studied the evolution of this
population, considering both its cross section for absorption and the
kinematic structure of the {\MgII} profiles.  They hypothesize that
the decline of the incidence of $W_r(2796) > 1$~{\AA} absorbers at
$z < 0.8$ is consistent with the decline in the global star formation
rate, and suggest a large contribution of galactic superwinds in shaping
the kinematics of the profiles of these systems.  \citet{nestor05} came
to similar conclusions based on a study of the SDSS data.

In this paper, we present 33 strong {\MgII} absorbers in the redshift
range $0.3 < z < 2.5$ observed with the Ultraviolet and Visual Echelle
Spectrograph (UVES) on the Very Large Telescope (VLT).  We will
quantify the absorption systems with the same statistics as CV01 in
order to make a fair comparison.  Our goal is to describe the kinematics
of high redshift strong {\MgII} absorbers and to infer any possible
evolution.

In \S~\ref{sec:data} we present the quasar observations, the reduction
of the spectra, and the absorber detection process.  We also define
the statistics that we used to describe the {\MgII} profiles.  We
describe each system in detail in \S~\ref{sec:systems}.
\S~\ref{sec:results} gives our results, comparing the kinematics of
all systems over the range $0.3 < z < 2.5$ in our sample to that
of CV01, and \S~\ref{sec:summary} is a summary of our results.
\S~\ref{sec:discussion} is our discussion of the nature of strong
{\MgII} absorbers and their evolution.

\section{Data and Survey Method}
\label{sec:data}

\subsection{VLT/UVES Data}
\label{sec:vlt}

We obtained 20 high quality UVES/VLT QSO spectra from the ESO archive.
The quasar names are provided in Table~\ref{tab:tab1} along with V
magnitude, quasar emission redshift (from Simbad), and the wavelength
range of the UVES spectrum.  This same set of spectra was used for an
analysis of the flux power spectrum of the {\Lya} forest \citep{kim04}.
Coverage breaks exist at about 5723 -- 5841~{\AA} and 8520 --
8665~{\AA} in all of these spectra.  Also present in all spectra were
telluric absorption features.  These absorption features can make
detection of the weakest subsystems difficult.  However, we address
this difficulty in our discussion of each absorber
(\S~\ref{sec:systems}).  The resolution is $R \sim 45,000$, or
$\sim 6.7$~{\kms} (the same resolution as in CV01).  
The signal to noise of the UVES/VLT spectra
is high ($\sim 20$--$100$ per pixel).  The CV01 signal to noise values
tend to fall in the lower half of this range.  We take this into
account in our analysis, and consider possible biases when relevant.
The data
reduction procedure can be found in \citet{kim04}, and the procedure
for continuum fitting is described in \citet{lynch06}.

Our quasar sample was originally selected for a study of the
properties of {\Lya} forest clouds \citep{kim04}.  The criteria for
selection from the archive for that study included high $S/N$ and
large wavelength coverage.  The result should be a relatively unbiased
distribution of strong {\MgII} absorbers.  However, the study also
avoided quasars that had known $z>1.5$ DLAs, based upon prior low
resolution spectra.  This will introduce a small bias, but we quantify
the effect in \S~\ref{sec:equivalentwidth} by comparing the equivalent
width distribution of our sample of strong {\MgII} absorbers to the
unbiased distribution obtained from a survey of Sloan quasars
\citep{nestor05}.

\subsection{Sample Selection}
\label{sec:selection}

We searched the normalized spectra for {\MgIIdblt} doublets and found
strong {\MgII} systems along 14 of the lines of sight (the list of all
lines of sight surveyed can be found in Table~\ref{tab:tab1}).  We
would have excluded associated systems (within $5000$~{\kms} of the
quasar redshift) from our sample, but in fact did not find any.
Although we used a $5\sigma$ search criterion for the
{\MgII}~$\lambda$2796, strong absorbers substantially exceed this
threshold in spectra of this quality.  Because of the high $S/N$ of
the spectra, in the process of the search we also detected the weak
systems from \citet{lynch06}.  While these systems are not of
particular interest in this study, our ability to find weak absorbers
is relevant because the equivalent width of some individual components
corresponding to strong absorbers is comparable to that of weak
systems.  After detecting the {\MgIIdblt}, we also searched the
expected locations of {\FeII}~$\lambda$2344, $\lambda$2374,
$\lambda$2383, $\lambda$2587, $\lambda$2600 and {\MgI}~$\lambda$2853.
The {\FeII} transitions can be used to better understand the
kinematics of systems with saturation in the {\MgII} profiles, and to
constrain ionization conditions and abundance pattern.  The
{\MgI}~$\lambda$2853 transition was studied for the same reasons, and
also to provide insight into properties of the lowest ionization
state.  Many other transitions were also detected but are not included
in this paper.  For example, {\SiII}~$\lambda$1260 and
{\CII}~$\lambda$1335 were not included because they are generally
detected in the {\Lya} forest, making their kinematics more difficult
to interpret.  Higher ionization transitions and Lyman series lines
were also detected for many of these systems, and they will be used to
place constraints on the physical properties of the systems in future
work.

\subsection{Defining Kinematic Subsystems}
\label{sec:subsystems}

A significant fraction of the strong absorption systems are comprised
of more than just a single component.  Thus, we define subsystems
using the definition from CV01, ``absorption features that are
separated by more than 3 pixels (i.e., a resolution element) of
continuum flux.''  We define these subsystems in order to examine the
kinematics of each system on the smallest scale so that we may extract
the most information possible from the systems.  The subsystems are
defined by their absorption in the {\MgII}~$\lambda$2796 transition,
which must be matched in the {\MgII}~$\lambda$2803 transition.  The
wavelength coverage of each subsystem is determined by the wavelength
on either side of the subsystem at which the equivalent width in that
pixel returns to within $1\sigma$ of the continuum value.  This
definition assures that features between subsystems and noise in the
spectra do not affect our interpretation of kinematics.

CV01 applies a uniform sensitivity cutoff for subsystems with
$W_r(2796) < 0.015${\AA}.  Although our sample is complete to a better
sensitivity, we adhere to this less sensitive cutoff in order to make
a fair comparison (see \S~\ref{sec:subsysprop}).
Table~\ref{tab:tab2} shows that although two absorbers approach this
limit (subsystem 2 in the $z_{abs} = 1.243967$ system toward Q0122-380
and subsystem 2 in the $z_{abs} = 1.149861$ system toward Q0453-423),
none are below it.  Therefore, we have no need to adjust the sample
due to this cutoff.  CV01 also exclude any high velocity material
above $v > 500$~{\kms}.  Our sample does not include any subsystems at
such high velocity; thus, we do not exclude any subsystems based on
this criterion.  We mention these cutoffs for completeness.

\subsection{Defining Absorption Properties}
\label{sec:Absorption}

We use several quantities to compare the shapes of the {\MgII}
profiles of the systems.  We formally define the redshift of a system
by the optical depth weighted mean of the {\MgII}~$\lambda$2796 profile.
The specific expression used to calculate redshift, as well as formal
definitions of equivalent width, subsystem velocity, apparent column
density, and velocity width, are given in Appendix A of CV01.  All of
these quantities are used here, but the velocity width, also known
as the kinematic spread, is quite important for this study and thus
deserves further comment.

The kinematic spread of a system is an optical depth weighted average of the
second moment of velocity.  The combination of a system's kinematic
spread and its equivalent width gives physical insight into the nature
of the system.  There is a maximum possible equivalent width for any
kinematic spread, produced by a fully saturated, ``square'' profile.
While there is no distinct opposite to these ``square profile
systems'', there are profiles with high kinematic spread and low
equivalent widths.  These systems generally have very few or no pixels
saturated and consist of multiple subsystems with a large kinematic
spread.

\section{Systems}
\label{sec:systems}

\begin{figure}
\figurenum{1a}
\centering
\vspace{0.0in}
\epsscale{0.45}
\plotone{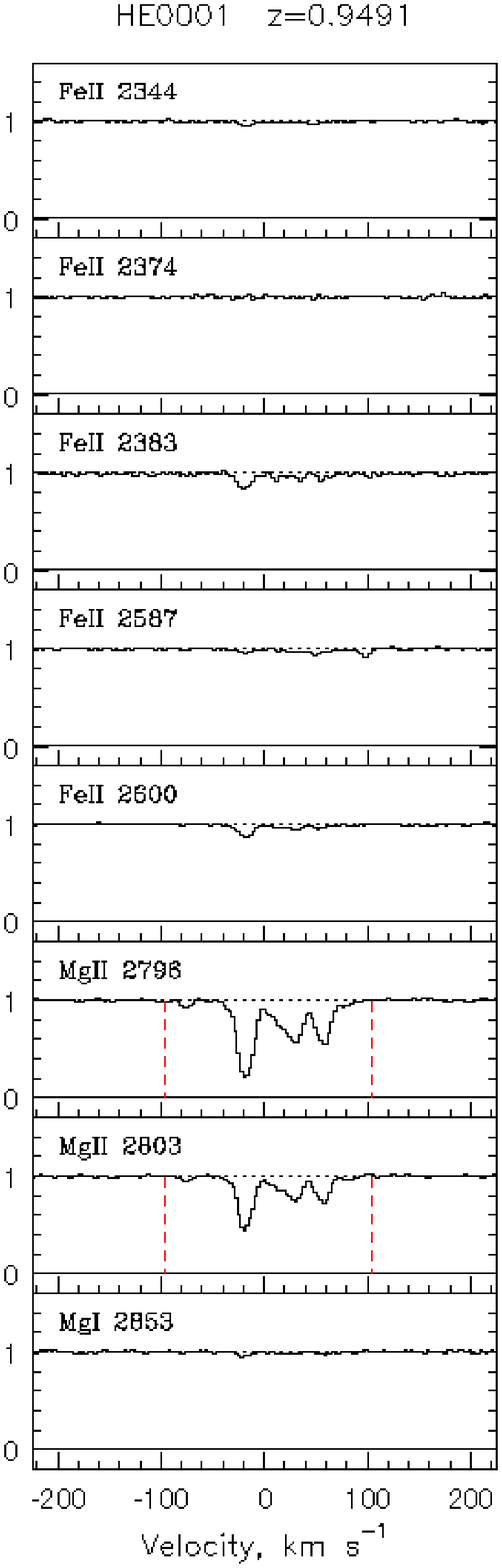}
\caption{VLT/UVES profiles of {\MgII}~$\lambda$2796, {\MgII}~$\lambda$2803, 
{\MgI}~$\lambda$2853, {\FeII}~$\lambda$2344, {\FeII}~$\lambda$2374,
{\FeII}~$\lambda$2383, {\FeII}~$\lambda$2587, and
{\FeII}~$\lambda$2600 (if detected) for the various strong {\MgII}
absorption lines systems.  The spectra are normalized and the
different transitions aligned in velocity space, with the zero-point
defined as the optical depth mean of the {\MgII}~$\lambda$2796
profile.  The velocity range, which differs from system to system, is
indicated by the label on the horizontal axis.  The vertical dashed
lines delineate the separate subsystems.  The crosses are above
features in the spectra that are not detections of the transition
highlighted in that window.  Figures~\ref{fig:fig1b}--ag are available
in the electronic edition of the Journal. The printed edition contains
only one example of this figure.}
\label{fig:fig1}
\end{figure}

We divide the systems into two redshift regimes, 0.3 $<$ z $<$
1.2 (low redshift) and 1.2 $<$ z $<$ 2.5 (high redshift).  We use the
low redshift regime for comparison to the CV01 data, and then combine
those samples for comparison to the high redshift sample.  Here we
present a short description of each strong {\MgII} absorber in the
UVES/VLT sample.  Descriptions of the CV01 absorbers were given in
\S~3.3 of that paper.  In Figures~\ref{fig:fig1}--ag 
we show the {\MgII}~$\lambda$2796, {\MgII}~$\lambda$2803,
{\FeII}~$\lambda$2344, {\FeII}~$\lambda$2374, {\FeII}~$\lambda$2383,
{\FeII}~$\lambda$2587, {\FeII}~$\lambda$2600, and {\MgI}~$\lambda$2853
transitions for each system if they are covered by the spectra.
Table~\ref{tab:tab2} lists each system with its kinematic spread, rest
frame equivalent width of {\MgII}~$\lambda$2796, and its doublet ratio,
$W_r(2796)/W_r(2803)$.  It also lists these quantities for the
individual subsystems identified for each of the systems.
Table~\ref{tab:tab3} gives the velocity ranges for the subsystems, and
lists the rest frame equivalent widths of the {\MgII}, {\FeII}, and
{\MgI} transitions.

\subsubsection{HE0001-2340 $z_{abs}=0.9491$}
\label{sec:0.9491}

This system, seen in Figure~\ref{fig:fig1}, has only a single
subsystem, with one central, dominant, but apparently unsaturated,
component, with weaker blended components to the red, and with one
blueward, weak component.  {\FeII} absorption is detected in the
strongest component and several of the weaker redward components.
Weak {\MgI}~$\lambda$2853 absorption is detected in the two strongest
components of the system.

\subsubsection{HE0001-2340 $z_{abs}=1.5862$}
\label{sec:1.5862}

\begin{figure}
\figurenum{1b}
\centering
\vspace{0.0in}
\epsscale{0.45}
\plotone{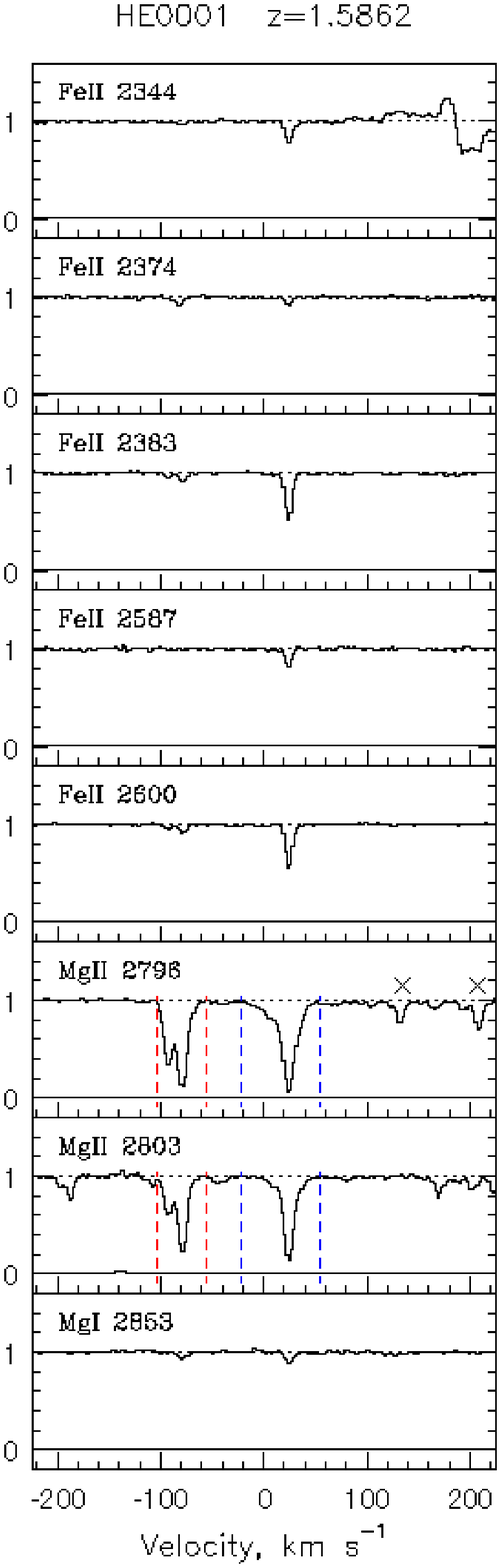}
\caption{}
\label{fig:fig1b}
\end{figure}

Figure~\ref{fig:fig1b} shows that there is no true ``dominant
subsystem'' in this absorber.  The two subsystems have roughly equal
equivalent widths (the blueward subsystem has $W_r(2796) = 0.17$~{\AA}
and the redder component has $W_r (2796) = 0.18$~{\AA}).  There is a small
feature between the two subsystems in both the {\MgII}~$\lambda$2796
and the {\MgII}~$\lambda$2803.  However, the {\MgII}~$\lambda$2803
equivalent width is greater, which is unphysical.  Consequently, we
exclude this region when calculating the kinematics of this system.
If there is any real absorption at this velocity, the paucity of the
{\MgII} is such that it would have no significant impact on the system
kinematics.  {\FeII} is detected in both subsystems, but it is
stronger in the redward one.  Two other features are apparent in the
{\MgII}~$\lambda$2796 velocity window redward of the system.
Neither is confirmed by {\MgII}~$\lambda$2803 in a
clean region of the spectrum, thus they can be cleanly rejected as
{\MgII} absorption.  A feature blueward of the system in the
{\MgII}~$\lambda$2803 velocity window has no corresponding detection in
{\MgII}~$\lambda$2796.

\subsubsection{HE0001-2340 $z_{abs}=2.1844$}
\label{sec:2.1844}

\begin{figure}
\figurenum{1c}
\centering
\vspace{0.0in}
\epsscale{0.45}
\plotone{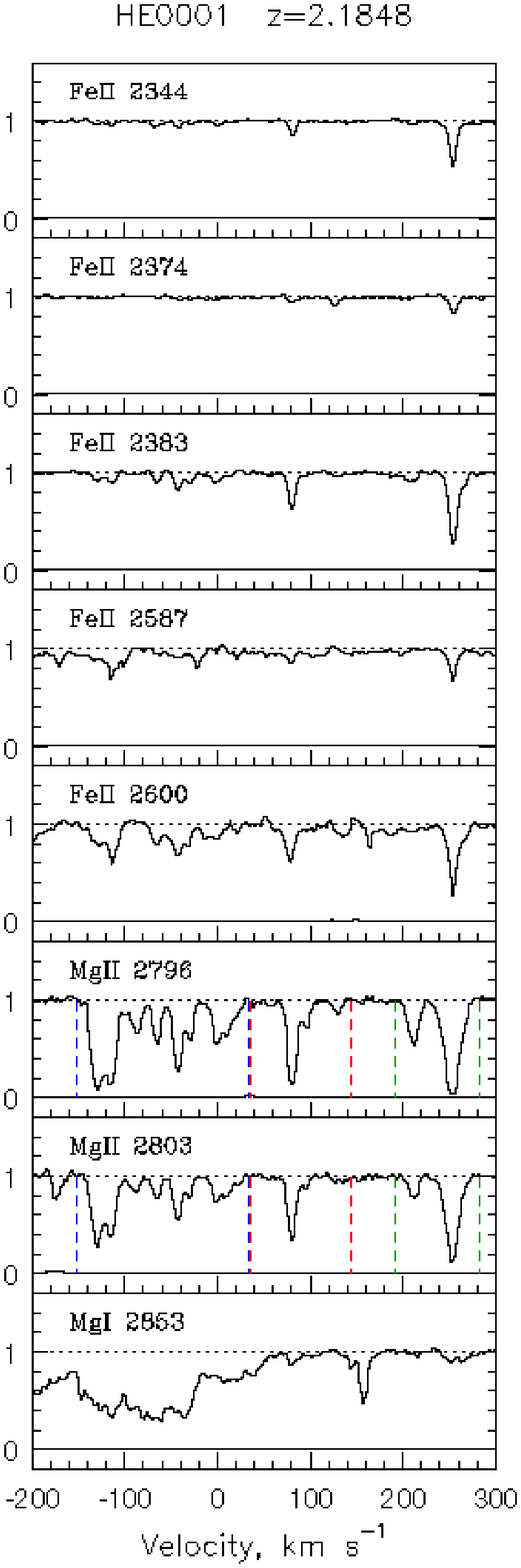}
\caption{}
\label{fig:fig1c}
\end{figure}

This unique system consists of multiple absorption features spread out
over a velocity range of $-140~{\kms} < v < 275$~{\kms} with no
apparent saturation; its profile is shown in Figure \ref{fig:fig1c}.
Generally, systems with such large velocity spreads do exhibit
saturation.  It is unusual in its large number of relatively weak
separated comonents over such a large velocity spread.  In most
systems, one subsystem can be classified as dominant because it
produces a great majority of the equivalent width; in this system,
none of the three subsystems is significantly stronger than all of the
rest.  The bluest subsystem consists of five components, and the
central velocity and red subsystems each consist of two components.
The {\MgI}~$\lambda$2853 transition is severely contaminated by a
blend, particularly to the blue, but useful limits may be obtained
toward the red.  {\FeII} is detected in several transitions for all
three subsystems, though {\FeII}~$\lambda$2587 and $\lambda$2600
suffer from blends due to telluric absorption.

\citet{richter05} have modeled this sub-DLA system based on these same
VLT data, having established that $\log N({\HI}) =19.7$~{\cmsq} based
upon a fit to the {\Lya} profiles.  They found an oxygen abundance of
$1/65$ solar, and a particularly low nitrogen content, suggesting
recent star formation.  They note that the large kinematic spread is
suggestive of an ongoing merger that has triggered recent star
formation.

\subsubsection{Q0002-422 $z_{abs}=0.8366$}
\label{sec:0.8366}

\begin{figure}
\figurenum{1d}
\centering
\vspace{0.0in}
\epsscale{0.45}
\plotone{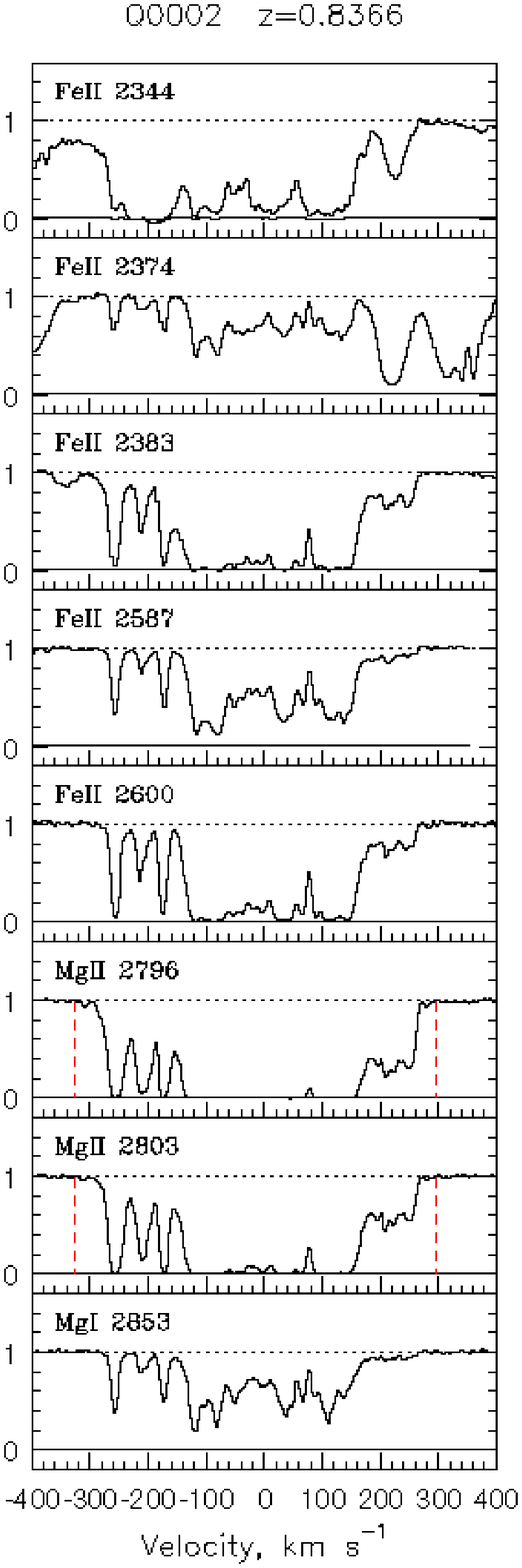}
\caption{}
\label{fig:fig1d}
\end{figure}

This extremely strong {\MgII} absorber is one of the two strongest in
our sample and has absorption over more than $500$~{\kms} and is fully
saturated in the range $\sim-130$~{\kms} to $160$~{\kms}, except for a small
break at $\sim 80$~{\kms}.  The absorption profile is shown in Figure
\ref{fig:fig1d}. Absorption in {\FeII} is extremely strong as well,
and provides useful constraints except for the redward region of the
{\FeII}~$\lambda$2374 transition.  The {\MgI}~$\lambda$2853 transtion
shows close to the same absorption features as the {\MgII} and
{\FeII}.  

It is very unusual to find a system this strong at $z<1$, based
upon the rapid evolution of $W_r(2796) > 2$~{\AA} systems as found by
\citet{nestor05}.  Statistically, based on comparisons of the
very strongest {\MgII} absorbers and their {\Lya} profiles, $42$\% of
systems with $W_r(2796)/W_r(2600) < 2$ and $W_r(2853) > 0.1$~{\AA} (this
system has $W_r(2796)/W_r(2600) = 1.48$~{\AA} and $W_r(2853) = 1.59$) are DLAs
\citep{rao06}.  However, this system is kinematically similar to the
$z=1.5541$ system toward Q$1213-0017$ (though much stronger), which is
suggested to be a ``superwind" absorber by \citet{bond01}, and is
known not to be a DLA based upon \citet{rao00}.

\subsubsection{Q0002-422 $z_{abs}=1.5418$}
\label{sec:1.5418}

\begin{figure}
\figurenum{1e}
\centering
\vspace{0.0in}
\epsscale{0.45}
\plotone{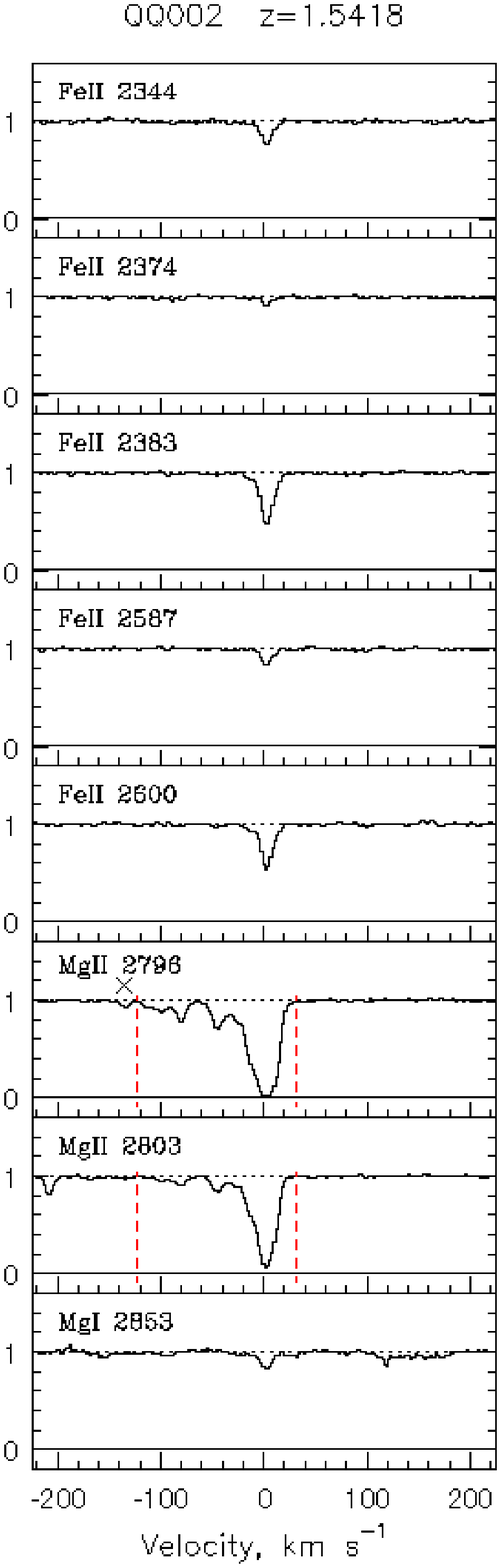}
\caption{}
\label{fig:fig1e}
\end{figure}

The {\MgII} profile shown in Figure \ref{fig:fig1e} consists of a
strong, nearly saturated, component with several weaker components to
the blue.  However, no absorption is detected redward of the strongest
absorption.  The feature $200$~{\kms} blueward of the system in the
{\MgII}~$\lambda$2803 window does not correspond to a 5-$\sigma$
detection in the {\MgII}~$\lambda$2796 and is, therefore, not {\MgII}
absorption.  The strongest absorption component is detected in five
{\FeII} transitions as well as in {\MgI}~$\lambda$2853.  The component
at $\sim -40$~{\kms} is also detected in two stronger {\FeII}
transitions.  The feature at $\sim 100$--$200$~{\kms} in the
{\MgI}~$\lambda$2853 window is a blend, and it lies outside of the
defined {\MgII} absorption region and thus has no effect on our
calculations.

\subsubsection{Q0002-422 $z_{abs}=2.1678$}
\label{sec:2.1678}

\begin{figure}
\figurenum{1f}
\centering
\vspace{0.0in}
\epsscale{0.45}
\plotone{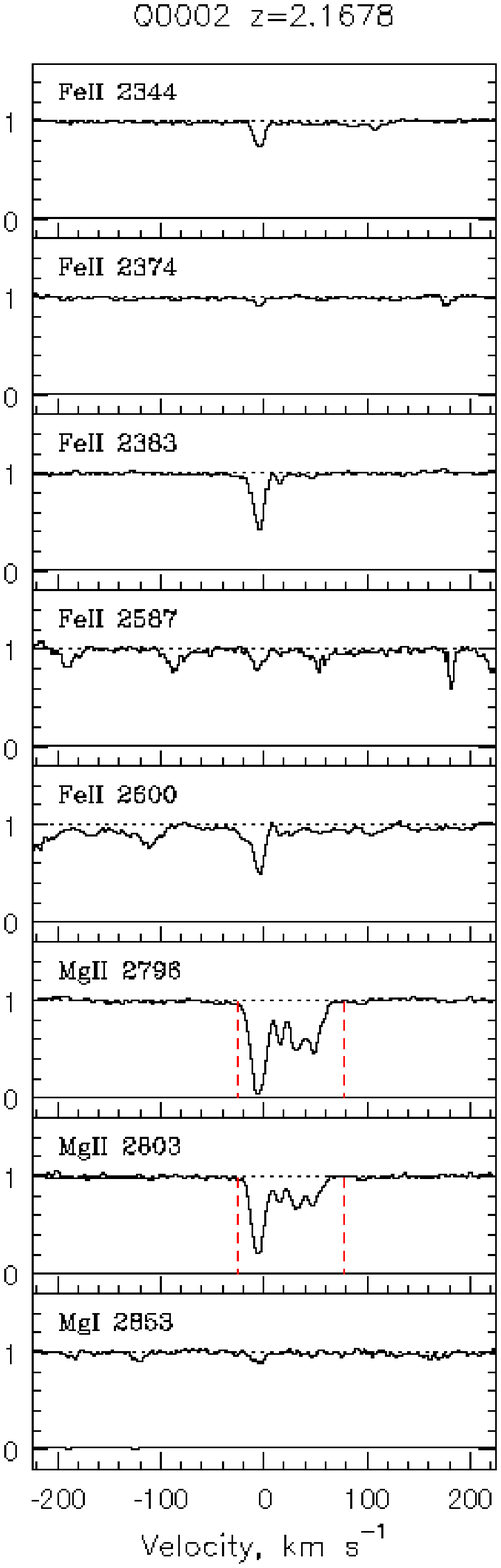}
\caption{}
\label{fig:fig1f}
\end{figure}

This system has four distinct minima in the {\MgII} profiles, the
strongest of which is nearly saturated in {\MgII}~$\lambda$2796.  The
profiles can be seen in Figure \ref{fig:fig1f}.  All components
appear to be detected in {\FeII}, though several transitions are in
regions of the spectrum contaminated by atmospheric absorption.  The
{\MgI} window contains a detection at the expected location of the
strongest absorption, but this portion of the spectrum is also
significantly contaminated by telluric features, so we cannot measure
{\MgI} accurately.

\subsubsection{Q0002-422 $z_{abs}=2.3019$}
\label{sec:2.3019}

\begin{figure}
\figurenum{1g}
\centering
\vspace{0.0in}
\epsscale{0.45}
\plotone{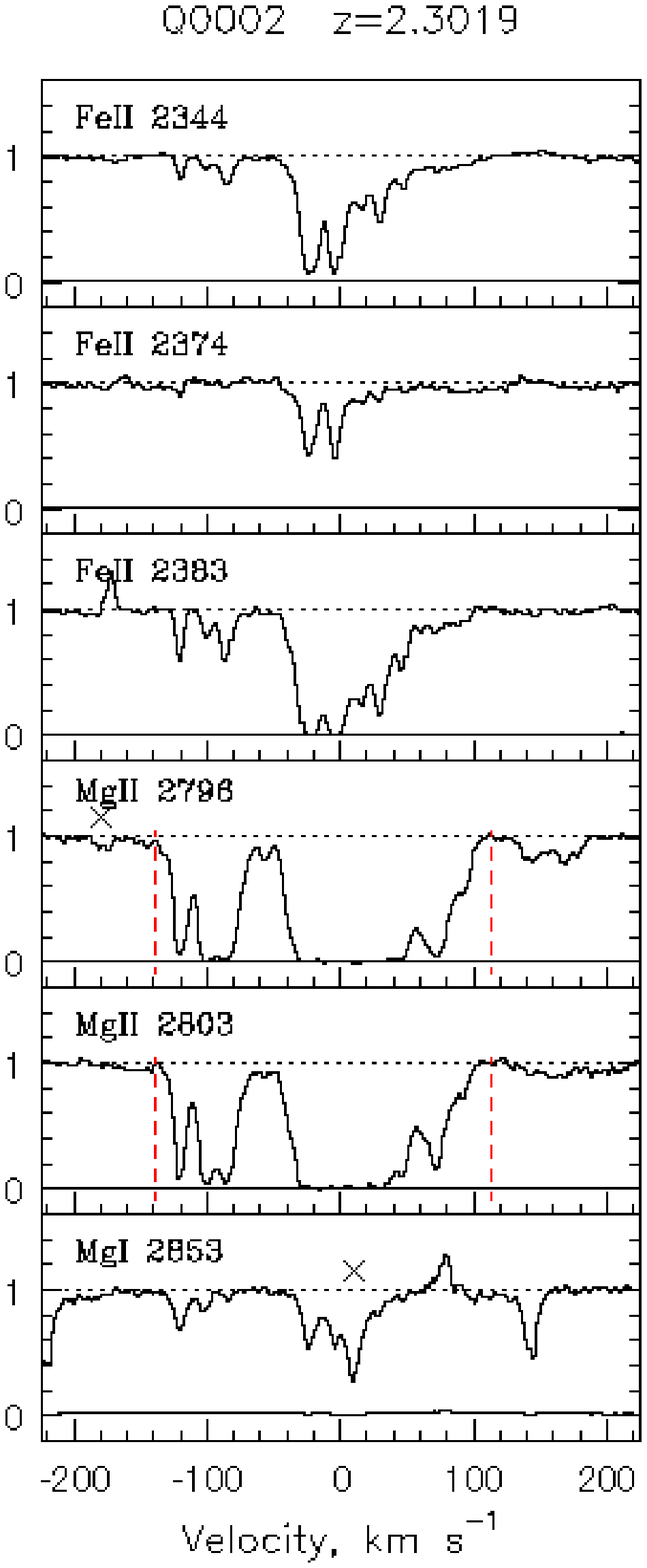}
\caption{}
\label{fig:fig1g}
\end{figure}

This very strong, saturated {\MgII} absorber resembles the
``superwind profiles" given in \citet{bond01}.  The profile can be seen
in Figure \ref{fig:fig1g}.  Kinematically, it has one very broad,
mostly saturated, region centered at $0$~{\kms}, and another saturated
region centered at $\sim -100$~{\kms}.  The absorption between these
two regions is very weak but does not fully recover.  Thus, this is
formally classified as having only one subsystem.  The
{\FeII}~$\lambda$2587 and $\lambda$2600 transitions were not covered
by the spectra.  The other three {\FeII} transitions show detections
in the expected velocity range.  {\MgI}~$\lambda$2853 is also detected
over most of the velocity range but is contaminated by atmospheric
absorption, particularly at $\sim 10$~{\kms}.

\subsubsection{Q0002-422 $z_{abs}=2.4641$}
\label{sec:2.4641}

\begin{figure}
\figurenum{1h}
\centering
\vspace{0.0in}
\epsscale{0.45}
\plotone{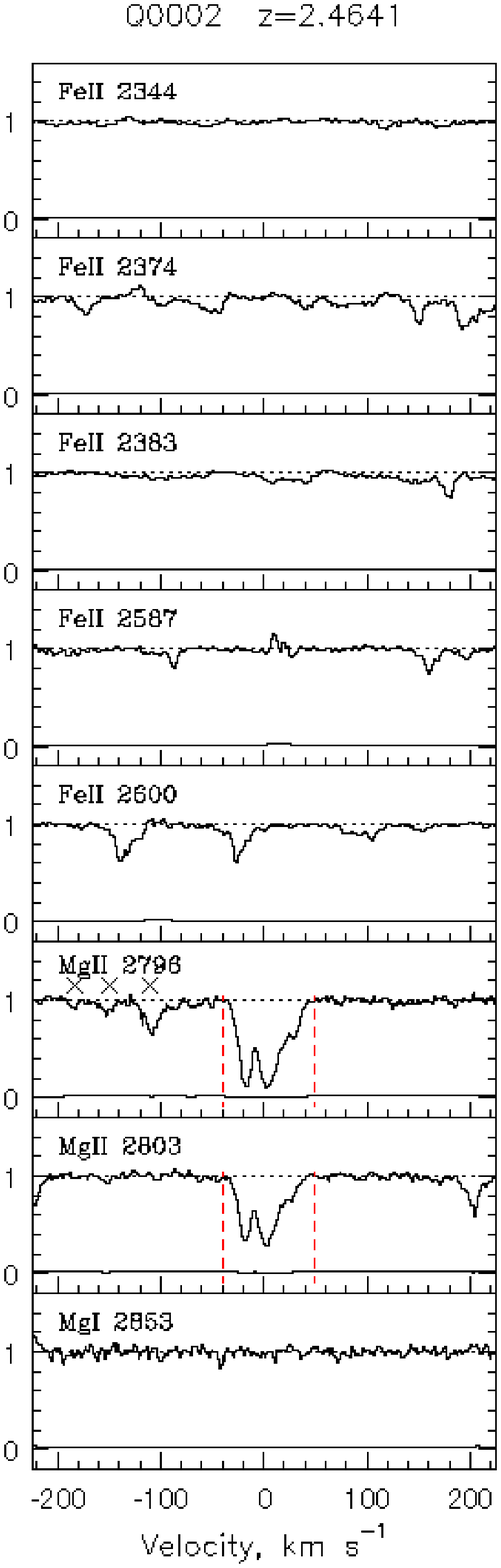}
\caption{}
\label{fig:fig1h}
\end{figure}

This double horned system, shown in Figure \ref{fig:fig1h}, consists
of just one subsystem.  There are several additional features, in the
range of $100$-$400$~{\kms} from the dominant absorption, in the
{\MgII}~$\lambda$2796 and in the {\MgII}~$\lambda$2803 windows.
However, the identity of each as {\MgII} is refuted by the other
transition.  Other very weak features are present at higher velocities
(more than $400$~{\kms} separated from $z_{abs}$).  In principle, they
could have contributions from {\MgII} absorption, however, they are
strongly blended with atmospheric absorption, and would be at
significantly higher velocities than is seen in any other systems.
These ``very high velocity" features are, therefore, believed to be
atmospheric and not {\MgII} associated with this system. Neither the
{\MgI}~$\lambda$2853, nor any of the {\FeII} transitions show clear
detections in somewhat contaminated spectral regions, but reasonable
limits are provided.

\subsubsection{Q0109-3518 $z_{abs}=1.3495$}
\label{sec:1.3495}

\begin{figure}
\figurenum{1i}
\centering
\vspace{0.0in}
\epsscale{0.45}
\plotone{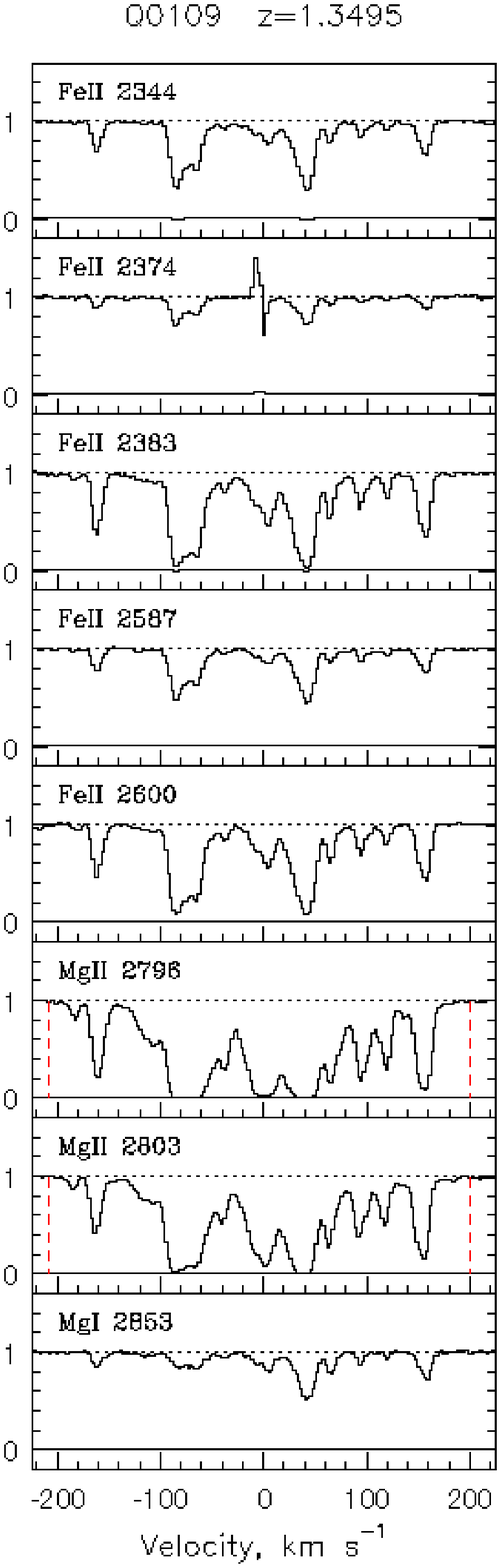}
\caption{}
\label{fig:fig1i}
\end{figure}

Kinematically, this system, which can be seen in Figure
\ref{fig:fig1i} bears a strong resemblence to the $z_{abs}=2.1844$
system toward HE$0001-2340$.  However, it consists of only one very
broad subsystem and it has saturation in some components.  Still,
there could be widespread, merger-induced star formation (as suggested
by \citet{richter05} for the $z_{abs}=2.1844$ system toward HE$0001-2340$)
causing the large kinematic spread in this case as well.  The
{\FeII}~$\lambda$2374 transition is affected by poor sky subtraction, but all
{\FeII} transitions show detections over the same velocity range as the
{\MgII}. {\MgI}~$\lambda$2853 is also detected over this velocity range.

\subsubsection{Q0122-380 $z_{abs}=0.4437$}
\label{sec:0.4437}

\begin{figure}
\figurenum{1j}
\centering
\vspace{0.0in}
\epsscale{0.45}
\plotone{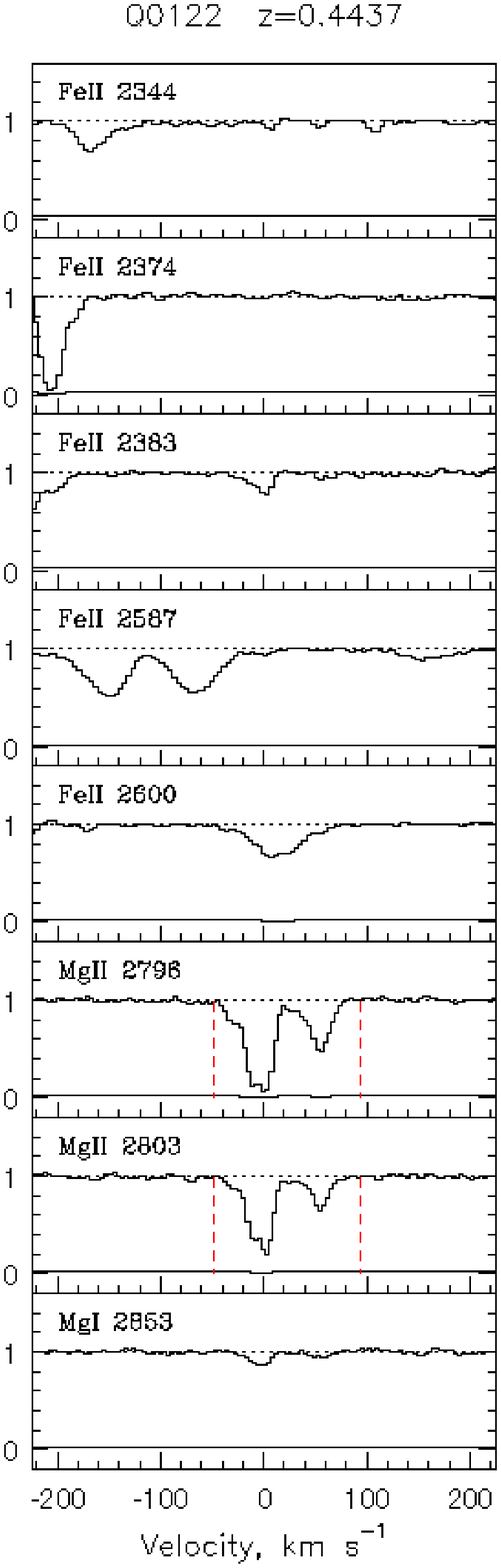}
\caption{}
\label{fig:fig1j}
\end{figure}

This system has a single subsystem that includes a strong, nearly
saturated component at zero velocity, with a blue wing, and a weaker
redward component.  {\FeII} and {\MgI} are detected in all components.
However, most of the {\FeII} transitions exhibit some blending with
only the $\lambda$2383 transition not affected in the absorbing region
and exhibitng absorption.  All of these transitions can be seen in
Figure
\ref{fig:fig1j}.

\subsubsection{Q0122-380 $z_{abs}=0.8597$}
\label{sec:0.8597}

\begin{figure}
\figurenum{1k}
\centering
\vspace{0.0in}
\epsscale{0.45}
\plotone{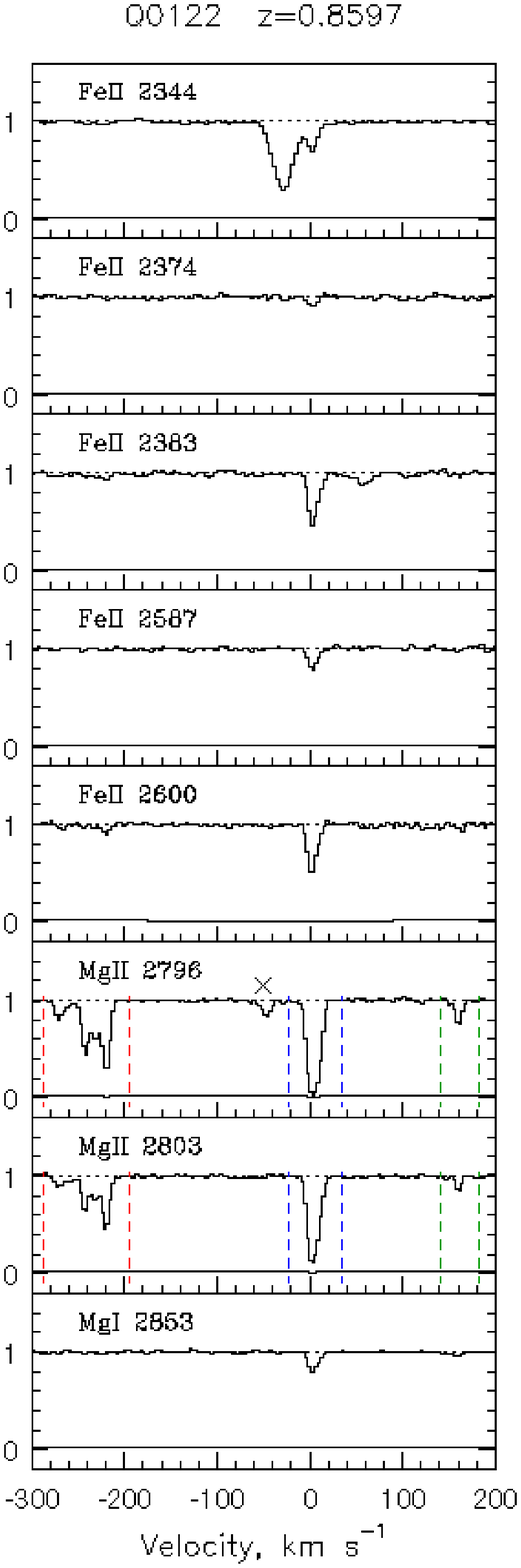}
\caption{}
\label{fig:fig1k}
\end{figure}

The central subsystem of this absorber is deep and narrow, and is
saturated in {\MgII}~$\lambda$2796, but not in {\MgII}~$\lambda$2803.
The system is shown in Figure \ref{fig:fig1k}.  The bluest subsystem
is shallow and wide with multiple, narrow components.  A third,
redward subsystem is significantly weaker than the other two.  A
feature just blueward of the central subsystem in
{\MgII}~$\lambda$2796 is not matched in the {\MgII}~$\lambda$2803 and
is, therefore, not physical.  The central subsystem is detected in
{\FeII} and {\MgI}.  The {\FeII}~$\lambda$2344 has a blend at
$\sim$-30 {\kms} due to {\CIV}~$\lambda$1548 at z = 1.8149.  The
blueward subsystem is detected in {\FeII}, but not in {\MgI}.  The
redward subsystem appears to be detected in {\MgI} and in {\FeII}, but
only at a $3\sigma$ level.  If it is a real detection, {\MgI} would be
quite strong relative to {\MgII}.  This system could be generated by a
similar physical process as gives rise to the $z_{abs}=2.1844$ system
toward HE$0001-2340$, but in this case it would have to produce fewer
components over the same velocity range.

\subsubsection{Q0122-380 $z_{abs}=1.2439$}
\label{sec:1.2439}

\begin{figure}
\figurenum{1l}
\centering
\vspace{0.0in}
\epsscale{0.45}
\plotone{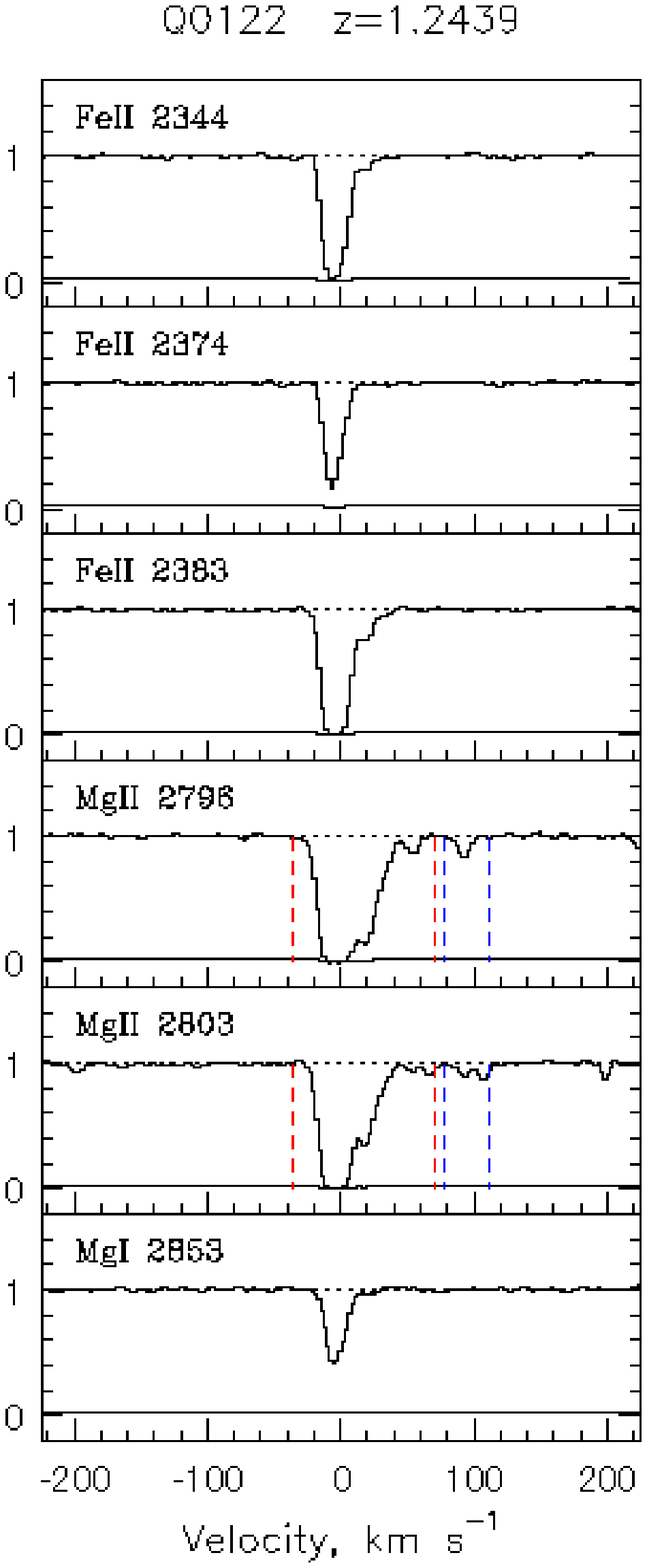}
\caption{}
\label{fig:fig1l}
\end{figure}

This system, displayed in Figure \ref{fig:fig1l}, has one dominant
subsystem and one smaller redward component, which classifies as a
separate subsystem.  The smaller component (at $\sim 90$~{\kms}) is
confirmed in the {\MgII}~$\lambda$2803.  The other detections in the
{\MgII}~$\lambda$2803 panel are known atmospheric absorption features.
The {\FeII}~$\lambda$2587 and $\lambda$2600 transitions are not covered by the
spectra.  The other three {\FeII} transitions as well as the {\MgI}
transition show detections for the dominant subsystem, but not for the
redward subsystem.

\subsubsection{HE0151-4326 $z_{abs}=0.6630$}
\label{sec:0.6630}

\begin{figure}
\figurenum{1m}
\centering
\vspace{0.0in}
\epsscale{0.45}
\plotone{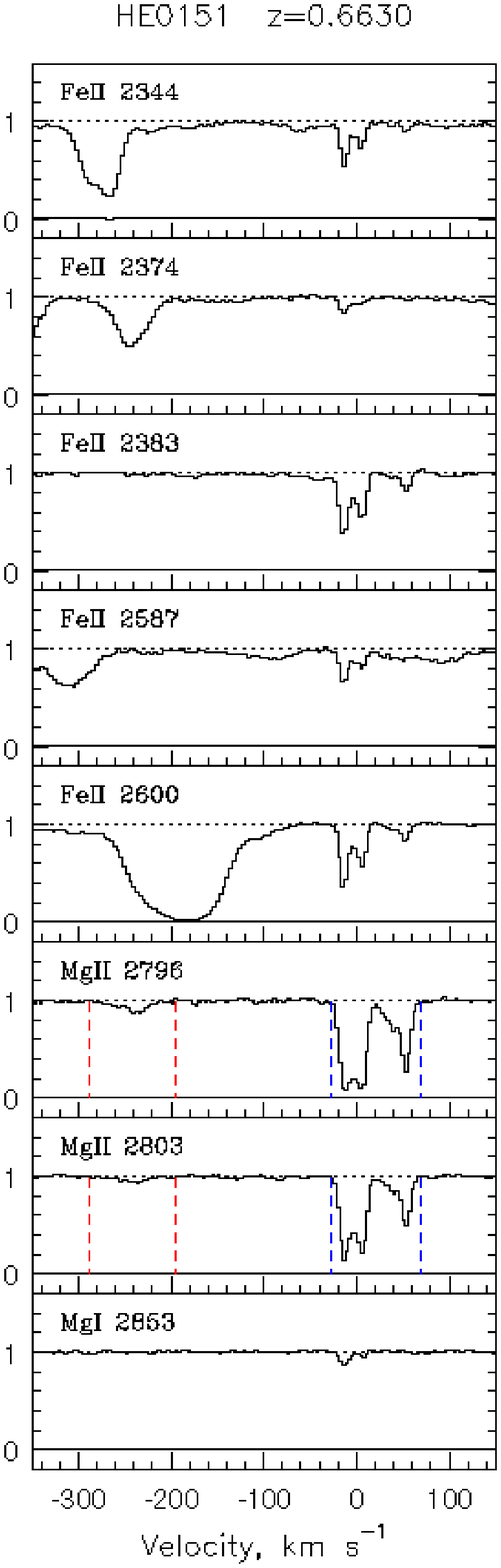}
\caption{}
\label{fig:fig1m}
\end{figure}

This {\MgII} profile has a subsystem with several strong components
(from $\sim-25$ to $\sim 60$~{\kms}), and a broad, weak component at
$\sim$-240~{\kms}.  All relevant transitions are shown in Figure
\ref{fig:fig1m}.  The {\FeII} transitions are in the {\Lya} forest of
this quasar, but by combining information from all of them we measure
{\FeII} over the full velocity range of this system.  While the strong
subsystem is detected in {\FeII}, the blueward subsystem is not.
{\MgI}~$\lambda$2853 absorption is detected only in the strongest two
components, and is considerably stronger in the blueward of those (at
$\sim -16$~{\kms}).  The broad component at $\sim -240$~{\kms} is of
particular interest.  Such broad, weak features, which may indicate
bulk motion, may be more common than we realize, since they are only
detectable in high $S/N$ spectra.

\subsubsection{PKS0237-23 $z_{abs}=1.3650$}
\label{sec:1.3650}

\begin{figure}
\figurenum{1n}
\centering
\vspace{0.0in}
\epsscale{0.45}
\plotone{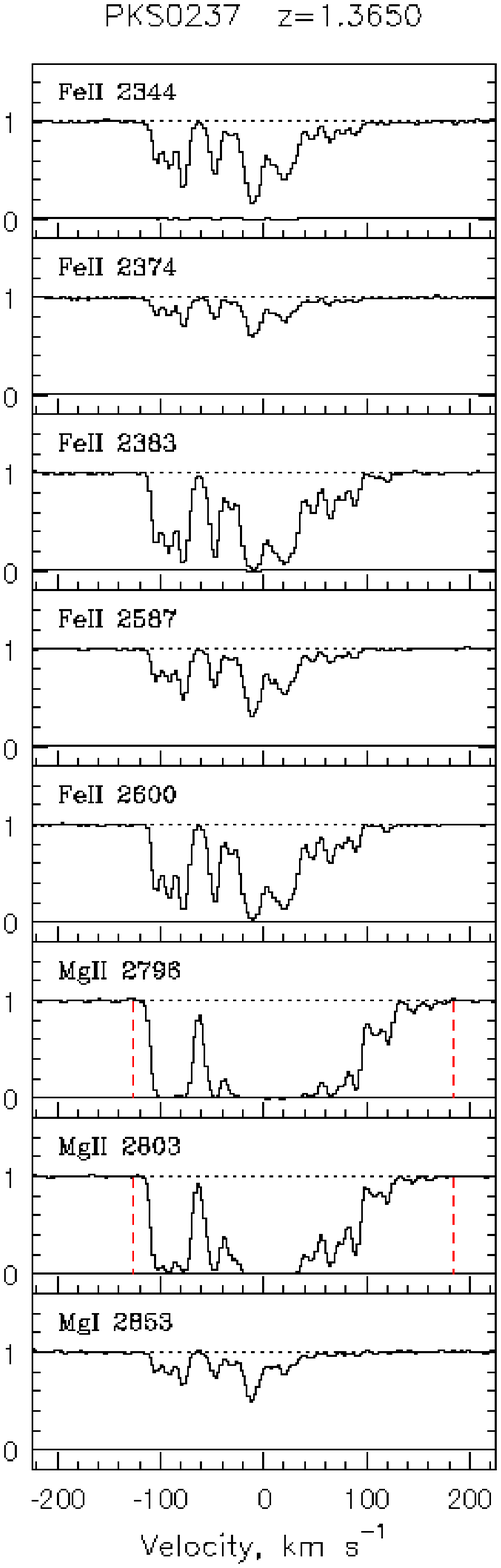}
\caption{}
\label{fig:fig1n}
\end{figure}

This system consists of one broad subsystem, with numerous components,
many of which are saturated.  The absorption profiles of {\MgII},
{\FeII}, and {\MgI} are shown in Figure \ref{fig:fig1n}.  Several very
weak components are apparent at $\sim 140$--$160$~{\kms}, the reddest
part of the single subsystem.  Both {\FeII} and {\MgI} are detected in all
but the weakest components of this absorber.

\subsubsection{PKS0237-23 $z_{abs}=1.6371$}
\label{sec:1.6371}

\begin{figure}
\figurenum{1o}
\centering
\vspace{0.0in}
\epsscale{0.45}
\plotone{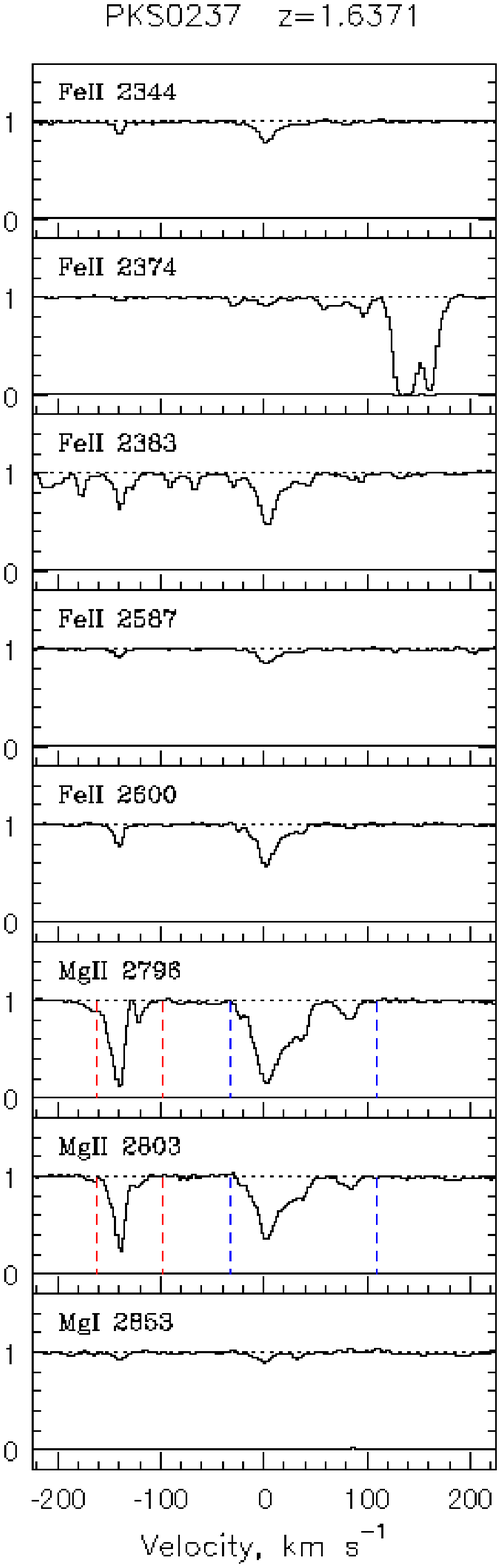}
\caption{}
\label{fig:fig1o}
\end{figure}

This system, shown in Figure \ref{fig:fig1o}, consists of one central,
broad, multicomponent subsystem, a blueward subsystem with one deeper
narrow component, and a weak component just redward.  Both subsystems
are detected in {\FeII} and {\MgI}.  The {\FeII}~$\lambda$2374
contains a blend outside of the absorbing region.  The
{\FeII}~$\lambda$2383 is affected by telluric absorption.  An
interesting note about this system: the zero velocity component is not
as close to being saturated as the blueward subsystem, which is
relatively strong in comparison.  This could signify a double galaxy
(perhaps giant/dwarf) configuration along the line of sight.

\subsubsection{PKS0237-23 $z_{abs}=1.6574$}
\label{sec:1.6574}

\begin{figure}
\figurenum{1p}
\centering
\vspace{0.0in}
\epsscale{0.45}
\plotone{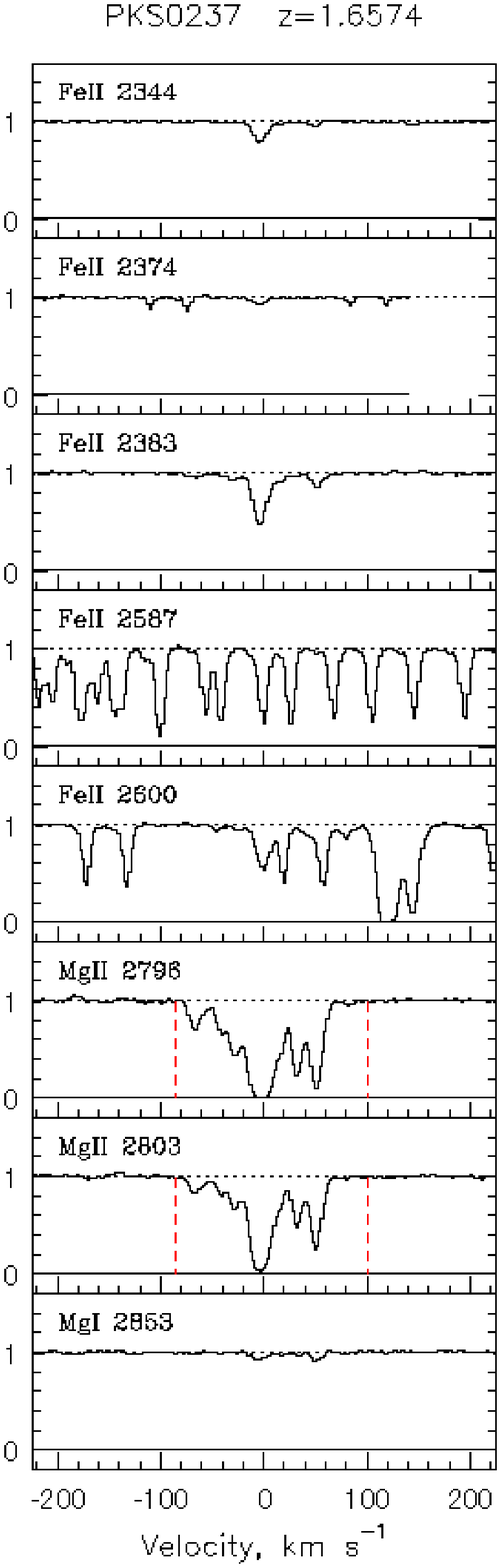}
\caption{}
\label{fig:fig1p}
\end{figure}

The one broad, saturated subsystem in this {\MgII} profile consists of
multiple components.  The profile is displayed in Figure
\ref{fig:fig1p}.  There are several components both blueward and redward of
the dominant one. The two strongest components are detected in {\MgI}
and {\FeII}.  However, the {\FeII}~$\lambda$2587 transition is heavily
affected by telluric absorption.  The {\FeII}~$\lambda$2600 contains a
feature outside of the absorbing region, which does not affect our
analysis but is noted for completeness.

\subsubsection{PKS0237-23 $z_{abs}=1.6723$}
\label{sec:1.6723}

\begin{figure}
\figurenum{1q}
\centering
\vspace{0.0in}
\epsscale{0.45}
\plotone{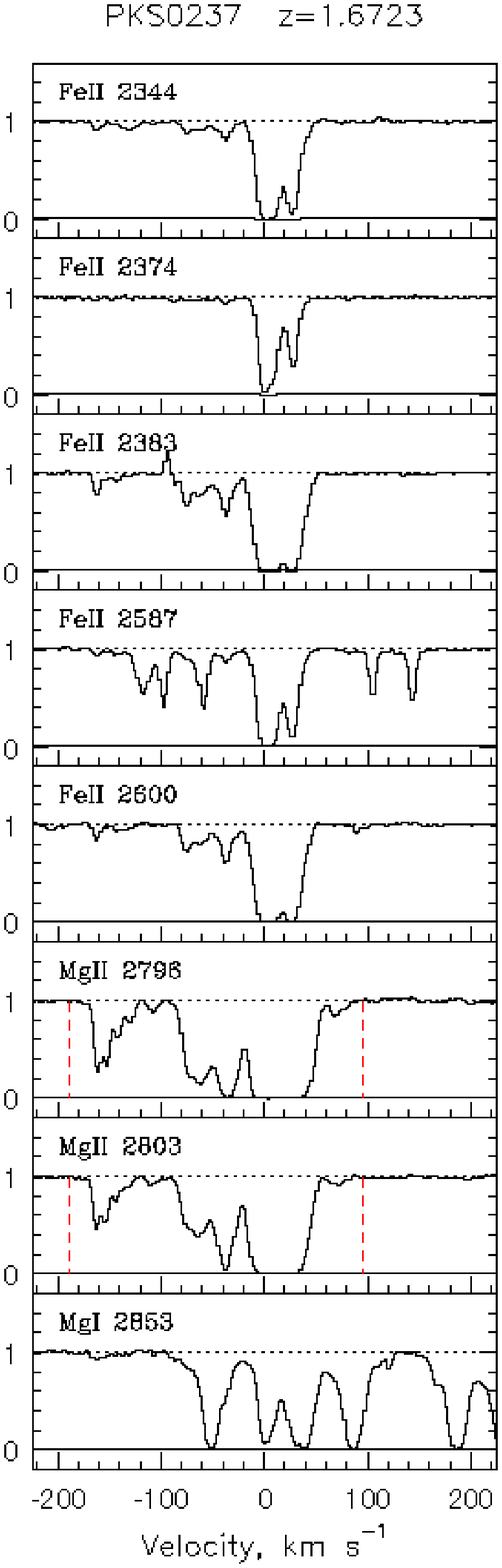}
\caption{}
\label{fig:fig1q}
\end{figure}

This system is saturated over the range $-10 < v < 35$~{\kms} and
consists of one subsystem with multiple components.  The system can be
seen in Figure \ref{fig:fig1q}.  This could be a ``superwind system",
with a weak blueward region.  While the {\MgI} transition is too
severely blended, the {\FeII} provides useful constraints despite
telluric absorption in the $\lambda$2587 transition.  For example, the
absorption in {\FeII} is strongest in the blueward portion of the
saturated region.

\subsubsection{Q0329-385 $z_{abs}=0.7627$}
\label{sec:0.7627}

\begin{figure}
\figurenum{1r}
\centering
\vspace{0.0in}
\epsscale{0.45}
\plotone{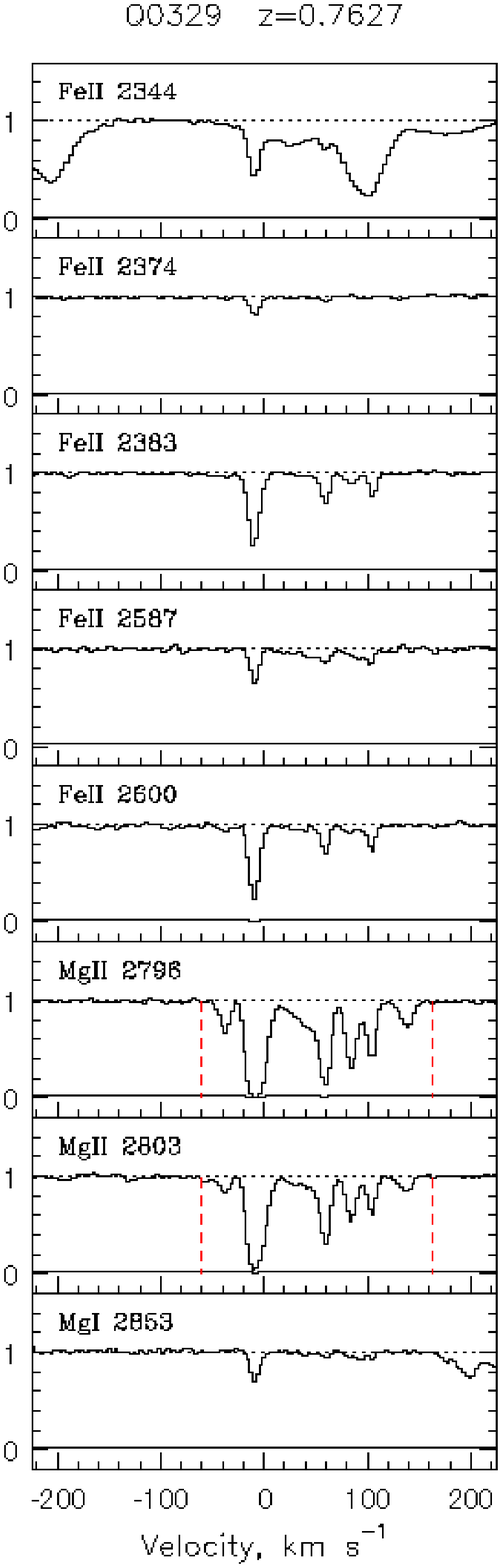}
\caption{}
\label{fig:fig1r}
\end{figure}

This system, shown in Figure \ref{fig:fig1r}, contains multiple
components with only one saturated component in {\MgII}~$\lambda$2796.
There is evidence of absorption in {\FeII} for all but the reddest
component.  The {\FeII}~$\lambda$2344 transition is blended due to
lines in the {\Lya} forest.  There is {\MgI}~$\lambda$2853 detected
for the strongest component.  The {\MgI} for other components would be
blended with the {\CIV}~$\lambda$1548 absorption from a system at
$z_{abs}=2.2513$.  The $z_{abs}=0.7267$ system is characterized by a dominant
(though narrow) component with nearly all of the weaker components
(save one) redward of this component.

\subsubsection{Q0329-385 $z_{abs}=1.4380$}
\label{sec:1.4380}

\begin{figure}
\figurenum{1s}
\centering
\vspace{0.0in}
\epsscale{0.45}
\plotone{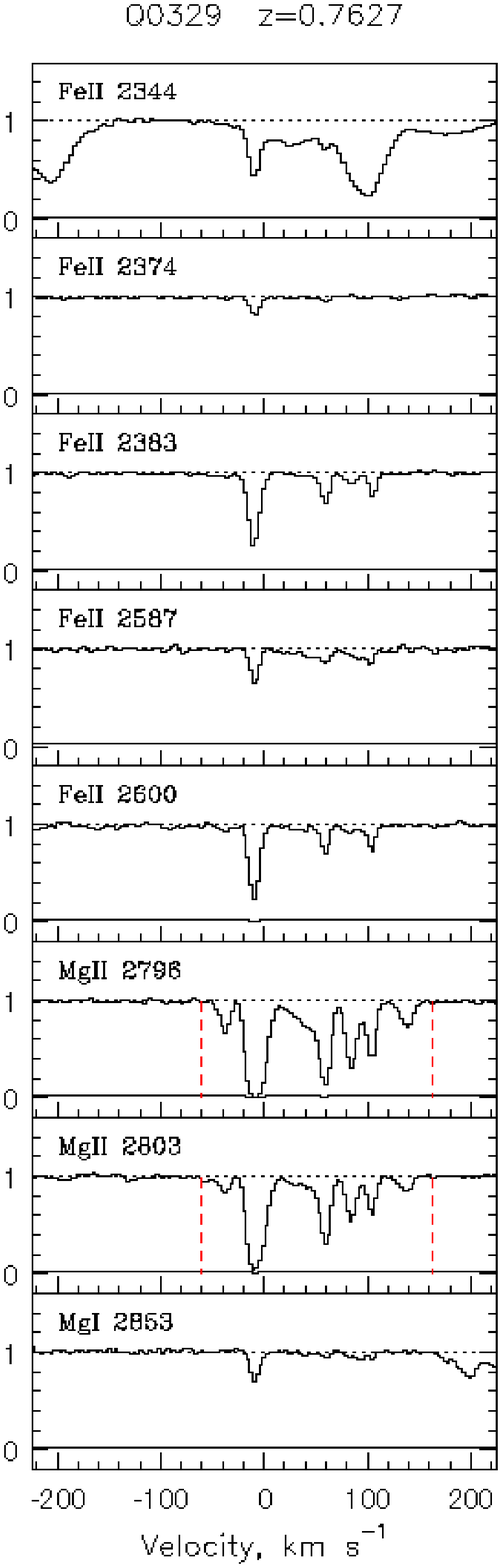}
\caption{}
\label{fig:fig1s}
\end{figure}

The central region of this absorber contains two components, both of
which are narrow.  The absorption profile can be seen in Figure
\ref{fig:fig1s}.  The bluest component of this system is also narrow
and classifies as a separate subsystem.  The reddest region of the
absorber consists of two narrow components combined with a broader
component or set of components. The {\MgII}~$\lambda$2803 transition
is affected by poor sky subtraction at $\sim 40$~{\kms} which does not
significantly impact our analysis, but is noted for completeness.  The
$\lambda$~2374 and $\lambda$~2383 transitions of {\FeII} are not
covered by the spectra.  All of the narrow components are detected in
{\FeII}.  However, {\MgI}~$\lambda$2853 is detected only in the
strongest component.  Limits for the weaker component are not strict
because this region of the spectrum is contaminated by atmospheric
absorption.

\subsubsection{Q0453-423 $z_{abs}=0.7261$}
\label{sec:0.7261}

\begin{figure}
\figurenum{1t}
\centering
\vspace{0.0in}
\epsscale{0.45}
\plotone{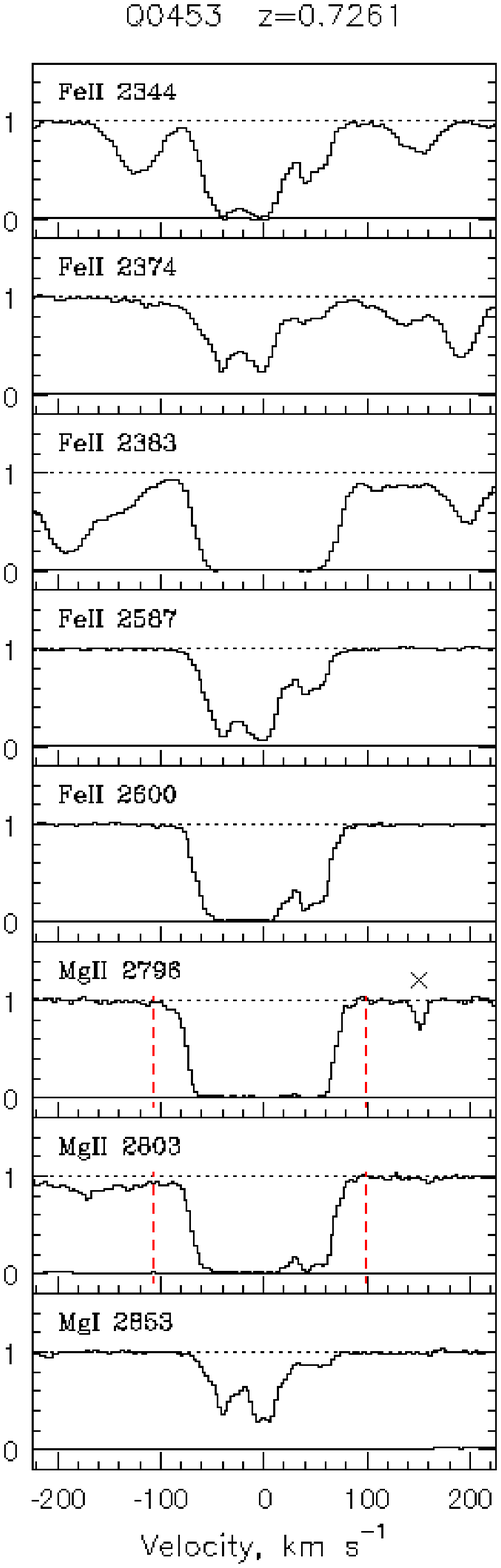}
\caption{}
\label{fig:fig1t}
\end{figure}

This nearly ``square profile'' has a velocity range of $\sim
-70$~{\kms} to $\sim 60$~{\kms}.  The feature in the
{\MgII}~$\lambda$2796 at $v\sim140$~{\kms} is due to
{\SiIV}~$\lambda$1402 at $z_{abs} = 2.44264$.  Similarly, although the very
weak feature in the {\MgII}~$\lambda$2803 panel, at $v\sim 162$~{\kms}
is a $5\sigma$ detection, the red portion of its profile does not have
sufficient corresponding absorption in {\MgII}~$\lambda$2796 to
confirm it as {\MgII}.  If either of these features were real, they
would represent an extra subsystem, but because of the lack of
confirmation with both doublet members, they are convincingly ruled
out.  The dominant subsystem is detected in {\MgI} and {\FeII}.  All
relevant transitions can be seen in Figure
\ref{fig:fig1t}.

In a study of metallicity and abundance
pattern, \citet{ledoux02} argue that this system is likely to be a
sub-DLA or DLA, though it is impossible to measure this directly due
to the full Lyman limit break from the $z_{abs}=2.3045$ system.

\subsubsection{Q0453-423 $z_{abs}=0.9085$}
\label{sec:0.9085}

\begin{figure}
\figurenum{1u}
\centering
\vspace{0.0in}
\epsscale{0.45}
\plotone{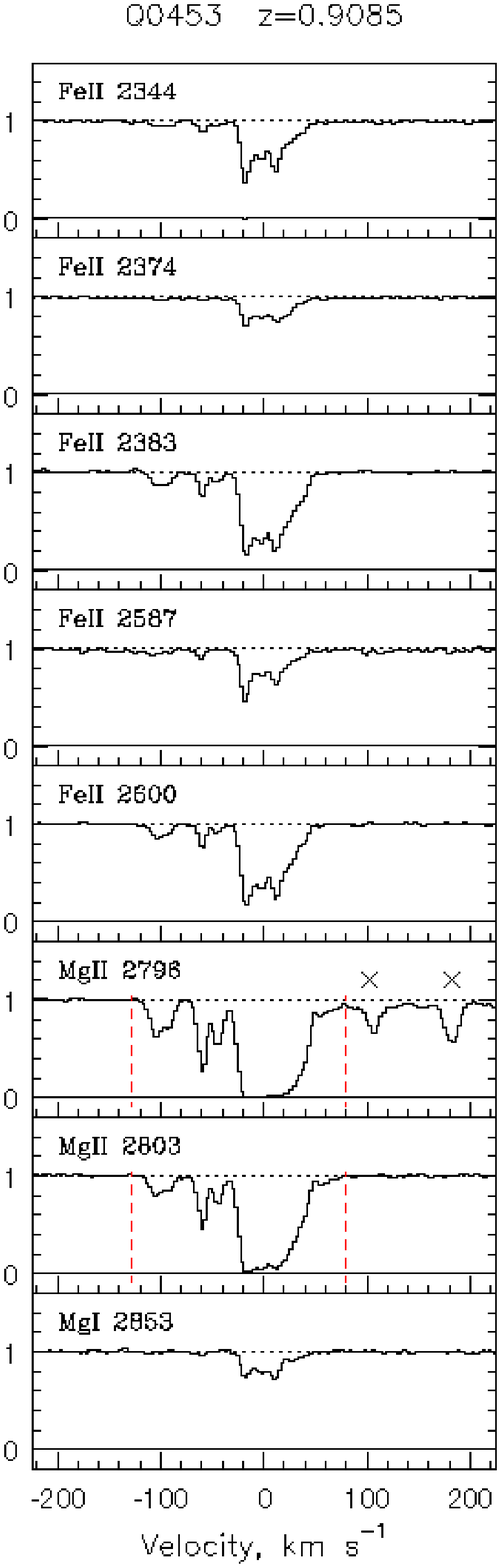}
\caption{}
\label{fig:fig1u}
\end{figure}

This multiple component system, displayed in Figure \ref{fig:fig1u},
consists of one strong broad saturated component, two closely spaced
narrow components blueward of the system, and a broad component
further blueward.  The two features redward of the system are due to a
{\CIV}~$\lambda$1550 system at z = 2.4435.  All components of the
system are detected in {\FeII}.  {\MgI} is detected in the saturated
region and in the strongest narrow component.

\subsubsection{Q0453-423 $z_{abs}=1.1498$}
\label{sec:1.1498}

\begin{figure}
\figurenum{1v}
\centering
\vspace{0.0in}
\epsscale{0.5}
\plotone{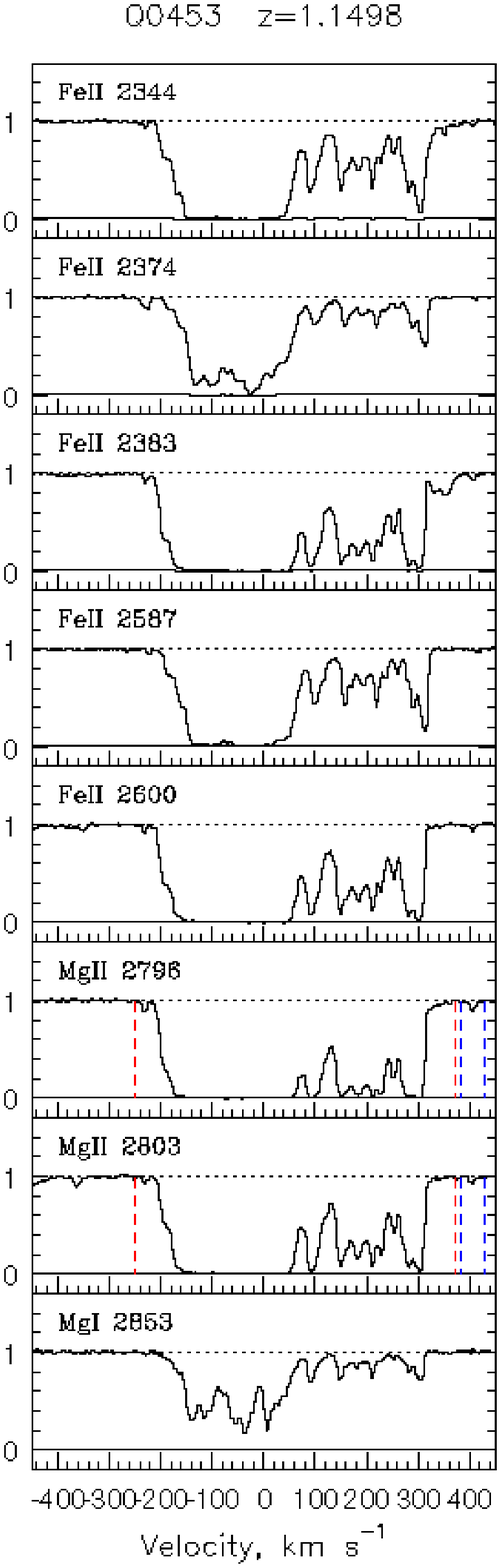}
\caption{}
\label{fig:fig1v}
\end{figure}

This single component system is shown in Figure \ref{fig:fig1v} and
has an unusually large equivalent width ($W_r(2796)\sim 4.5$~{\AA}).
All components are detected in {\FeII} as well as the {\MgI}
transition.  The bluer half of this system contains the majority of
the metals as evidenced by its saturation in the weaker {\FeII}
transitions, in which the redder half of the absorber becomes
unsaturated.  \citet{ledoux02} have determined metallicities and
abundance patterns for this system, finding a super-solar metallicity.
It is likely to be a sub-DLA or DLA system, but the Lyman break region
cannot be measured due to the full Lyman limit break from the
$z_{abs}=2.3045$ system.

\subsubsection{Q0453-423 $z_{abs}=1.6299$}
\label{sec:1.6299}

\begin{figure}
\figurenum{1w}
\centering
\vspace{0.0in}
\epsscale{0.45}
\plotone{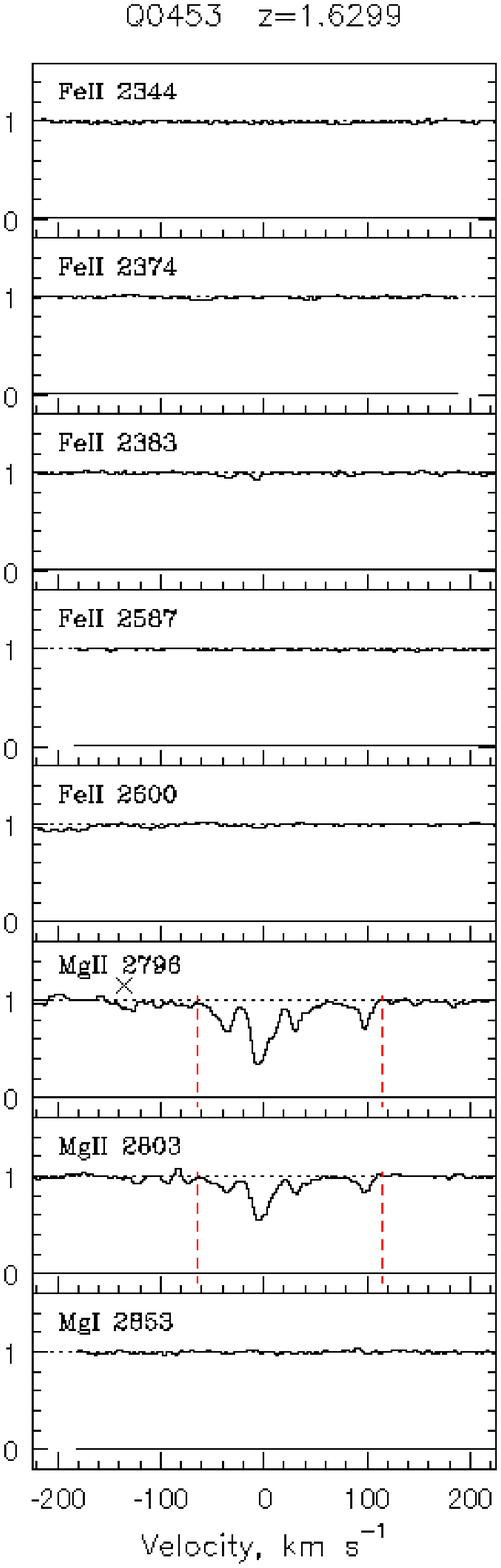}
\caption{}
\label{fig:fig1w}
\end{figure}

This single subsystem absorber consists of four narrow components in
{\MgII}, and perhaps some blended, weaker broad ones.  The absorption
profile can be seen in Figure \ref{fig:fig1w}.  The region of the
spectrum where {\MgII} is detected is contaminated by atmospheric
absorption features, leading to a number of detected features around
both the {\MgII}~$\lambda$2796 and {\MgII}~$\lambda$2803 profiles.
However, all of these features could be eliminated as {\MgII} by
examining the corresponding position in the other transition. Also,
in the region at 170 $< v <$ 185~{\kms}, the {\MgII}~$\lambda$2803
profile is a bit weak relative to {\MgII}~$\lambda$2796, so there
could actually be two subsystems here, though it is more
likely that there is just one.  {\FeII}
is detected in only the two strongest components while {\MgI} is not
detected in this system.

\subsubsection{Q0453-423 $z_{abs}=2.3045$}
\label{sec:2.3045}

\begin{figure}
\figurenum{1x}
\centering
\vspace{0.0in}
\epsscale{0.45}
\plotone{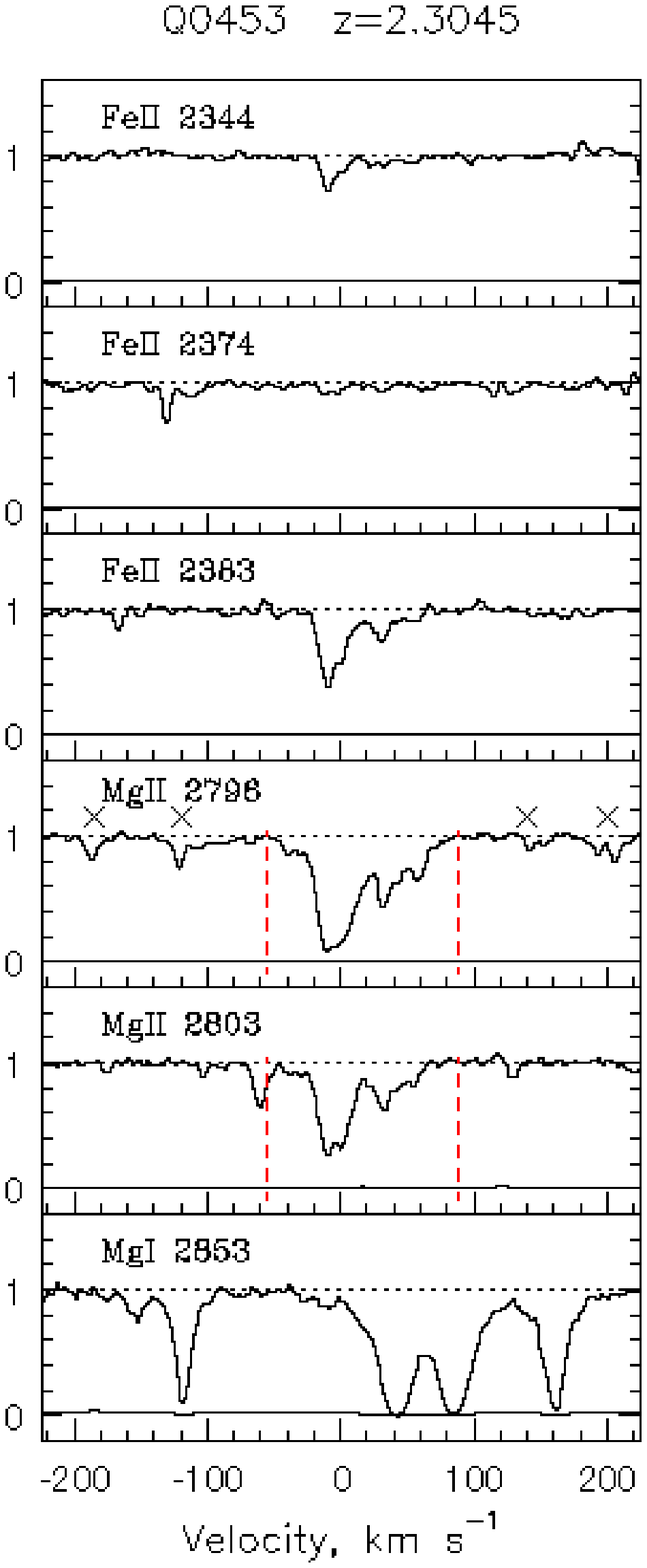}
\caption{}
\label{fig:fig1x}
\end{figure}

This system consists of one multiple component subsystem and is
displayed in Figure \ref{fig:fig1x}.  Based on an {\it HST}/FOS
spectrum, the system produces a Lyman limit break
\citep{ledoux02}, but the {\Lya} profile is not damped, with
$\log N({\HI})\sim19.2$~{\cmsq}.
Of the features both redward and blueward of the
system in the {\MgII}~$\lambda$2796 panel, none are detected in the
{\MgII}~$\lambda$2803, indicating that these do not represent
additional {\MgII} absorption and, therefore, do not represent
additional subsystems; these features are due to atmospheric
absorption.  The system is detected in {\FeII}, however the
{\FeII}~$\lambda$2587 and $\lambda$2600 transitions are not covered by
the spectra.  The {\MgI} is too severely blended to provide a useful
measurement.

\subsubsection{HE0940-1050 $z_{abs}=1.7891$}
\label{sec:1.7891}

\begin{figure}
\figurenum{1y}
\centering
\vspace{0.0in}
\epsscale{0.45}
\plotone{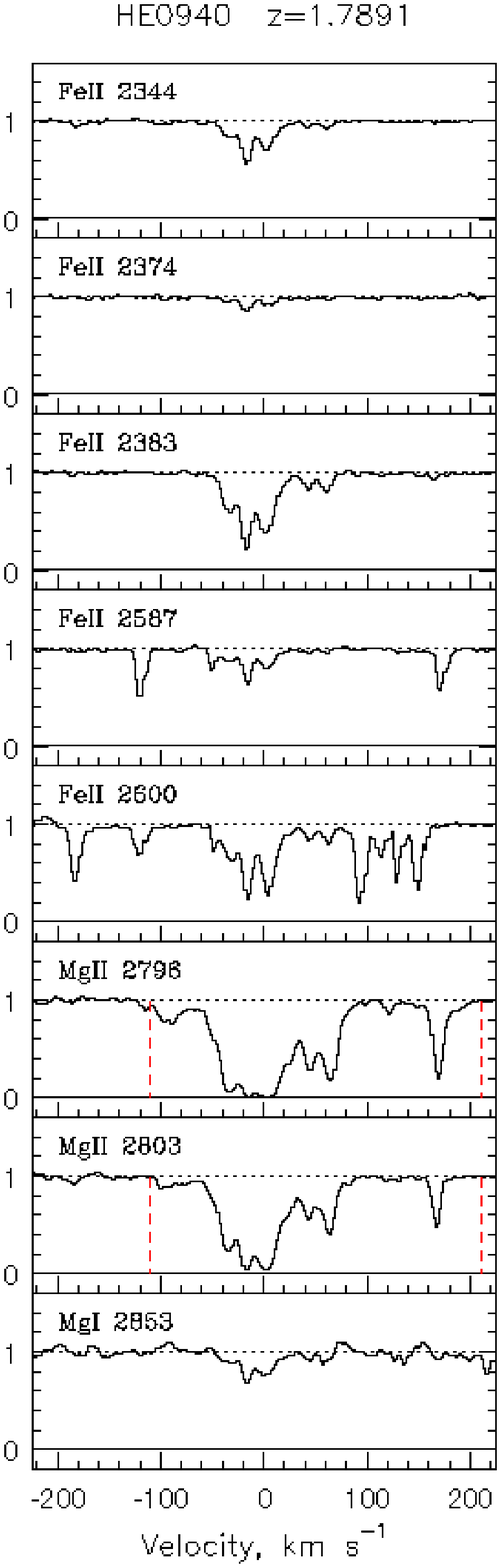}
\caption{}
\label{fig:fig1y}
\end{figure}

This system, seen in Figure \ref{fig:fig1y}, consists of only one
subsystem, but it contains numerous components.  The kinematics of the
components is of particular interest.  For example, a narrow component
is situated at $v\sim 170$~{\kms}, redward of a multiple component
region (which is saturated over a small velocity range).  There is
just enough weak absorption between these two regions that this system
classifies as having only one subsystem.  It appears that the
component at $v\sim 120$~{\kms} in {\MgII}~$\lambda$2796 may not be
matched in the $\lambda$~2803 transition, however there is very weak
absorption detected in that velocity range of the
{\MgII}~$\lambda$2803 panel.  The $v\sim 170$~{\kms} would be a
separate subsystem if the connecting absorption is not real.  The
central region of the system and the $v\sim 170$~{\kms} component are
detected in {\FeII} despite telluric features in the $\lambda$2587 and
$\lambda$2600 transitions.  Only the saturated region is detected
in {\MgI}.  There is no information about the {\Lya} line
for this system because of a Lyman limit break from a system at
$z_{abs}=2.9170$.

\subsubsection{HE1122-1648 $z_{abs}=0.6822$}
\label{sec:0.6822}

\begin{figure}
\figurenum{1z}
\centering
\vspace{0.0in}
\epsscale{0.45}
\plotone{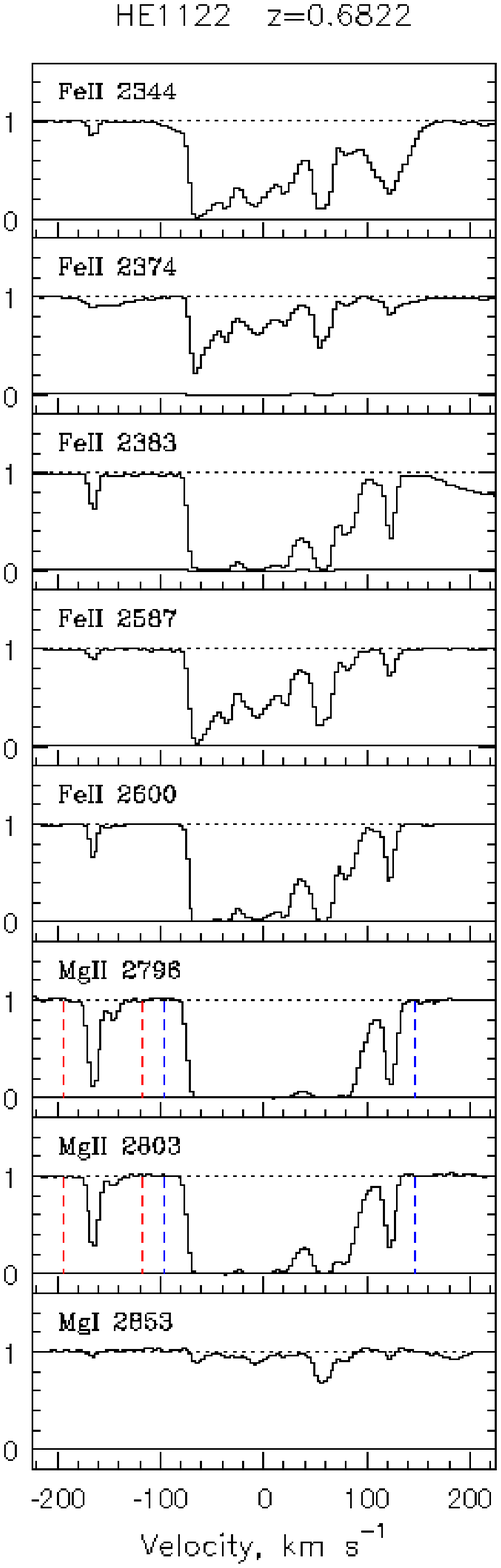}
\caption{}
\label{fig:fig1z}
\end{figure}

This system is similar to the $z_{abs} = 1.7891$ system along the
HE0940-1050 line of sight in that it contains a broad saturated region
with an offset narrow component.  The absorption profile is shown in
Figure \ref{fig:fig1z}.  In this case, the offset component is
classified as a separate subsystem because the flux fully recovers
between the two regions.  All components of the system are detected in
both {\FeII} and {\MgI}.  The {\FeII}~$\lambda$2587 transition
provides the best picture of the true absorption characeristics of the
system because it shows the relative strengths of each component with
only mild saturation in one component.  This system is clearly a
damped {\Lya} absorber \citep{varga00}.  \citet{varga00} used a
Keck/HIRES spectrum to determine that the system has low dust content
and an abundance pattern consistent with an old, metal-poor stellar
population. \citet{ledoux02} further refined the abundance pattern
determinations using the same VLT/UVES spectrum as used here.

\subsubsection{HE1341-1020 $z_{abs}=0.8728$}
\label{sec:0.8728}

\begin{figure}
\figurenum{1aa}
\centering
\vspace{0.0in}
\epsscale{0.45}
\plotone{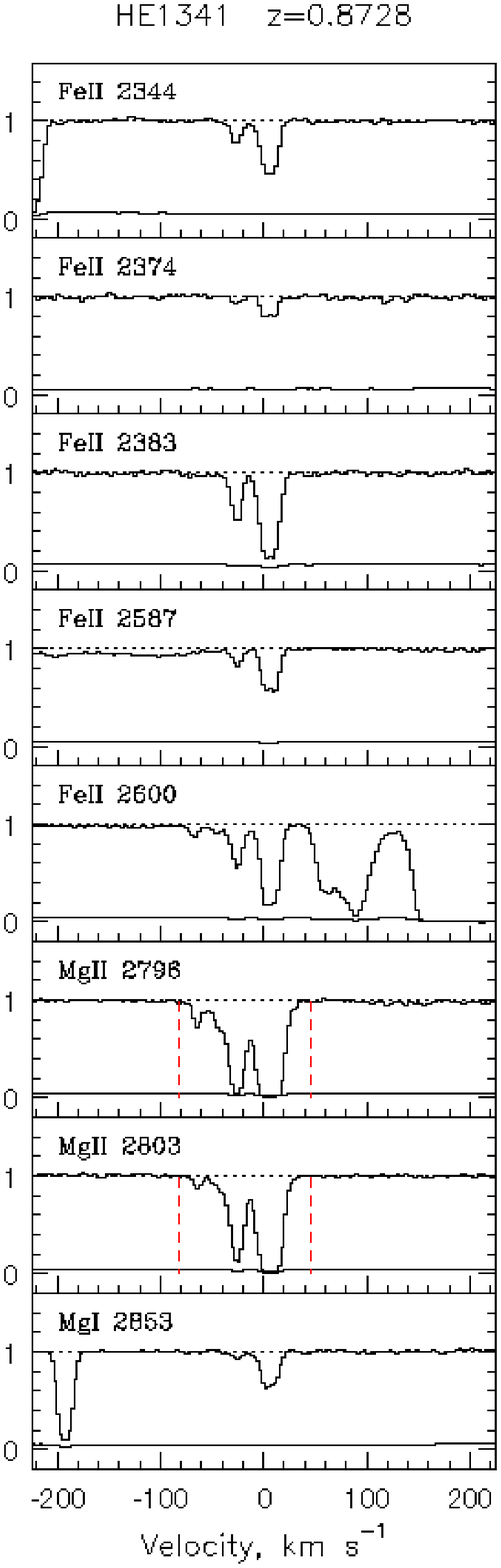}
\caption{}
\label{fig:fig1aa}
\end{figure}

In this system, two of the three well separated components are
saturated.  The system is displayed in Figure \ref{fig:fig1aa}.  The
{\MgII} profile is simple with all of the absorption (not in the
strongest component) located blueward of the strongest component.  The
two strongest components are detected in {\FeII} and {\MgI}.  The
blend in the {\FeII}~$\lambda$2600 is due to a {\CIV}~$\lambda$1548
system at $z_{abs} = 2.1474$.  The feature at $\sim$ -200 {\kms} in the {\MgI}
is due to {\FeII}~$\lambda$2344 at $z_{abs} = 1.2788$.  The weakest component
is detected in {\FeII} but not in {\MgI}.

\subsubsection{HE1341-1020 $z_{abs}=1.2767$}
\label{sec:1.2767}

\begin{figure}
\figurenum{1ab}
\centering
\vspace{0.0in}
\epsscale{0.45}
\plotone{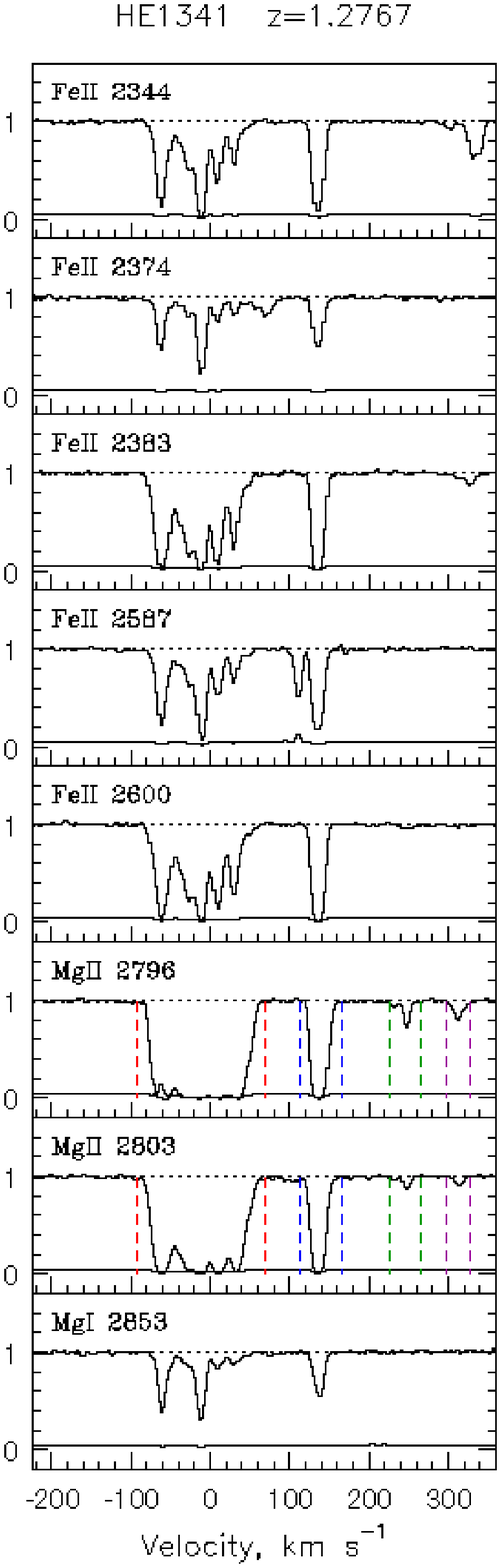}
\caption{}
\label{fig:fig1bb}
\end{figure}

This four subsystem absorber, shown in Figure \ref{fig:fig1bb},
consists of one broad saturated subsystem centered at $\sim 0$~{\kms},
a narrow saturated component centered at $\sim140$~{\kms}, and two
small weak components at $\sim 245$~{\kms} and $\sim 310$~{\kms}.  The
two saturated subsystems are detected in both {\FeII} and {\MgI}, but
the two weaker components are detected in {\MgII} only.  The feature
in the {\FeII}~$\lambda$2383 transition redward of two stronger
subsystems is due to {\AlIII}~$\lambda$1862 at $z_{abs} = 1.9155$.  The
feature at $\sim 280$~--~$360$~{\kms} in the {\FeII}~$\lambda$2344
panel is actually {\MgI}~$\lambda$2853 from the $z_{abs}=0.8728$
absorber.

\subsubsection{PKS2126-158 $z_{abs}=2.0225$}
\label{sec:2.0225}

\begin{figure}
\figurenum{1ac}
\centering
\vspace{0.0in}
\epsscale{0.45}
\plotone{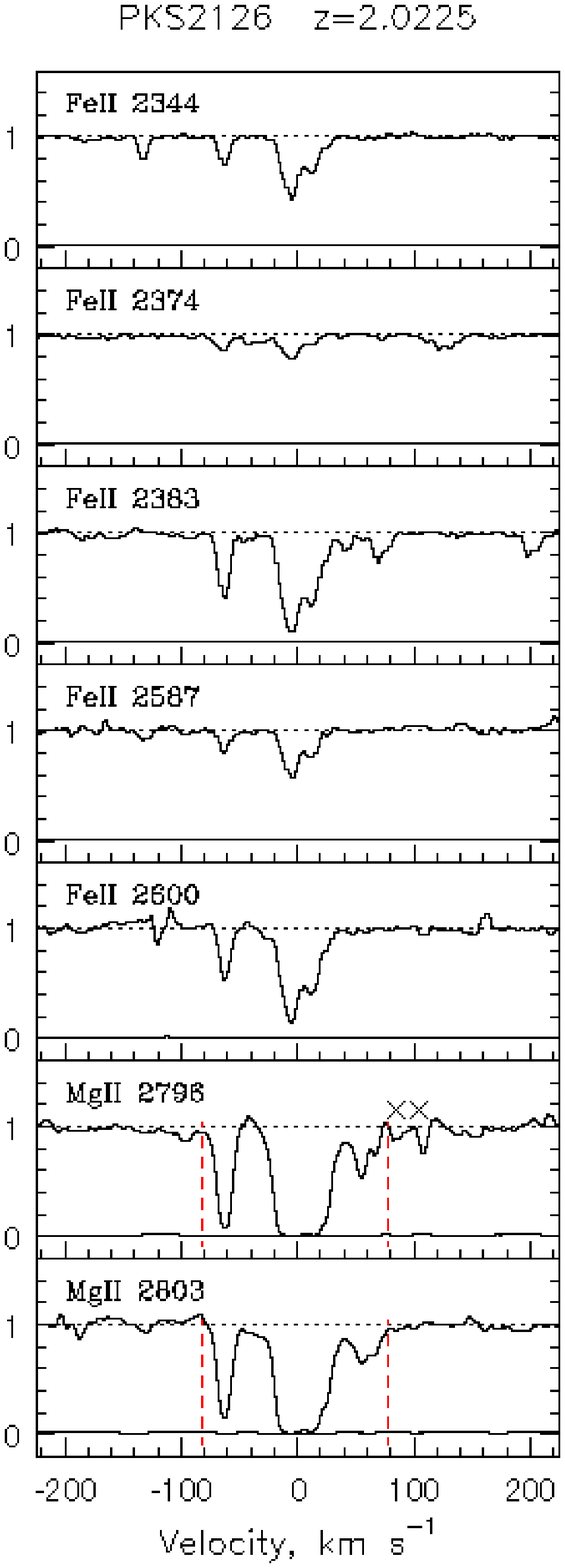}
\caption{}
\label{fig:fig1cc}
\end{figure}

This system can be seen in Figure \ref{fig:fig1cc} and consists of a
narrow component, a multiple component saturated region, and a two
(weak) component region.  Although the system is classified as being
only one subsystem, there are two features redward of the system that
could be due to {\MgII} absorption that can neither be confirmed or
denied because of atmospheric absorption.  One feature is located at
$\sim300$~{\kms} and another at $\sim565$~{\kms}.  Neither is likely
to be due to {\MgII} because of differences in profile shapes at the
expected positions of the $\lambda$2796 and $\lambda$2803 transitions.
Also, it would be quite unusual to have a subsystem separated by such
a large velocity from the subsystem.  The {\MgI}~$\lambda$2853
transition was not covered by the spectra.  {\FeII} is detected in all
but the weakest components.  The features at $\sim 30$--$80$~{\kms} in
the {\FeII}~$\lambda$2383 panel are likely to be atmospheric
absorption, since they are not confirmed by {\FeII}~$\lambda$2600.

\subsubsection{HE2217-2818 $z_{abs}=0.9424$}
\label{sec:0.9424}

\begin{figure}
\figurenum{1ad}
\centering
\vspace{0.0in}
\epsscale{0.45}
\plotone{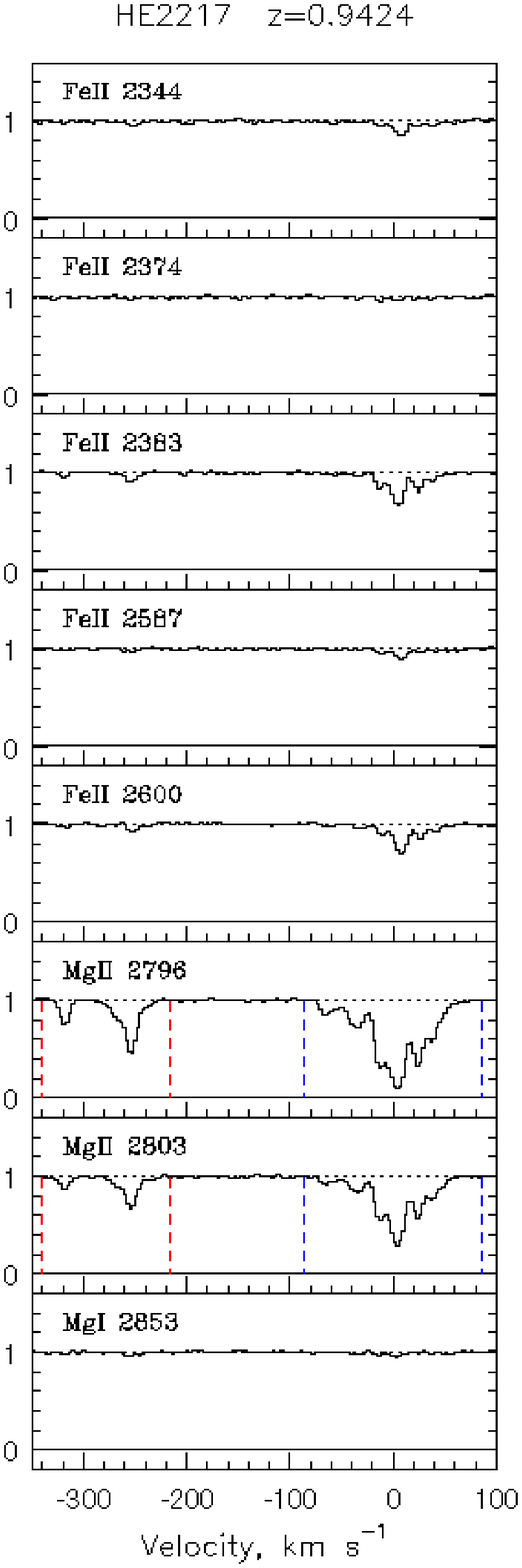}
\caption{}
\label{fig:fig1dd}
\end{figure}

This system's dominant subsystem is broad and unsaturated, and has
many components.  The system is displayed in Figure \ref{fig:fig1dd}.
It also has two narrower features blueward of this broad dominant
subsystem, which together constitute another subsystem.  Both
subsystems are detected in {\FeII} and {\MgI}.  The large velocity
separation between the two subsystems makes this system unusual and
interesting.  Perhaps this could be a case of having very high
velocity clouds along this sightline.

\subsubsection{HE2217-2818 $z_{abs}=1.6278$}
\label{sec:1.6278}

\begin{figure}
\figurenum{1ae}
\centering
\vspace{0.0in}
\epsscale{0.45}
\plotone{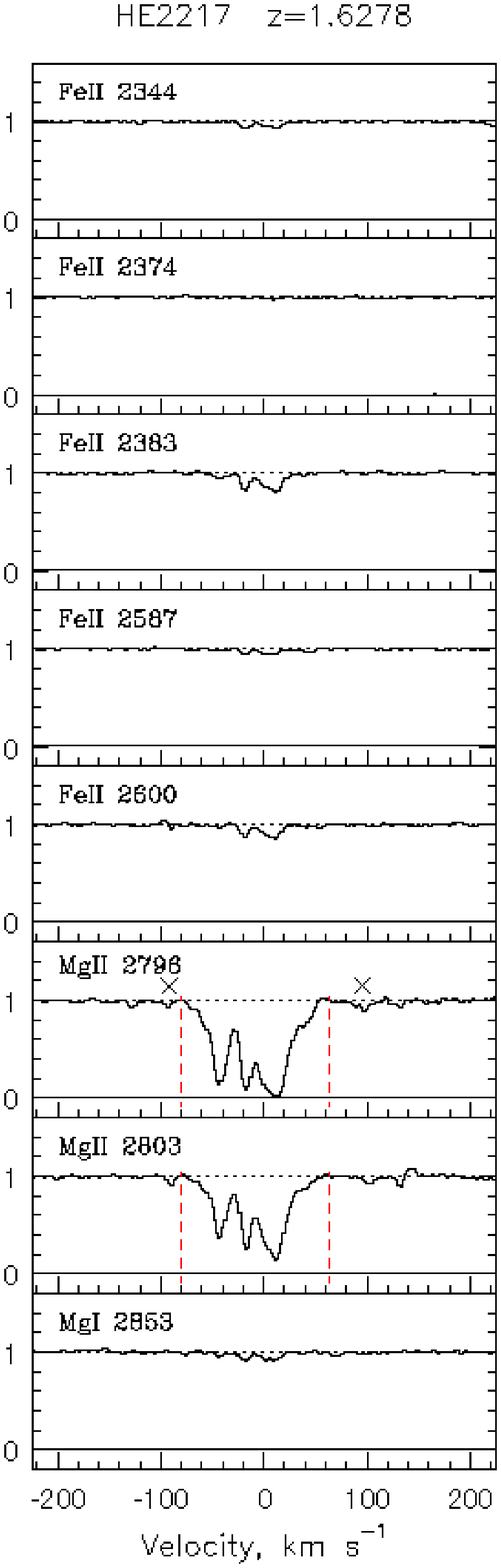}
\caption{}
\label{fig:fig1ee}
\end{figure}

This system, which is shown in Figure \ref{fig:fig1ee}, consists of
one broad subsystem with multiple components, only one of which may
have unresolved saturation.  Although the weak feature immediately
blueward of the system (at $\sim -90$~{\kms} in the
{\MgII}~$\lambda$2796 panel) has corresponding absorption in the
{\MgII}~$\lambda$2803 panel, the $\lambda$2803 absorption is
relatively too strong, and its minimum is not aligned.  We cannot rule
out a small $\lambda$2796 absorption feature, but the
{\AlII}~$\lambda$1670 and {\CII}~$\lambda$1334 for this system
indicate that the feature is due to atmospheric absorption.  The
features redward of the system, centered at $\sim 100$ and $\sim
130$~{\kms}, are also likely to be atmospheric absorption.  This is
supported by the appearance of uncertain sky line subtraction (``small
emission lines"), particularly in the {\MgII}~$\lambda$2803 panel.
Also, the minima in these features do not coincide in the
$\lambda$2796 and $\lambda$2803 panels, indicating that there must be
at least some contribution from blends.  If these features are really
{\MgII} absorption, they would be the smallest subsystems in our
sample.  The system is detected in both {\FeII} and {\MgI}.

\subsubsection{HE2217-2818 $z_{abs}=1.6921$}
\label{sec:1.6921}

\begin{figure}
\figurenum{1af}
\centering
\vspace{0.0in}
\epsscale{0.45}
\plotone{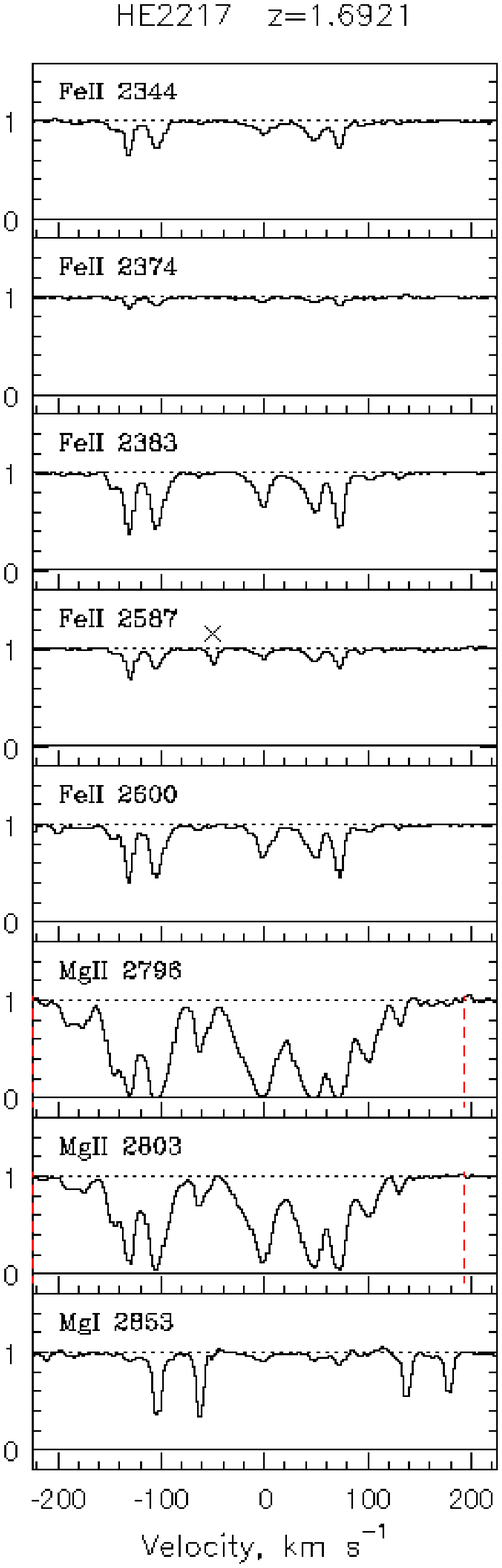}
\caption{}
\label{fig:fig1ff}
\end{figure}

This system, like the $z_{abs} = 2.1844$ system toward HE$0001-2340$,
consists of multiple components with little saturation.  Both systems
have the appearance of merging substructures, but this one is more
compact, and could be composed of just two merging objects.  This
suggestion is based partially on the kinematics of the {\FeII}
absorption, which is detected in many of the components, but only
weakly from $-80$ to $-20$~{\kms}.  The absorption profile can be seen
in Figure \ref{fig:fig1ff}.  The {\MgII} kinematics would also
classify this system as resembling a superwind.  The {\MgI} has
detections in some components of the system, but they are too severaly
blended with atmospheric absorption to use as more than just upper
limits in our analysis.

\subsubsection{B2311-373 $z_{abs}=0.3398$}
\label{sec:0.3398}

\begin{figure}
\figurenum{1ag}
\centering
\vspace{0.0in}
\epsscale{0.45}
\plotone{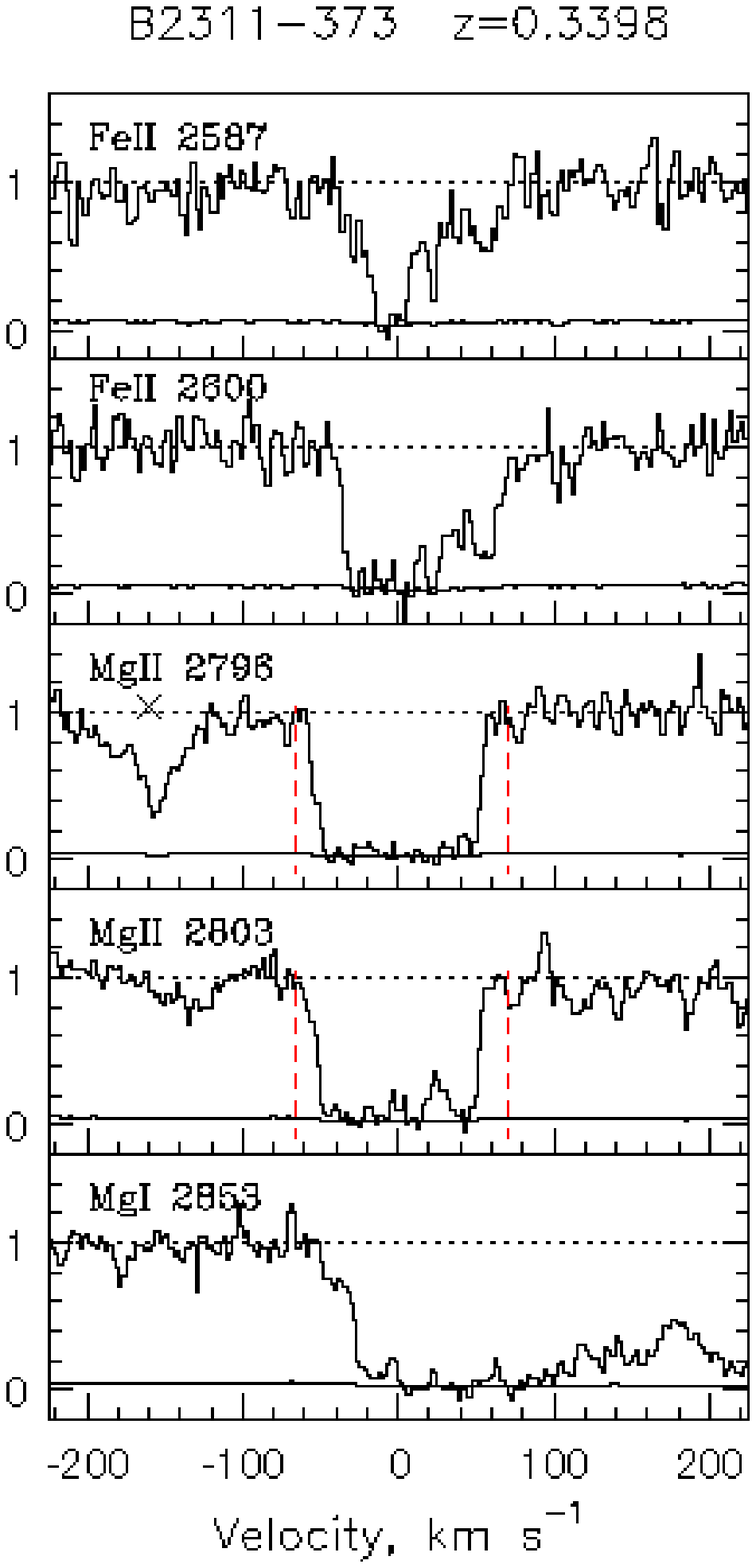}
\caption{}
\label{fig:fig1gg}
\end{figure}

This ``square profile" system is displayed in Figure \ref{fig:fig1gg}
and is located in the {\Lya} forest of this quasar.  The
{\FeII}~$\lambda$2344, $\lambda$2374, and $\lambda$2383 transitions
were not covered because they are blueward of the available
spectra. The system is detected in {\FeII}.  The MgI is too severely
blended to provide a useful constraint.  This system was observed in
the radio by CORALS to study the effect of dust on DLAs
\citep{akerman05}

\section{Results}
\label{sec:results}

Here we consider the properties of strong {\MgII} absorbers at high
$z$ and compare with those at low $z$ (both from our sample and from
CV01).  We consider whether our low $z$ sample is consistent with that
of CV01 (as we would expect).  In order to quantify possible
evolutionary trends, we must evaluate the absorption strength and
kinematic properties of the {\MgII} profiles.  We generally use the
same statistics to describe the profiles as defined in CV01.

We rely on the Kolmogorov-Smirnov (K-S) test to consider whether
differences between samples are significant.  This test takes the
cumulative distributions of a quantity for the two different samples,
finds the maximum difference between them, the ``KS statistic'', and computes the
probability that the two samples are drawn from the same distribution,
$P(KS)$.  This probability should be less than a few percent in order
that we can consider that there is a significant difference.  We must
look further than just this single statistic to consider the nature
of the difference.

\subsection{System Properties: Equivalent Width}
\label{sec:equivalentwidth}

We divide the sample into four subsamples based on equivalent width.
Sample A consists of all of our absorbers.  Sample B consists of
absorbers with $0.3~{\le}~W_r(2796) < 0.6$~{\AA}, and Sample C of
those with $0.6~{\le}~W_r(2796) < 1.0$~{\AA}. Sample D absorbers
have $W_r(2796)~{\ge}~0.6$~{\AA}, and Sample E absorbers have
$W_r(2796)~{\ge}~1.0$~{\AA}.  The equivalent width ranges are
identical to those used by CV01 in order to directly compare the
subsamples, and to identify any differences between them.  CV01 chose
these ranges based on cosmological evolution found by \citet{SS92}, and in
order to consider possible kinematic differences.  In
Table~\ref{tab:tab4} we present the statistical information for each
subsample, including number of absorbers in each subsample, average
rest frame equivalent width, average redshift of the absorbers,
average doublet ratio, and which absorbers belong to that subsample.

\begin{figure}
\figurenum{2}
\centering
\vspace{0.0in}
\epsscale{1.0}
\plotone{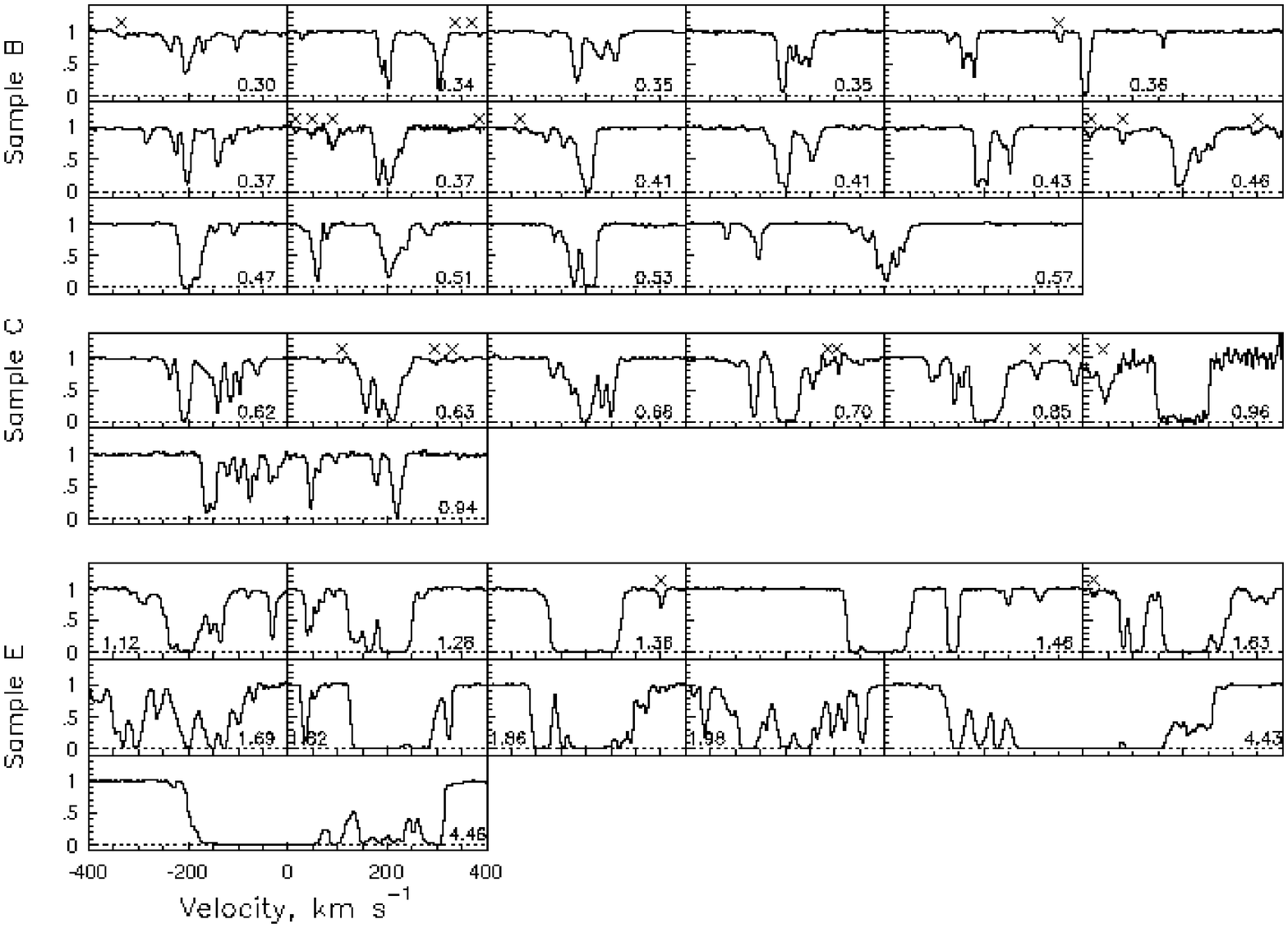}
\caption{{\MgII}~$\lambda$2796 profiles (normalized flux) for systems
from the VLT/UVES sample, in velocity space, ordered by increasing
equivalent width.  The equivalent width, in {\AA}, is displayed in a lower corner
of each panel (usually the lower right).  The profiles are also
separated into subsamples B, C, and E, as discussed in
\S~\ref{sec:equivalentwidth}.  The profiles that occupy a single panel
have a velocity spread of 400~{\kms}, and those that are displayed
in a double panel have a velocity spread of 800~{\kms}.}
\label{fig:plot2}
\end{figure}

Figure \ref{fig:plot2} displays the {\MgII}~$\lambda$2796 profiles for
our absorbers, divided into the subsamples B, C, and E.  The
absorbers shown are only those from the VLT sample; we do not include
any absorbers from the CV01 sample (see Fig.~13 of that paper).  Also,
we note that each of the subsamples contains absorbers from both the
low and high redshift ranges.  The profiles are shown in velocity
space; the vertical axis is the normalized continuum flux, and the number
within each window is the rest frame equivalent width of the absorber.
As expected, the amount of saturation in the system increases as the
equivalent width $W_r(2796)$ increases from sample B to C to E.

Figure \ref{fig:ewrest} shows the binned equivalent width distribution
of our data and that of the CV01 data.  Panel {\it a} shows our high
redshift data, panel {\it b} our low redshift data, and panel {\it c}
the CV01 low redshift data.  Qualitatively, our data (both the low and
high redshift regimes) are similar to the CV01 data with two
exceptions: our data include a few systems with 1.5~{\AA} $< W_r(2796)
<$ 2~{\AA}, and our low redshift data have two outliers with
$W_r(2796) \sim $ 4.5~{\AA}.  Quantitatively, a
K-S test shows that our low redshift data and the CV01 data are
consistent with being drawn from the same distribution, with a
probability of $P(KS) = 0.77$ and a KS statistic of 0.23.  A K-S test
between our low and high redshift samples yields $P(KS)=0.99$ (KS stat
= 0.15).  Finally, a KS test between our low redshift sample combined
with the CV01 sample and our high redshift sample yields a
$P(KS)=0.68$ (KS statistic = 0.22).  Since the VLT data that we have used
in this study were obtained as part of a study of the {\Lya} forest,
rather than for our purposes, we must consider biases that may have
been introduced by the selection criteria for that study
\citep{kim04}.  \citet{kim04} tended to avoided quasars with known DLA's at
$z>1.5$, so there should be no bias for lower redshift systems.  The
fact that we see no significant difference between the VLT samples and
the more homogeneous CV01 sample confirms this, so that we can
consider the two low redshift samples as equivalent.  There could,
however, be a small bias against large equivalent width systems at
high redshift, introduced by the selection criteria for the {\Lya}
forest study.

In Figure \ref{fig:ewrestcum}, we compare the cumulative equivalent
width distribution function for our sample to that determined from the
much larger Sloan Digital Sky Survey (SDSS) database, which covers a
similar redshift interval \citep{nestor05}.  Since the equivalent
width distribution evolves for $W_r(2796) > 2$~{\AA}, we make the
comparison separately for our low redshift and high redshift samples.
We find that both samples are consistent with being drawn from the
same distribution as found for the much larger SDSS sample which
should provide an accurate equivalent width distribution for
$W_r(2796) > 0.3$~{\AA}.  The probability is $P(KS)=0.70$ (KS stat =
0.16) that the low redshift sample equivalent widths were
drawn from the Nestor function with $\left< z \right>$ = 0.84.
Similarly, the probability is $P(KS)=0.51$ (KS stat = 0.18) that the
high redshift sample equivalent widths are consistent with
being drawn from the $\left< z \right>$ = 1.65 Nestor function.

\subsection{System Properties: Redshift}
\label{sec:redshift}

Figure \ref{fig:red} shows the binned redshift distributions of our
data and the CV01 data.  Our systems have redshifts ranging from
$z=0.33$ to $z=2.47$, with a mean of $\left< z \right>$ = 1.37.  Our
data contain a greater number of systems in the high redshift regime
($1.2 < z < 2.5$) than in the low ($0.3 < z < 1.2$), 20 versus 13.
The number of systems in the high redshift regime is roughly the same
as the number of systems at low redshift studied by CV01.  Ideally, we
would have a larger number of systems in the low redshift regime.
However, the systems that we do have are used primarily to verify that
our low redshift sample is similar to CV01.  

Figure \ref{fig:plot1alt} shows all absorbers in our sample, as well
as those from CV01, in redshift order.  We divide the plot into the
low and high redshift regimes to highlight the differences between the
two regimes.  We include the CV01 profiles (they are noted with a
star) to increase the low redshift sample size and to better
illustrate trends within the low redshift sample; all other systems
are from our VLT/UVES sample. 

We find that the most notable difference between the
two redshift regimes is the smaller number of subsystems at high $z$.
Also, in the low redshift regime, we see a larger fraction of profiles
with a dominant subsystem and one or more weaker subsystems.  At high
redshift, the smaller number of subsystems appears to be due to a
larger number of absorbing components that blend together in velocity
space.  Thus, the different absorbing components appear ``connected"
at high redshift, but ``separated'' at low redshift.  These
impressions, gained from careful inspection of the profiles, will also
be considered quantitatively in \S~\ref{sec:kinematics}.

Figure \ref{fig:zplots}a shows no evolution in equivalent width with
increasing redshift.  CV01 also saw no evolution in the equivalent
width, but over a smaller redshift range.  This is consistent with the
much larger SDSS survey of \citet{nestor05}, who find no evolution for
$W_r(2796) < 2$~{\AA}.  They do find a smaller number of $W_r(2796) >
2$~{\AA} systems at $z<1$, compared to the expectations for
cosmological evolution.  In our small sample, we do not have enough
very strong {\MgII} absorbers to make a comparison.  The agreement of
our equivalent width distribution with an unbiased sample confirms
that our strong {\MgII} sample does not suffer from any significant
biases.

Figure \ref{fig:zplots}b plots redshift versus the {\MgII} doublet
ratio, ($W_r(2796)$/$W_r(2803)$).  We see no evolution in the
distribution of doublet ratio over the full redshift range, combining
our data with that of CV01.  The plot does show a trend of decreasing
doublet ratio with increasing rest frame equivalent width, as
evidenced by the medians of the three subsample types seen in
Table~\ref{tab:tab4}.  This is as expected due to saturation in the
strongest systems.

\subsection{System Properties: Kinematics}
\label{sec:kinematics}

In Figures \ref{fig:kinematics}a and \ref{fig:kinematics}b we consider
the dependence of the kinematic spread, $\omega_v$, on system
equivalent width, and the evolution of $\omega_v$.  In Figure
\ref{fig:kinematics}a we see that the systems tend to cluster near two
envelopes.  Both the low and high redshift samples show this
dichotomy.  One envelope represents the ``saturation line" (solid
line); the other lies nearly vertically at $W_r(2796) = 0.3$~{\AA} due
to the larger number of small $W_r(2796)$ absorbers combined with the
sharp cutoff we applied to select only strong absorbers.  Systems
along these two envelopes represent two different absorption profile
types.  The envelope at $W_r(2796)=0.3$~{\AA} contains mostly systems
with multiple intermediate and high velocity subsystems, with a high
probability that the dominant subsystem is not heavily saturated.  The
saturation envelope contains mostly absorbers with a heavily saturated
dominant subsystem with few, if any, additional subsystems.  This
envelope is derived from the minimum kinematic spread at a given
equivalent width, corresponding to a ``square'' profile, saturated
over its full velocity range.  The strongest {\MgII} absorbers tend to
lie near this envelope, with small higher velocity components causing
them to rise above it.  The fact that few of the strongest absorbers
have much higher $\omega_v$ implies that equal strength subsystems
separated by large velocities are rare.

Figure \ref{fig:kinematics}b shows no significant evolution in the
kinematic spread.  A KS test between the combined low redshift sample
(ours and CV01) and the high redshift sample yields a KS statistic of
0.17, with a probability 0.84 of being drawn from the same
distribution.  The Spearman/Kendell rank order test also shows no
correlation between $\omega_v$ and $z$.  We do note that six of the
seven $\omega_v$ values $>100$~{\kms} fall in the low redshift sample.
Because of this, we considered several alternative KS tests,
e.g. dividing at the median $z$ or at the median $\omega_v$ in order
to define the two samples for comparison.  However, we still find no
significant differences between these samples.  We therefore conclude
that, statistically, the distributions of $\omega_v$ for the low and
high redshift samples are indistinguishable.  Thus, the evolution that
we noted in \S~\ref{sec:redshift}, based upon visual inspection of the
absorption profiles in Figure~\ref{fig:plot1alt}, is not seen in the
$\omega_v$ statistic.  Consequently, we consider additional
statistics.

Even though it is the second moment of velocity, since the kinematic
spread is weighted by optical depth, it is not maximally sensitive to
weak intermediate and high velocity components.  In order to emphasize
possible evolution of these weak components, we consider another
statistic, $\Delta v$, calculated by identifying the maximum and
minimum velocity of pixels that are included in the detected regions
of a system.

Figure~\ref{fig:kinematics}c shows the full velocity range for each
absorber (spanning all the subsystems) versus redshift.  A K-S test
between the two low redshift samples (VLT/UVES and CV01) yields
$P(KS)=0.36$ (KS stat = 0.33).  K-S tests for the $\Delta v$
distributions yield the same general result as for the kinematic
spread: there is no significant difference between the combined low
redshift sample and the high redshift sample $P(KS)=0.94$ (KS stat =
0.14). The Spearman/Kendell test also shows no significant correlation
between $\Delta v$ and $z$.  Therefore, it appears that there is no
deficit of high velocity {\it components} at high redshift.  This
confirms that the evolution of kinematics shown in
Fig.~\ref{fig:plot1alt} and discussed in section
\S~\ref{sec:redshift}, is not due to an increase in the overall
velocity range for systems.

Since we suggested, in \S~\ref{sec:redshift}, that the observed
kinematic evolution of the {\MgII} profiles is due to additional weak
absorbing components at high redshift, we now consider a statistic
that is sensitive to these.  We define the ``absorbed pixel fraction''
as the fraction of pixels with any absorption (in detected regions)
over the entire velocity range ($\Delta v$ from
Fig.~\ref{fig:kinematics}c) of an absorbing system.
Figure~\ref{fig:kinematics}d shows the absorbed pixel fraction as a
function of redshift.  We see that the majority of high redshift
systems (70\%; $14/20$) have absorption over the full velocity range,
while only half of the low redshift systems (50\%; $18/36$) do.  By
bootstrapping, we consider the significance of this result.  If 20
absorbed pixel fractions (the size of the high redshift sample) are
drawn at random from the absorbed pixel fractions of the combined low
redshift sample (VLT/UVES and Keck/HIRES) there is only a 5.6\% chance
of finding $14/20$ or more systems that are fully absorbed.  This is
an indication of the tendency for high redshift systems to be more
likely to have weak components connecting the stronger absorbing
regions.  We note that this result is not significantly affected
by bias due to the higher signal to noise of the VLT/UVES data.
This is because nearly all absorbed pixels in all of the spectra have
strong enough absorption to be detected even in the noisiest spectrum.

We can also compare the observed distributions of absorbed pixel fraction
for the high and low redshift samples.  However, because of the tendency
to have a large number of values of absorbed pixel fraction equal to 1,
the K-S test is not expected to be very sensitive to differences,
and we apply the Anderson-Darling test.
While the K-S test is most sensitive to differences near the center
of the distribution, the Anderson-Darling test gives more weight to
the tails.  We first compare the absorbed pixel fractions of the
VLT/UVES low redshift sample and the CV01 low redshift sample.
We find the Anderson-Darling statistic, $A^2$ is equal to 9.9,
and by bootstrapping assess that this value or a lower value
arise 53\% of the time if the absorbed pixel fraction of the
two samples are drawn from the same distribution.
Next, we compare the absorbed pixel fraction distributions for the
high and combined low redshift samples, finding $A^2 = 38.7$.
Here, from bootstrapping, we find a smaller chance of
16\% that the absorbed pixel fraction for the two samples are
drawn from the same distribution.

These results are suggestive of a statistical difference between the
high and low redshift samples, with the tendency to have a larger
fraction of absorbed pixels in systems at high redshift than in those
at low redshift.  Although there is such a statistical effect, the
difference is not large because there are significant numbers
of individual systems at high and low redshift that are similar
in properties to one another.

\subsection{Subsystem Properties}
\label{sec:subsysprop}

Figure \ref{fig:subsysplots}a shows the cumulative distribution of the
$5\sigma$ rest frame equivalent width detection limit at the position
of {\MgII}~$\lambda$2796 for our absorbers.  This plot is indicative
of how sensitive our survey is to the weakest subsystems, which are
likely to be single, unresolved components.  Our spectra have the same
resolution as those of CV01; however, because our signal to noise is
typically higher and because of the higher redshift of our sample, we
have a significantly better rest-frame detection sensitivity.  While
CV01 is 100\% complete down to 0.017~{\AA}, our sample is 100\%
complete down to 0.0087~{\AA}.  Our increased sensitivity indicates
that we are able to detect weaker subsystems more easily than CV01.
As mentioned in \S~\ref{sec:subsystems}, none of our subsystems are below
the 90\% cutoff of CV01 at $W_r < 0.015$.

Figure~\ref{fig:subsysplots}b displays the equivalent width
distribution of intermediate and high ($> 40$ {\kms}) velocity
subsystems.  CV01 compare their data to a power law distribution and
identify a turnover at $W_r(2796)$ $\sim$ 0.08~{\AA}.  Our data, for
$W_r(2796)>0.08$~{\AA} are consistent with the same power law, $n(W_r)
\propto W_r^{-1.6}$, as shown in Figure~\ref{fig:subsysplots}b, and also with
the turnover, although we are not as sensitive due to our smaller
number of detected subsystems.
CV01 also compare their turnover with that of weak {\MgII} absorbers
($W_r(2796) < 0.3$~{\AA}) at $0.02$~{\AA}.  These turnovers indicate that 
these two classes of absorbers are different.

In \S~\ref{sec:discussion}, we will compare the distribution of
equivalent widths for intermediate and high velocity subsystems to
the distribution of equivalent widths of weak {\MgII} absorbers,
and consider implications.

In Figure~\ref{fig:subsys}, we plot the number of subsystems per
absorber versus systemic redshift.  The mean of the number of low
redshift subsystems (including VLT/UVES and CV01 data) is 1.86; the
mean of the number of high redshift subsystems is 1.43.  Because the
data are in the form of only five integer values, neither the K-S or
the Anderson-Darling test gives meaningful results.  Instead, we
simply focus on the fraction of the systems that have more than one
subsystem, $6/20$ in the high redshift sample, and $18/36$ in the low
redshift sample (VLT and CV01).  If, using bootstrapping techniques,
we repeatedly draw at random $20$ values from the $36$ values in the
low redshift sample, we find only a 5.6\% chance of finding $6/20$ or
fewer systems with $2$ or more subsystems.  We thus conclude that the
high redshift sample differs from the low redshift sample in the sense
that systems at high redshift have a smaller number of subsystems.
This is closely related to the result of comparing the fraction of
absorbing pixels in the high and low redshift samples, in
Figure~\ref{fig:kinematics}d, since the two statistics are similarly
defined.

The number of subsystems per absorber is sensitive to the signal to
noise of the spectrum.  Thus we conducted a simulation to consider
whether the higher signal to noise in our high redshift sample leads
to the tendency to have a smaller number of subsystems per absorber at
high redshift.  We added noise to each high redshift system, at a
level comparable to the lowest quartile of the noise distribution for
low redshift systems, $S/N =35$ per pixel.  Only a few high redshift
systems (the $z_{abs} = 1.7891$ system toward HE0940-1050, the
$z_{abs} = 1.6299$ system toward Q0453-423, and the $z_{abs} = 1.6723$
system toward PKS0237-23) split into more subsystems even at this
relatively high level of noise.  Taking into account the noise
distribution for the low redshift sample, we estimate that
statistically 0.5 - 1 more high redshift systems would have two or
more subsystems.  Increasing that number by 1 so that $7/20$ of the
high redshift systems have two or more subsystems still leads to an
11\% chance that the high and low redshift samples are different in
this sense.

\subsection{Apparent Column Densities of MgII, MgI, and FeII} 
\label{sec:acd}

Figure \ref{fig:naod} consists of four plots showing different
relationships for the apparent column densities of intermediate ($v >
40$~{\kms}) and high velocity ($v > 165$~{\kms}) subsystems.
The apparent column densities were computed using the apparent optical
depth method \citep{savage91}, as was described in Appendix A of CV01. 
The column densities of {\MgII}, {\FeII}, and {\MgI} as well as velocity
ranges for each system are presented in Table~\ref{tab:tab5}.  Panel
{\it a} of Figure~\ref{fig:naod} shows the column density of {\MgII} for each of these
subsystems versus the subsystem's centroid velocity.  CV01 found that
subsystem {\MgII} column density decreases with increasing subsystem
velocity.  This trend is also visible in both of the VLT/UVES samples,
with no apparent difference between the low and high redshift samples.
Finally, CV01 found a paucity of subsystems with $\log N({\MgII}) <
11.6$~{\cmsq}, despite the fact the they were complete to a lower
limit than that for many of their quasars.  Because the sensitivity of
the VLT/UVES sample is better (complete to at least $\log N({\MgII}) =
11.3$ in all quasars, with much better sensitivities in most), we are
able to test this claim.  We also find no subsystems with $\log
N({\MgII}) < 11.6$, supporting the existence of a physical mechanism
that prohibits subsystems weaker than this.  This is directly related
to the equivalent width turnover described in \S~\ref{sec:subsysprop}.

In panels {\it b} and {\it c} of Figure~\ref{fig:naod}, we plot the
logarithmic ratio of $N({\FeII})$ to $N({\MgII})$ as a function of
subsystem velocity and $\log N({\MgII})$.  These plots are closely
related, because large velocity subsystems tend to have
small $\log N({\MgII})$, as seen in panel {\it a}.  There is a large spread
of $\log N({\FeII})/N({\MgII})$, with values ranging from $-1.2$ to
$0.2$.  CV01 noted that many of the smallest $\log
N({\FeII})/N({\MgII})$ values occur for subsystems with small $\log
N({\MgII})$, which include many of the highest velocity subsystems.
The VLT/UVES sample does not have any subsystems with such small $\log
N({\FeII})/N({\MgII})$ values, either for weak or for strong
subsystems.  We note, however, that there are not enough medium and
high velocity subsystems to consider in the VLT/UVES data because it
contains a significant fraction of high redshift absorbers, which tend
not to have as many subsystems (see \S~\ref{sec:subsysprop}).  Thus,
we can neither refute or confirm the interesting suggestion of CV01
that some of the highest velocity subsystems are preferentially
$\alpha$-enhanced.

In panel {\it d} of Figure~\ref{fig:naod} we examine possible
evolution in $\log N({\FeII})/N({\MgII})$.  This plot again
illustrates the lack of subsystems at high $z$, discussed in
\S~\ref{sec:subsysprop}.  While the high $z$ sample is roughly
two-thirds the size of the low $z$ sample, it has only one-third the
number of subsystems with $v > 40$~{\kms}.  Furthermore, the three
highest redshift datapoints are not independent; they are all
subsystems of the $z=2.1844$ system toward HE0001-2340.  We note that
for $z>1.2$, there are no subsystems with $\log N({\FeII})/N({\MgII})
< -0.7$.  Small $\log N({\FeII})/N({\MgII})$ values would signify
either $\alpha$-enhancement or a higher ionization parameter at low
redshift.  The latter is unlikely in view of the known evolution of
the EBR (see \S~\ref{sec:intro}).  The existence of $\alpha$-enhanced
gas at low redshifts implies recent star formation in some fraction of
the absorber population.  The lack of such $\alpha$-enhanced absorbers
in the high redshift sample would be surprising but could be due to
small number statistics.  We repeatedly chose 10 subsystems (the
number in the high redshift sample) at random from the full sample of
40 subsystems with $v > 40$~{\kms}, and found a probability of $0.11$
that there would be none with $\log N({\FeII})/N({\MgII}) < -0.7$.

Figure \ref{fig:figure9} is a summary plot in which we show the
{\MgII}, {\MgI}, and {\FeII} column densities (from top to bottom)
of the subsystems of each system.   The systems are displayed from
left to right by increasing redshift and subsystems with different
velocities are designated by different symbols.  The small fraction of
subsystems at $1.65 < z < 2.5$ with {\MgI} data is due to frequent
contamination of that region of the spectrum by atmospheric
absorption.  There is no indication of evolution in $N({\MgII})$,
$N({\MgI})$, or $N({\FeII})$.

\subsection{Summary of Results}
\label{sec:summary}

\begin{figure}
\figurenum{3}
\centering
\vspace{0.0in}
\epsscale{.8}
\plotone{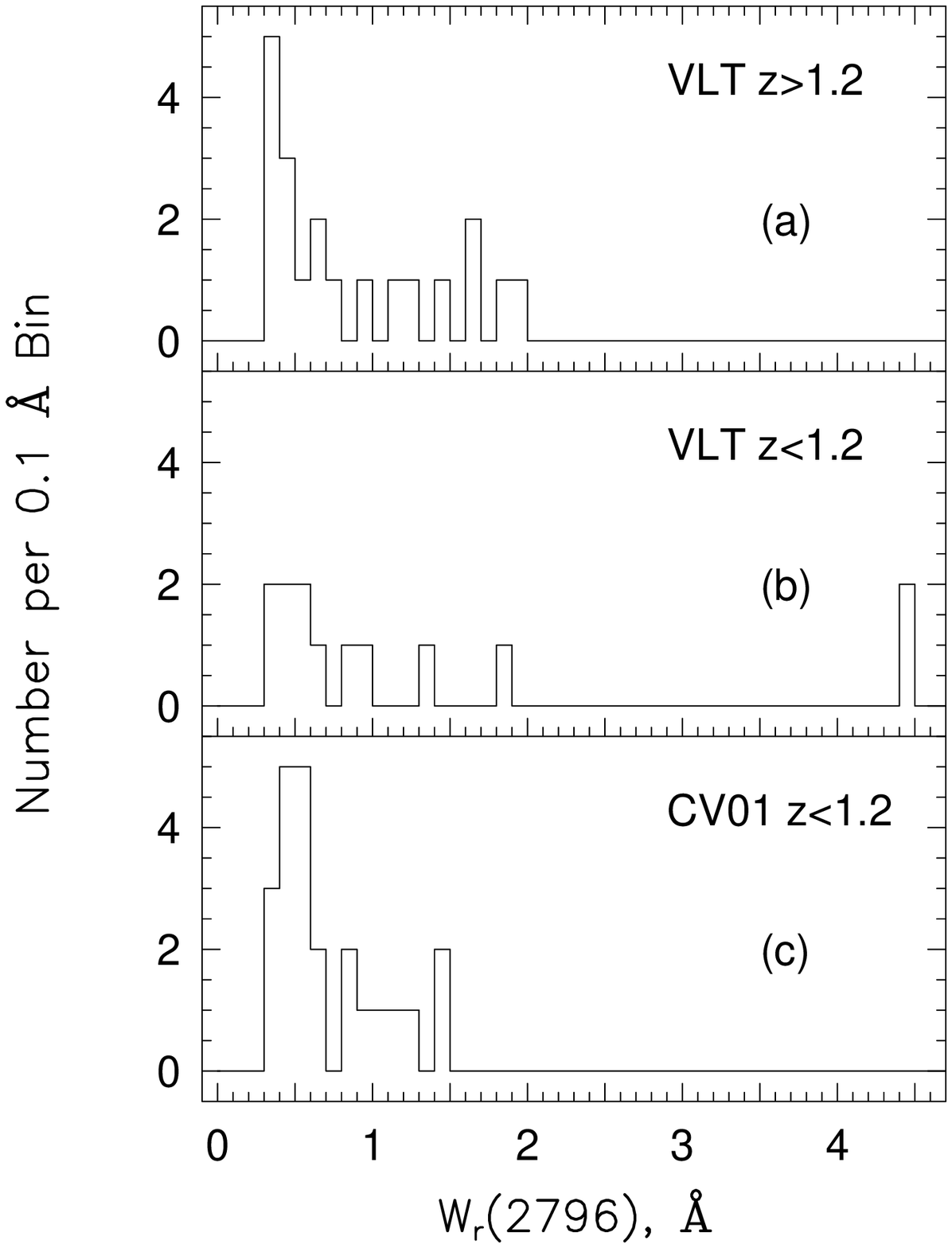}
\epsscale{1.0}
\caption{The rest-frame {\MgII}~$\lambda$2796 equivalent width
distributions for the different samples.  (a) The VLT/UVES systems with redshifts
less than 1.2. (b) The VLT/UVES systems with redshifts greater than
1.2. (c) All absorbers from the CV01 dataset.}
\label{fig:ewrest}
\end{figure}

\begin{figure}
\figurenum{4}
\centering
\vspace{0.0in}
\epsscale{1.0}
\plotone{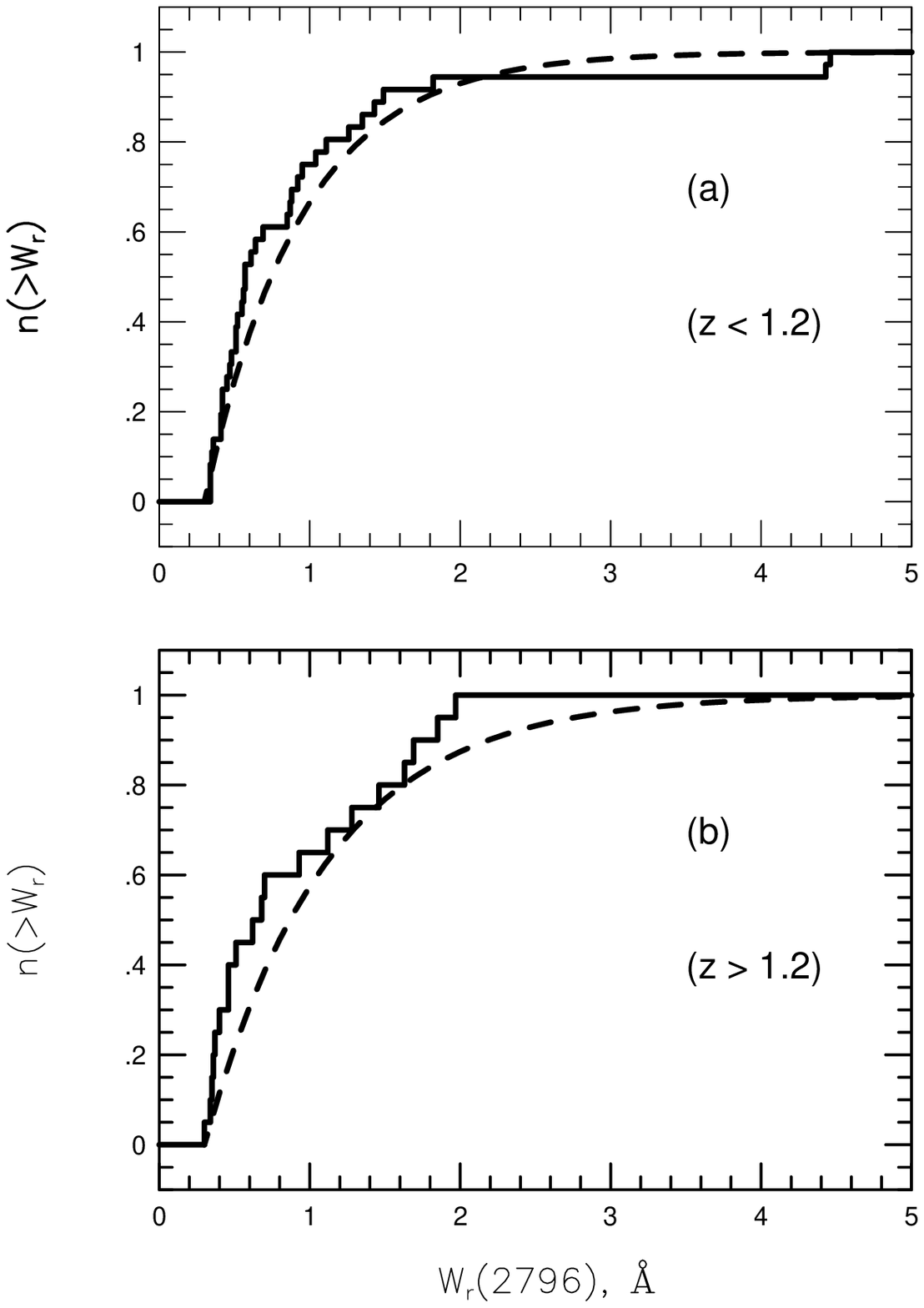}
\caption{(a) The cumulative {\MgII}~$\lambda$2796 equivalent width
distribution for $z<1.2$ systems (from CV01 and VLT/UVES data), given
as a solid line.  The dotted line represents an unbiased sample of
strong {\MgII} absorbers as predicted from the larger sample of
\citet{nestor05} using a $\left< z \right>$ = 0.84. (b) The cumulative 
{\MgII}~$\lambda$2796 equivalent width distribution for $z>1.2$
systems, given as a solid line.  The dotted line represents an
unbiased sample of strong {\MgII} absorbers as predicted from the
larger sample of \citet{nestor05} using a $\left< z \right>$ = 1.65.}
\label{fig:ewrestcum}
\end{figure}

\begin{figure}
\figurenum{5}
\centering
\vspace{0.0in}
\epsscale{1.2}
\plotone{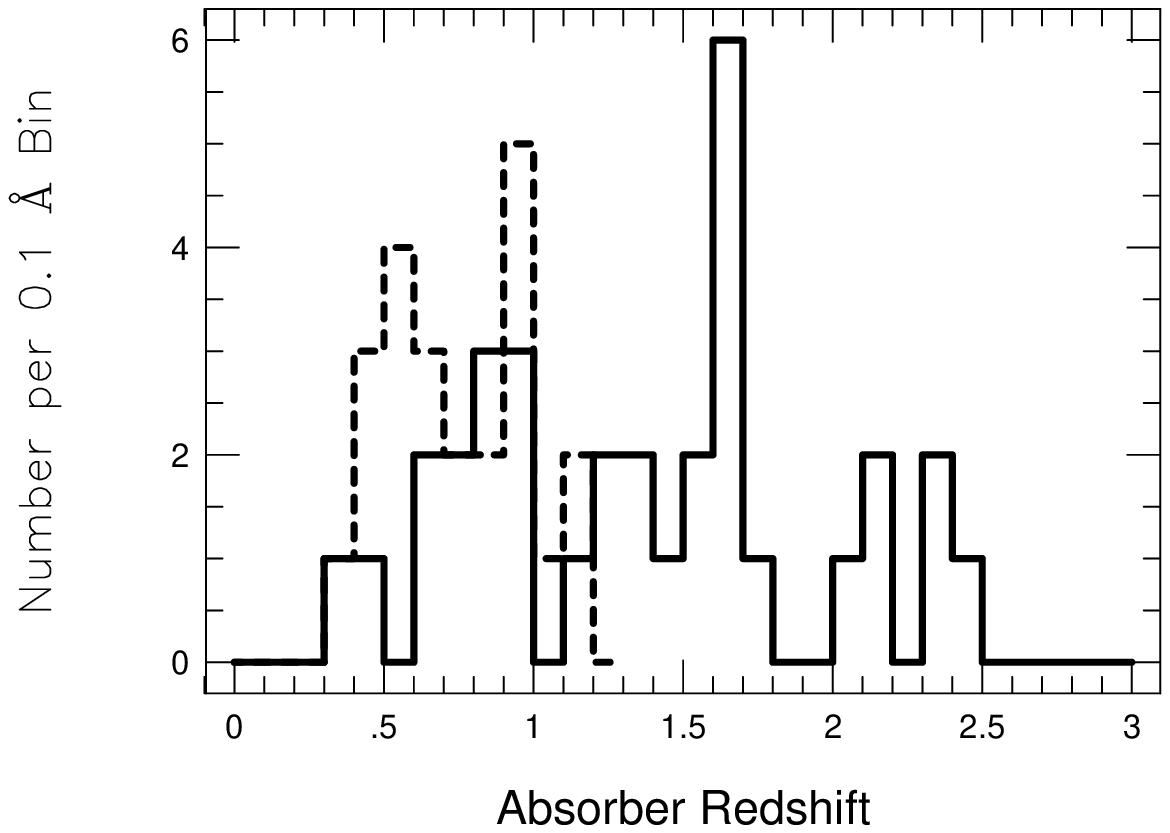}
\caption{Redshift distribution of strong {\MgII} absorbers.  The solid 
histogram represents the VLT/UVES data; the dashed histogram represents the
CV01 data.}
\label{fig:red}
\end{figure}

\begin{sidewaysfigure}
\figurenum{6}
\centering
\hglue -2.0in
\epsscale{0.9}
\plotone{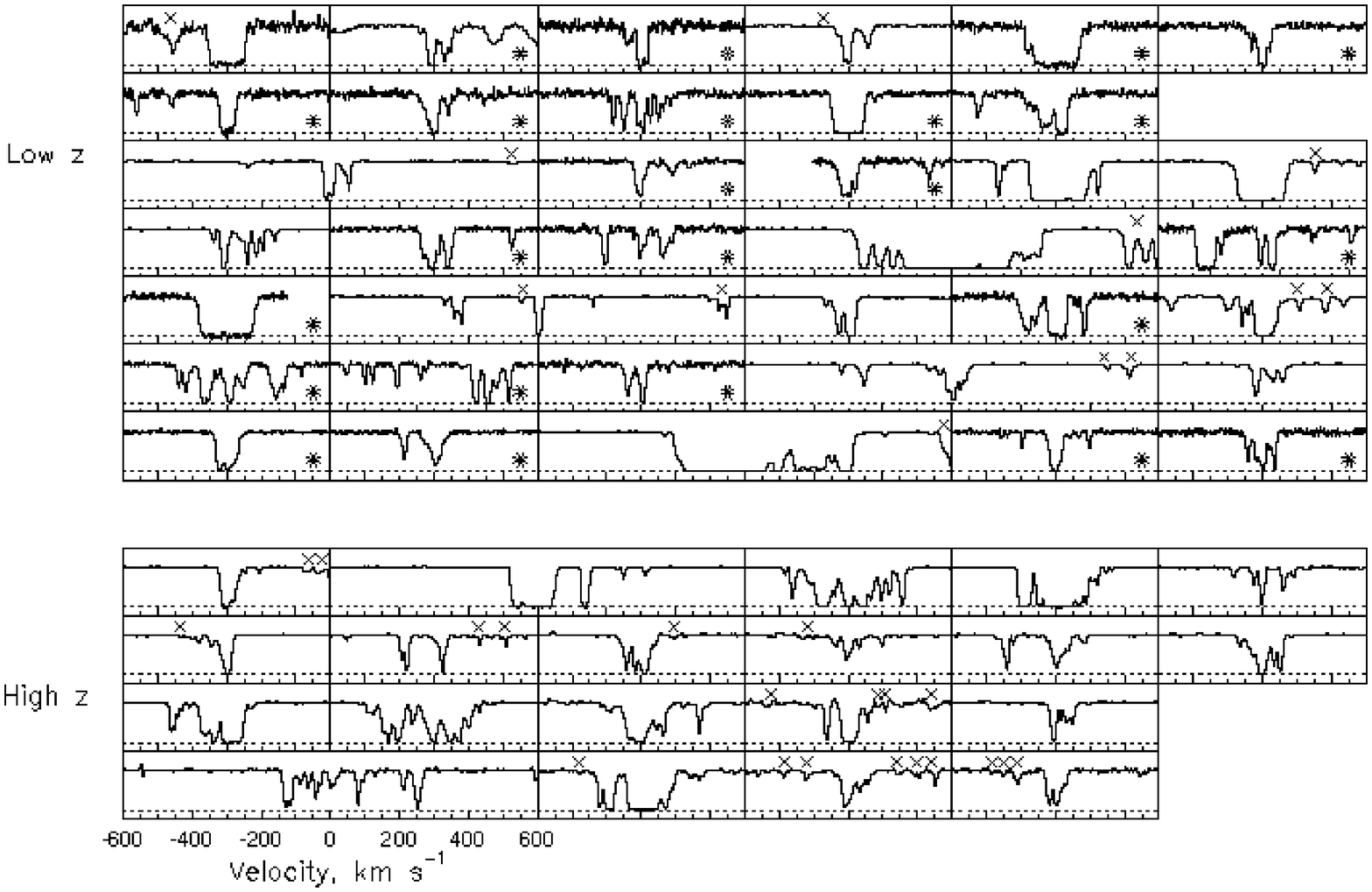}
\caption{{\MgII}~$\lambda$2796 absorption profiles for all
absorbers in our data set.  We separated the profiles into high and
low redshift samples to more clearly illustrate differences.  We
include both the VLT/UVES profiles and the CV01 profiles in the low
redshift sample.  The CV01 profiles are marked with an asterisk in the
lower right corner.  The crosses above the normalized flux indicate
features not associated with {\MgII} absorption at the system
redshift.  The velocity range of the single panels is -300~{\kms} $< v
<$ 300~{\kms}, and the range of the double panels is -600~{\kms} $< v
<$ 600~{\kms}, with two exceptions.  These exceptions are the
$z_{abs}=0.9276$ system toward Q1206+706 (second panel in the sixth
row), which ranges from -450~{\kms} $< v <$ 150~{\kms}, and the
$z_{abs}=0.8519$ toward Q0002+051 (last panel in the fourth row),
which ranges from -150~{\kms} $< v <$ 450~{\kms}.  }
\label{fig:plot1alt}
\end{sidewaysfigure}

\begin{figure}
\figurenum{7}
\centering
\vspace{0.0in}
\epsscale{1.0}
\plotone{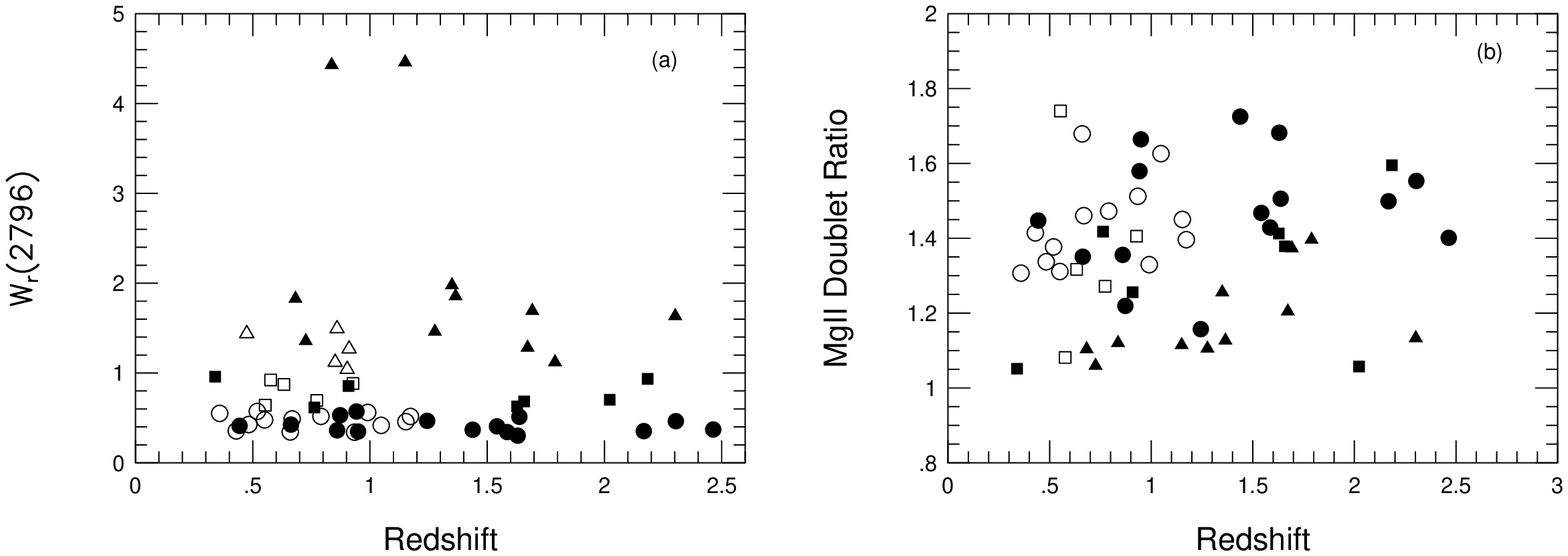}
\caption{(a) Rest frame equivalent width of {\MgII}~$\lambda$2796 versus
the redshift of the absorber.  The circles represent absorbers from
sample B, squares absorbers from sample C, triangles absorbers from
sample E.  The open shapes are from the CV01 data, the filled in are
the VLT/UVES data. (b) The doublet ratio of {\MgII} versus the
redshift of the absorber.  Points are as defined in panel (a).}
\label{fig:zplots}
\end{figure}

\begin{figure}
\figurenum{8}
\centering
\vspace{0.0in}
\epsscale{1.0}
\plotone{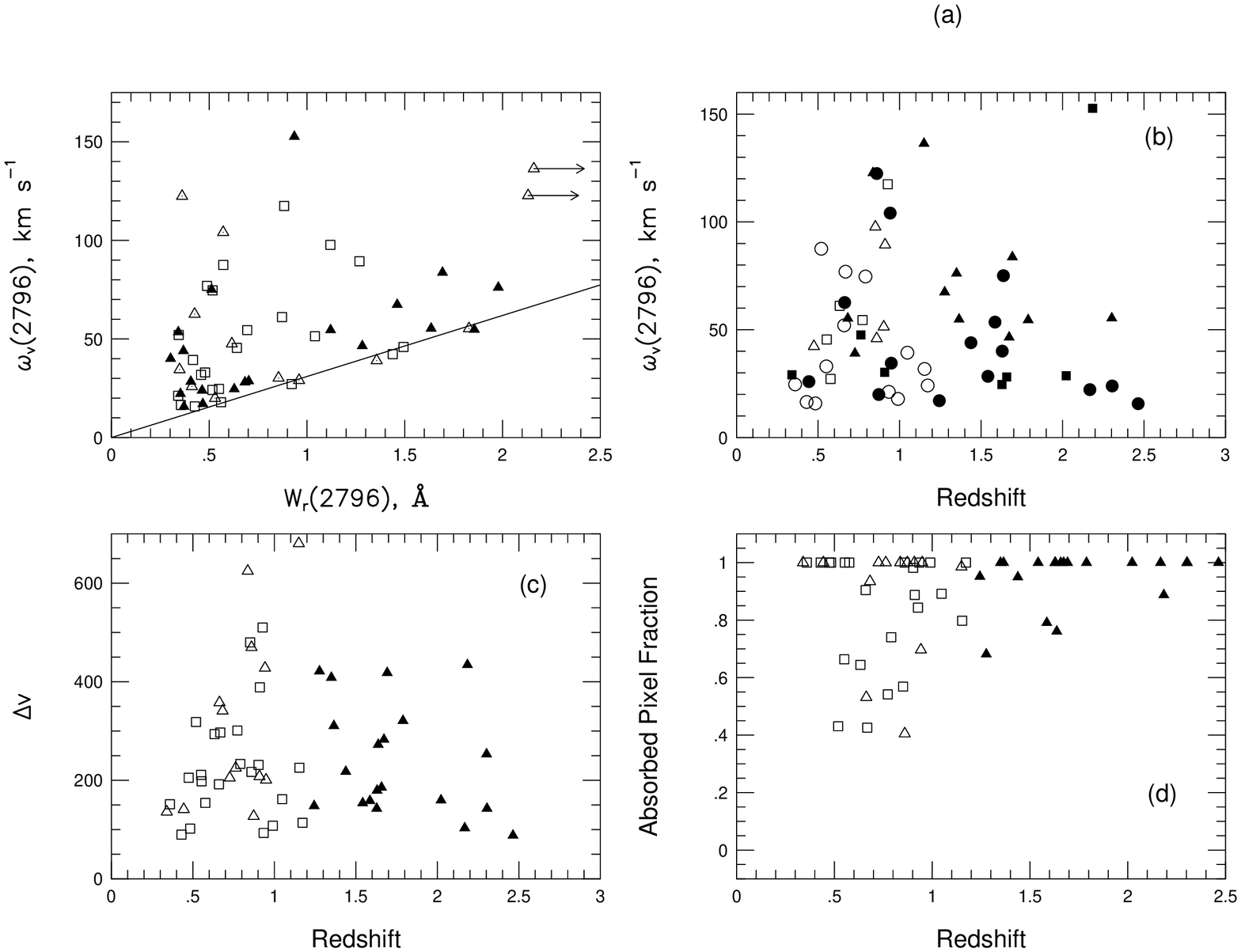}
\caption{ (a) Kinematic spread ($\omega_v$) versus rest frame 
equivalent width of {\MgII}~$\lambda$2796. The line represents the
minimum kinematic spread for a given equivalent width as explained in
\S~\ref{sec:equivalentwidth}.  The two points with corresponding 
arrows represent two absorbers with $W_r(2796) > 4$ {\AA}.  While we
preserve their kinematic spread values in this plot, we place them at
lower values for rest frame equivalent width in order to better see
trends in the rest of the data.  In this panel, and panels c and d,
the triangles represent data from VLT/UVES, the filled triangles for
high redshift systems and the open triangles for low redshift systems.
The squares represent systems from CV01.  (b) Kinematic spread
($\omega_v$) of the {\MgII}~$\lambda$2796 profile versus redshift for
the absorbers.  The different data point symbols are defined based
upon the rest frame equivalent width of {\MgII}~$\lambda$2796, as in
Fig.~\ref{fig:zplots}.  (c) $\Delta v$ versus system redshift.
$\Delta v$ is the maximum velocity range over which an absorber shows
any detected absorption, i.e. the velocity range from the bluest pixel
of the bluest subsystem to the reddest pixel of the reddest subsystem.
All data points are as defined for panel (b).  (d) Percentage of
pixels over the entire velocity range of the {\MgII}~$\lambda$2796
profile in which absorption is detected versus system redshift.  Data
points are as defined for panel (b). }
\label{fig:kinematics}
\end{figure}

\begin{figure}
\figurenum{9}
\centering
\vspace{0.0in}
\epsscale{1.0}
\plotone{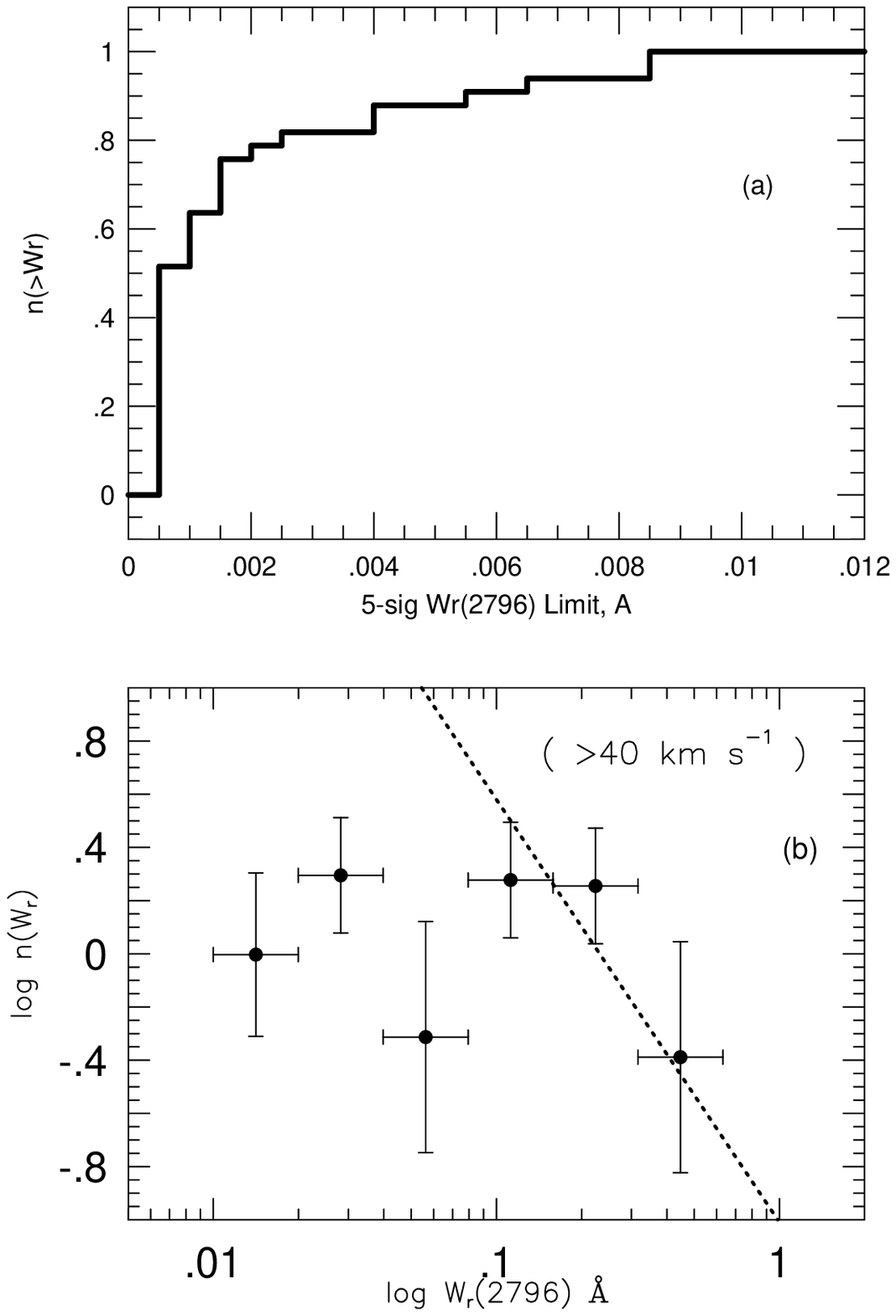}
\caption{(a) Cumulative distribution of the
$5\sigma$ rest frame equivalent width detection limits at
the position of {\MgII}~$\lambda$2796 for all strong {\MgII} systems
from VLT/UVES spectra. (b) Equivalent width
distribution of subsystems with velocities $> 40$ {\kms} (intermediate
and high velocity subsystems).  We include $1\sigma$ error bars on the vertical
axis.  Error bars on the horizontal axis represent the bin size.  The dotted line
reproduces a power law, with slope of $-1.6$, as found by CV01 for intermediate
and high velocity subsystems with $W_r > 0.08$~{\AA}.}
\label{fig:subsysplots}
\end{figure}

\begin{figure}
\figurenum{10}
\centering
\vspace{0.0in}
\epsscale{1.0}
\plotone{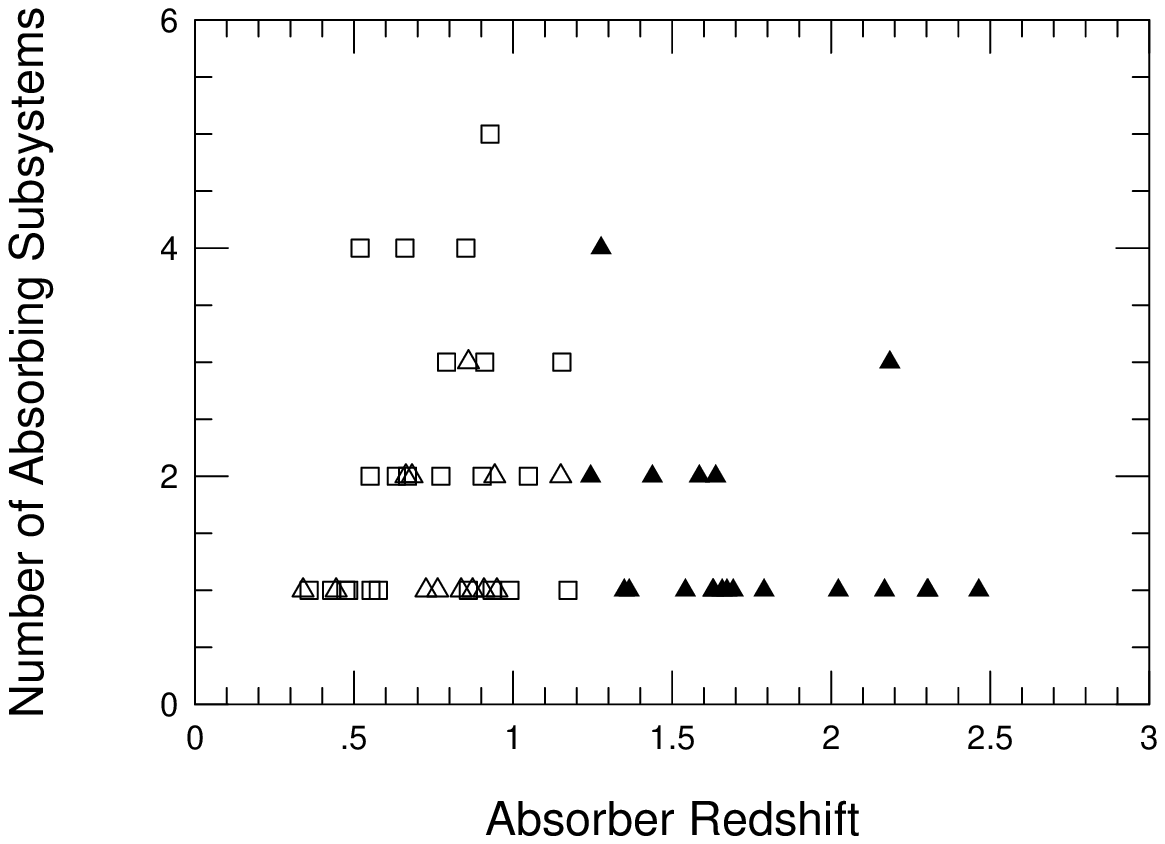}
\caption{This plot shows the number of subsystems per strong {\MgII} system
as a function of system redshift.  The filled triangles represent high
redshift systems from VLT/UVES, and the open triangles are low redshift
systems from VLT/UVES. The squares represent systems from CV01.}
\label{fig:subsys}
\end{figure}

\begin{figure}
\figurenum{11}
\centering
\vspace{0.0in}
\epsscale{1.0}
\plotone{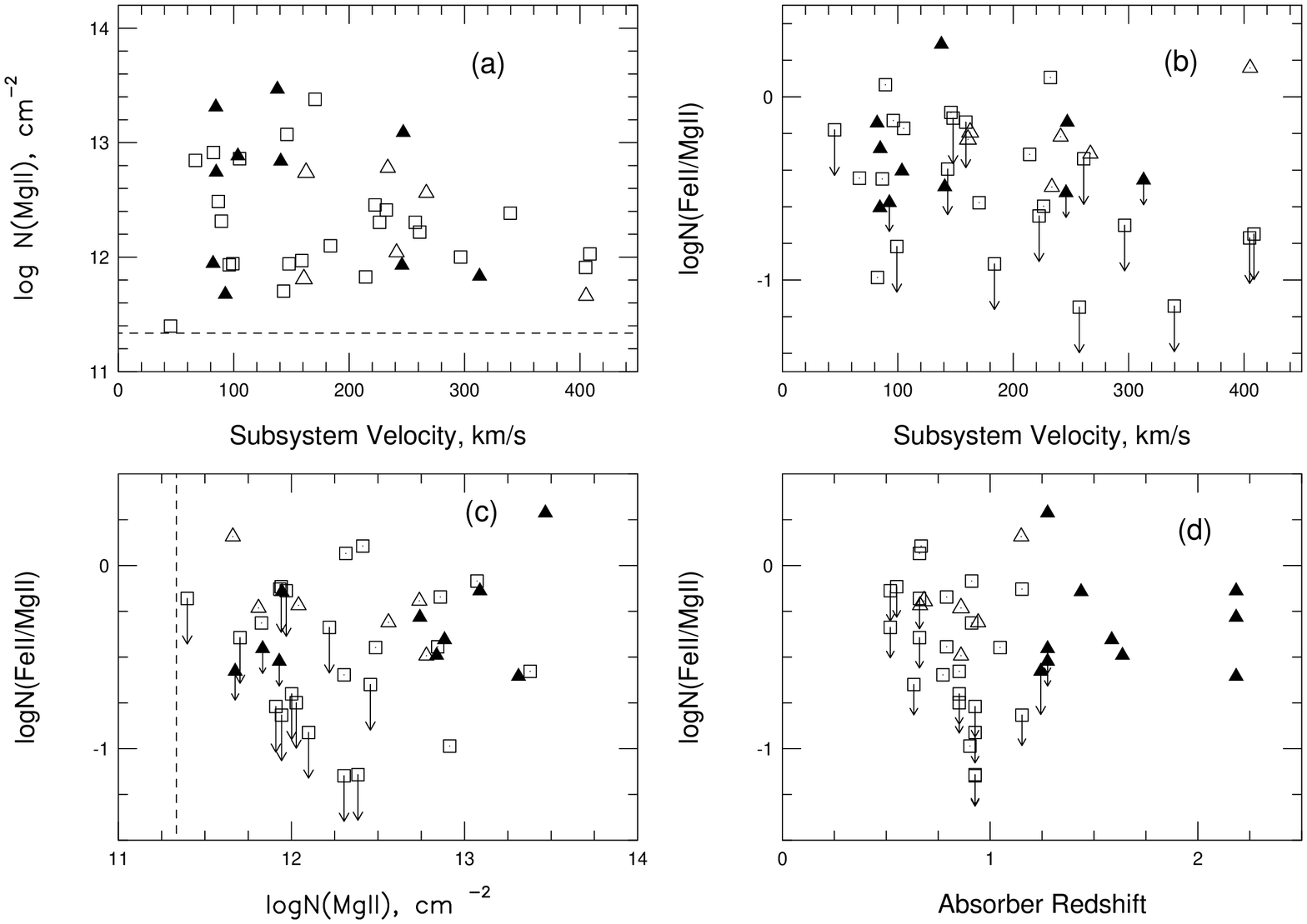}
\caption{ Data points in all panels of this figure are as defined for
Fig.~\ref{fig:subsys}.  These points all represent subsystems with $v~{\ge}~
40$~{\kms}.  The downward arrows in panels (b), (c), and (d) represent
upper limits resulting from lack of a $3\sigma$ detection. (a)
Logarithmic column density of {\MgII} of a subsystem versus the
subsystem's centroid velocity.  The dashed line at the bottom of the
figure indicates where the sample drops below 100\% completeness. (b)
Logarithmic ratio of $N({\FeII})$ to $N({\MgII})$ versus subsystem
velocity. (c) Logarithmic ratio of $N({\FeII})$ to $N({\MgII})$ versus
$\log N({\MgII})$.  The vertical line represents the initial drop from
100\% completeness. (d) The logarithmic ratio of $N({\FeII})$ to
$N({\MgII})$ versus system redshift.}
\label{fig:naod}
\end{figure}

\begin{figure}
\figurenum{12}
\centering
\vspace{0.0in}
\epsscale{1.0}
\plotone{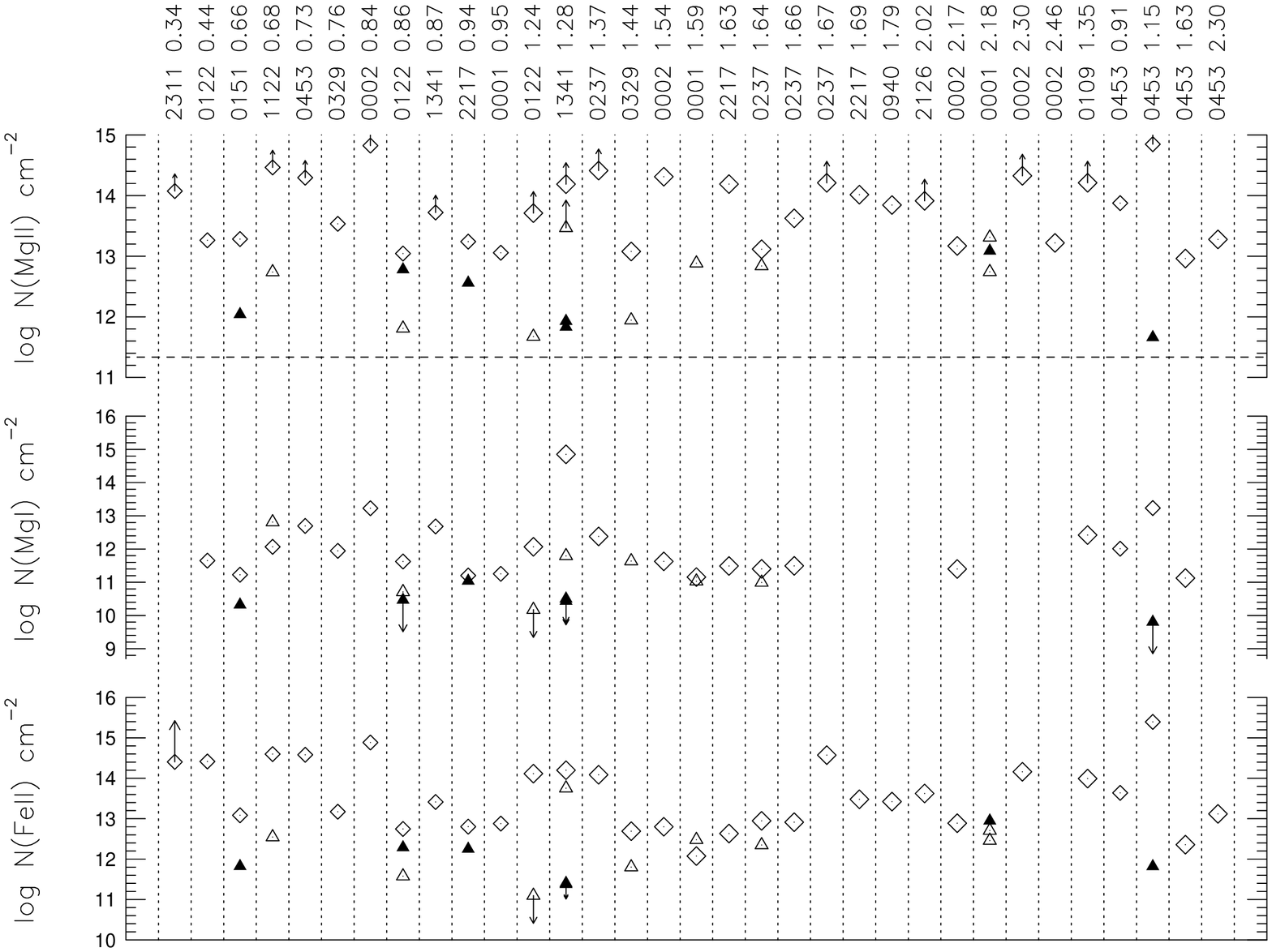}
\caption{{\MgII}, {\MgI}, and {\FeII} column densities for
each subsystem of each absorber.  The vertical dotted lines separate
different absorbers.  The absorbers are labeled first by the quasar
line of sight (without catalog designations which can be found in
Table~\ref{tab:tab1}) then by redshift.  The dashed line across the
{\MgII} panel represents the initial drop from 100\% completeness.
The diamonds represent subsystems with velocities v $< 40$~{\kms}, the
open triangles subsystems with $40~{\kms} {\le}$ v ${\le}~165$~{\kms}, and the
filled triangles subsystems with v $> 165$~{\kms}.  The arrows facing
upward represent lower limits resulting from saturation; the arrows
facing downward represent $3\sigma$ upper limits.}
\label{fig:figure9}
\end{figure}

Here we summarize the results of our comparisons of the $0.3 < z <
1.2$ (low redshift) and $1.2 < z < 2.5$ (high redshift) strong {\MgII}
absorber samples:

\begin{enumerate}

\item{We find no significant differences between {\MgII} rest-frame
 equivalent widths (Figure~\ref{fig:ewrest}) and kinematics
 (Figure~\ref{fig:kinematics}) for the VLT/UVES low redshift sample
 and the low redshift sample of CV01 from Keck/HIRES.  This indicates
 that we can legitimately group these two samples together and compare
 to the high redshift sample from VLT/UVES without concern about bias
 due to the different sample selection and instrumental
 configurations.}

\item{The cumulative equivalent width distribution function for strong 
{\MgII} absorbers is statistically indistinguishable for the high and
low redshift samples, as shown in Figures ~\ref{fig:ewrest} and
\ref{fig:zplots}a.  This is consistent with the much larger survey of
\citet{nestor05}, who found no evolution for $W_r(2796)<2$~{\AA}.}

\item{The kinematic spread of strong {\MgII} absorbers also does not 
show significant evolution over the interval $0.3 < z < 2.5$
(Figure~\ref{fig:kinematics}a).  At both low and high redshifts we see
the full range of spreads, from saturated, single trough profiles, to
those with many unsaturated components distributed over a large
velocity range.}

\item{The average number of separate subsystems in strong {\MgII} systems
 increases with decreasing redshift, from 1.43 at $1.2 < z < 2.5$ to
 1.86 at $0.3 < z < 1.2$ (see Figure~\ref{fig:subsys}).  This
 difference is significant.  The majority of the high redshift systems
 have only one subsystem.}

\item{The small number of subsystems in strong {\MgII} systems at high 
redshift is not due to an absence of high velocity components, as
evidenced by a lack of evolution in total system velocity spread (see
Figure~\ref{fig:kinematics}d).  Instead, it is due to weak components
``filling'' the velocity space between the regions that would be
separate subsystems at lower redshift.  This leads to a larger
fraction of absorbing pixels in high redshift systems (shown in
Figure~\ref{fig:kinematics}c).}

\item{We carefully consider whether the differences in the number of 
subsystems and in the fraction of absorbing pixels between high and
low redshift could be due to larger contamination of the high redshift
system profiles by atmospheric absorption.  This possibility is ruled
out after inspecting each {\MgII} doublet and finding matching
absorption for weaker components (those that ``fill in'' the velocity
space) in {\MgII}~$\lambda$2803 (see Figure~\ref{fig:fig1}--ag and the
system descriptions in \S~\ref{sec:1.6299} and \S~\ref{sec:1.7891},
where we note the two controversial cases).}

\item{Another way to understand the differences between the strong 
{\MgII} systems at high and low redshift is to visually examine the
profiles in Figure~\ref{fig:plot1alt}.  CV01 found that many profiles
at low redshift have strong central components and one or more weaker,
outlying intermediate or high velocity components. We also see some
profiles like this at high redshift, but they are not as common.
Instead we see complex kinematics, with more components spanning
roughly the same velocity range.}

\item{The distribution of {\MgII}~$\lambda$2796 rest frame equivalent 
width for intermediate and high velocity subsystems is generally
consistent with a $n(W_r) \propto W_r^{-1.6}$ power law, however there
is a turnover, with few subsystems with $W_r(2796) < 0.08$~{\AA},
shown in Figure~\ref{fig:subsysplots}a.  This is not due to
incompleteness, since our survey is 100\% complete to a limit of
$W_r(2796) = 0.009$~{\AA}.  This confirms the results of CV01.}

\item{CV01 found that high velocity subsystems of $0.3 < z < 1.2$ strong 
{\MgII} absorbers tend to have smaller $N({\MgII})$ than low velocity
subsystems.  We find the same trend at $1.2 < z < 2.5$ (see
Figure~\ref{fig:naod}a).}

\item{There is no clear evolution in the ratio of apparent column 
densities of {\FeII} and {\MgII} for intermediate ($40 < v <
165$~{\kms}) and high ($>165$~{\kms}) velocity subsystems
(Figure~\ref{fig:naod}).  This is, however, hard to evaluate because
of the smaller number of separate subsystems at high redshift.  We see
tentative evidence of $\alpha$-enhanced gas in low redshift
subsystems.}

\end{enumerate}

\section{Discussion}
\label{sec:discussion}

We find that there is significant overlap in the kinematic properties
of strong {\MgII} absorbers in the low ($0.3 < z < 1.2$) and high
($1.2 < z < 2.5$) redshift regimes.  However, there is a systematic
trend for high redshift systems to have weaker absorpting components
connecting stronger regions of absorption.  We expect that this
indicates galactic structures at $1.2 < z < 2.5$ tend to have a larger
number of {\it accreting} or {\it outflowing} clouds.  In fact, we see
very few high redshift examples of the classic disk/halo structures so
common in the low redshift CV01 sample, evidenced by a dominant
subsystem, with weaker subsystems to one side.  The accreting or
outflowing gas clouds are only evident in high resolution {\MgII}
profiles, and because their absorption is so weak, do not produce an
evolution in the equivalent width distribution of the population.

This evolutionary trend in the kinematics of strong {\MgII} absorbers
may be the first systematic evidence for hierarchical structure
formation seen in the gas.  It thus provides a long-sought link to the
volume of work on the evolution of galaxy morphology in deep imaging
surveys.  Logically, this sort of evolution should be seen in the gas
at high redshift, where images show that a large fraction of the solid
angle around a luminous galaxy is covered by material.  An extreme
example of this is the ``Spiderweb Galaxy'', a central cluster galaxy
at $z=2.2$, with many accreting, star forming companions
\citep{miley06}.  However, a large fraction of $z>1.2$ galaxies exhibit
morphological peculiarities, including the ``clump-cluster'' morphologies
\citep{Elm05} that would give rise to the absorption signature that we
find to be evident at those redshifts.

Having made a connection between strong {\MgII} absorbers and
structures seen in images, it is also important to consider possible
connections with other types of absorbers.  Qualitatively, weak
{\MgII} absorbers have absorption profiles similar to the weak
outlying subsystems that are common in strong {\MgII} absorbers at
$0.3 < z < 1.2$.  However, we noted in \S~\ref{sec:subsysprop} that
there is a significant difference between the equivalent width
distributions of these two populations of absorbers.  The weak {\MgII}
absorbers have an equivalent width distribution that follows a power
law, with $n(W_r) \propto W_r^{-1.6}$ down to $0.02$~{\AA}, while the
outlying subsystems show a turnover in this power law below $W_r =
0.08$~{\AA}.  CV01 also saw this trend, and argued that there might
not be a close connection between weak {\MgII} absorbers and outlying
subsystems of strong {\MgII} absorbers.  However, there have been
significant new findings revealing more information about the location
of weak {\MgII} absorbers relative to luminous galaxies. 

Through direct \citep{Church05} and indirect arguments \citep{Milni06} it is
now known that many weak {\MgII} absorbers are $\sim 30$-$100
h^{-1}$~kpc from luminous galaxies, and not usually at larger
distances as previously believed (e.g., \citet{Rigby02}).  Since weak
{\MgII} absorbers are clustered around the hosts of strong {\MgII}
absorption and since the two populations cover comparable fractions of
the sky, this would argue that it would be common for a line of sight
to pass through both a strong and a weak {\MgII} absorber. The
difference in equivalent width distributions of the weak {\MgII}
absorbers and the outlying subsystems of strong {\MgII} absorbers
could be explained simply by stronger clustering of the higher
equivalent width, weak {\MgII} absorbers (those with $W_r(2796) >
0.08$~{\AA}) around galaxies.  The $W_r(2796) < 0.08$~{\AA} weak
{\MgII} absorbers would then be at a larger separation from the
galaxy, and it would thus be less probable to pass through such an
absorber {\it and} through a galaxy.  Thus, it appears plausible that
outlying subsystems of strong {\MgII} absorbers are produced by the
same structures as weak {\MgII} absorbers.

Based on photoionization models, \citet{lynch06} found a similar
multi-phase structure for weak {\MgII} absorbers as for Milky Way HVCs
and concluded that they could have common origins.  Also,
\citet{Milni06} considered the observed cross-sections of high and low
ionization systems at $z<1$ and concluded that a filametary or
sheet-like geometry is favored for weak {\MgII} absorbers.  This is
consistent with the distribution of {\OVI} around the Milky Way
\citep{Sem03}.

We have suggested that outlying subsystems of strong {\MgII} absorbers
and weak {\MgII} absorbers may be produced by the same structures, and
that weak {\MgII} absorbers and HVCs may be related.  Therefore, we
should also directly compare subsystems of strong {\MgII} absorbers
and HVCs.  As described in \S~\ref{sec:intro}, the HVCs that are
detected in 21-cm surveys are only a subset of the population, with a
larger fraction detected in {\OVI} absorption \citep{Sem03}.  In fact
the kinematics of low ionization absorption detected looking out
through the Milky Way (see, e.g., \citet{Gang05},
\citet{Fox05}, \citet{Col05}, and \citet{Jen03})
would be indistinguishable from those of some strong {\MgII} absorbers at
$0.3 < z < 1.2$.  Based upon all of these considerations, there is
a compelling case for a three-way connection between outlying subsystems
of strong {\MgII} absorbers, weak {\MgII} absorbers, and extragalactic
analogs to HVCs.

Based on this connection, we can extend our interpretation of the
evolution in the kinematics of strong {\MgII} absorbers.  In this
context, at high redshifts ($1.2 < z < 2.5$) the structures that
give rise to weak {\MgII} absorption and HVCs are present, but they 
appear more connected, by unsettled gas, to the strong {\MgII} absorbing
structures.  As we approach the current epoch, this gas settles into
the weak/HVC regions, or more likely, into the strong {\MgII} absorbing
region.

To test this idea, it will be important to make direct comparisons of
the physical properties (particularly metallicities and densities)
of weak {\MgII} absorbers and of the outlying subsystems of strong
{\MgII} absorbers.  This will require access to high resolution coverage
of the Lyman series lines associated with these absorbers, and to other
key constraining transitions, such as {\OVI} and {\CIII}.  This can
be achieved with optical spectra for $z>2$ absorbers, but at such high
redshifts high velocity subsystems are not common.  To study the
crucial low redshift regime, high resolution ultra-violet spectra will
be essential.  

If the connection is real, we finally have a handle on the relationship
between absorption signatures and familiar structures we see in the
local universe and in deep imaging studies.  With the tool of quasar
absorption lines, we then have the ability to learn things about the
gaseous assembly of galaxies that we cannot learn from imaging alone.
More specifically, through observations of the gas, we can study the
relative importance of gas accretion and of major mergers in
determining the evolution of galaxy morphology over the past
$\sim10$~Gyrs.

This work was funded by the National Science Foundation grant NSF
AST-07138, by NASA through grant NNG04GE73F, and by an NSF REU Supplement.
We are indebted to the ESO Archive for making this work possible.
We also thank an anonymous referee who helped us to clarify the results
and improve the presentation of this paper.

\begin{deluxetable}{lrcc}
\tablewidth{0pc}
\tablenum{1}
\tablecolumns{4}
\tablecaption{List of Observed Quasars}
\tablehead
{
\colhead{QSO} &
\colhead{V} &
\colhead{$z_{emit}$} &
\colhead{Observed Wavelength Range} 
}
\startdata
HE0001-2340 & 16.7 & 2.259 & 3056 - 10074\\
Q0002-422 & 17.5 & 2.659 & 3056 - 10073\\
Q0109-3518 & 16.6 & 2.405 & 3056 - 10074\\
Q0122-380 & 17.1 & 2.192 & 3064 - 10198\\
HE0151-4326 & 17.2 & 2.784 & 3056 - 10075\\
PKS0237-23 & 16.8 & 2.223 & 3063 - 10075\\
PKS0329-255* & 17.1 & 2.707 & 3056 - 10074\\
Q0329-385 & 17.2 & 2.435 & 3063 - 8507\\
Q0420-388* & 16.9 & 3.117 & 3401 - 10074\\
Q0453-423 & 17.3 & 2.657 & 3057 - 10074\\
HE0940-1050 & 16.6 & 3.083 & 3101 - 10082\\
B1055-301* & 19.5 & 2.523 & 3293 - 5756\\
HE1122-1648 & 17.7 & 2.405 & 3051 - 10083\\
HE1158-1843* & 16.9 & 2.448 & 3056 - 10074\\
HE1341-1020 & 17.1 & 2.135 & 3052 - 10415\\
PKS1448-232* & 17.0 & 2.218 & 3062 - 10073\\
PKS2126-158 & 17.3 & 3.280 & 3352 - 9609\\
HE2217-2818 & 16.0 & 2.413 & 3052 - 9898.8\\
B2311-373 & 18.5 & 2.476 & 3292 - 6686\\
HE2347-4342* & 16.3 & 2.880 & 3063 - 10095\\
\tablecomments{*No strong {\MgII} absorbers were found in these spectra.}
\enddata
\label{tab:tab1}
\end{deluxetable}

\begin{deluxetable}{llrrrcc}
\tablewidth{0pc}
\tablenum{2}
\tablecolumns{7}
\tablecaption{Kinematic Subsystem Properties and System Totals}
\tablehead
{
\colhead{QSO} &
\colhead{$z_{abs}$} &
\colhead{Subsys} &
\colhead{$\left< v \right>$} &
\colhead{$\omega_{v}$} &
\colhead{$W_{\rm r}(2796)$} &
\colhead{DR} \\
 &
 &
 &
\colhead{[{\kms}]} &
\colhead{[{\kms}]} &
\colhead{[{\AA}]} &
}
\startdata
B2311-373 & 0.339862 & 1 & \nodata & $ 29.1\pm  0.4$ & $0.959\pm0.005$ & $1.05\pm0.01$ \\
\hline
Q0122 & 0.443791 & 1 & \nodata & $ 25.9\pm  0.2$ & $0.412\pm0.003$ & $1.45\pm0.02$ \\
\hline
HE0151 & 0.663069 & 1 & $-241.1\pm  0.8$ & $ 17.2\pm  0.7$ & $0.042\pm0.001$ & $1.80\pm0.10$ \\
 & & 2 & $  8.1\pm  0.1$ & $ 23.9\pm  0.1$ & $0.384\pm0.001$ & $1.32\pm0.01$ \\
 & & Total & \nodata & $ 62.6\pm  0.7$ & $0.425\pm0.001$ & $1.35\pm0.01$ \\
\hline
HE1122 & 0.682246 & 1 & $-162.7\pm  0.1$ & $  7.3\pm  0.2$ & $0.126\pm0.001$ & $1.41\pm0.02$ \\
 & & 2 & $  7.4\pm  0.5$ & $ 48.7\pm  0.3$ & $1.702\pm0.001$ & $1.09\pm0.00$ \\
 & & Total & \nodata & $ 55.3\pm  0.3$ & $1.828\pm0.002$ & $1.10\pm0.00$ \\
\hline
Q0453 & 0.726110 & 1 & \nodata & $ 39.1\pm  0.1$ & $1.356\pm0.002$ & $1.06\pm0.00$ \\
\hline
Q0329 & 0.762783 & 1 & \nodata & $ 47.6\pm  0.4$ & $0.616\pm0.005$ & $1.42\pm0.02$ \\
\hline
Q0002 & 0.836643 & 1 & \nodata & $122.8\pm  0.3$ & $4.431\pm0.002$ & $1.12\pm0.00$ \\
\hline
Q0122 & 0.859777 & 1 & $-233.5\pm  0.5$ & $ 15.8\pm  0.5$ & $0.177\pm0.003$ & $1.50\pm0.05$ \\
 & & 2 & $  4.1\pm  0.1$ & $  4.9\pm  0.2$ & $0.162\pm0.002$ & $1.21\pm0.03$ \\
 & & 3 & $160.8\pm  1.2$ & $  4.8\pm  1.8$ & $0.022\pm0.002$ & $1.50\pm0.30$ \\
 & & Total & \nodata & $122.4\pm  1.0$ & $0.361\pm0.005$ & $1.36\pm0.03$ \\
\hline
HE1341 & 0.872808 & 1 & \nodata & $ 19.9\pm  0.4$ & $0.530\pm0.007$ & $1.22\pm0.03$ \\
\hline
HE2217 & 0.942415 & 1 & $-266.9\pm  0.1$ & $ 24.9\pm  0.1$ & $0.121\pm0.001$ & $1.60\pm0.01$ \\
 & & 2 & $  3.0\pm  0.1$ & $ 23.7\pm  0.1$ & $0.450\pm0.001$ & $1.57\pm0.00$ \\
 & & Total & \nodata & $104.1\pm  0.2$ & $0.571\pm0.001$ & $1.58\pm0.00$ \\
\hline
HE0001 & 0.949148 & 1 & \nodata & $ 34.5\pm  0.2$ & $0.348\pm0.001$ & $1.66\pm0.01$ \\
\hline
Q0122 & 1.243967 & 1 & $  2.5\pm  0.3$ & $ 13.5\pm  0.3$ & $0.449\pm0.003$ & $1.18\pm0.01$ \\
 & & 2 & $ 92.7\pm  1.2$ & $  5.3\pm  1.4$ & $0.019\pm0.002$ & $0.77\pm0.11$ \\
 & & Total & \nodata & $ 17.1\pm  0.5$ & $0.467\pm0.004$ & $1.16\pm0.01$ \\
\hline
HE1341 & 1.276756 & 1 & $-10.9\pm  1.0$ & $ 33.4\pm  0.5$ & $1.177\pm0.007$ & $1.09\pm0.01$ \\
 & & 2 & $137.8\pm  0.3$ & $  6.2\pm  0.2$ & $0.225\pm0.004$ & $1.06\pm0.03$ \\
 & & 3 & $245.8\pm  1.5$ & $  7.7\pm  1.3$ & $0.032\pm0.004$ & $1.85\pm0.54$ \\
 & & 4 & $313.0\pm  1.3$ & $  6.1\pm  1.1$ & $0.027\pm0.004$ & $2.40\pm0.90$ \\
 & & Total & \nodata & $ 67.4\pm  1.2$ & $1.461\pm0.010$ & $1.11\pm0.01$ \\
\hline
PKS0237 & 1.365055 & 1 & \nodata & $ 54.8\pm  0.3$ & $1.856\pm0.001$ & $1.13\pm0.00$ \\
\hline
Q0329 & 1.438025 & 1 & $-82.1\pm  0.5$ & $  6.3\pm  0.6$ & $0.034\pm0.002$ & $2.09\pm0.23$ \\
 & & 2 & $ 19.5\pm  0.4$ & $ 36.9\pm  0.2$ & $0.334\pm0.003$ & $1.69\pm0.03$ \\
 & & Total & \nodata & $ 44.0\pm  0.3$ & $0.368\pm0.003$ & $1.73\pm0.03$ \\
\hline
Q0002 & 1.541850 & 1 & \nodata & $ 28.4\pm  0.2$ & $0.406\pm0.001$ & $1.47\pm0.01$ \\
\hline
HE0001 & 1.585523 & 1 & $ -2.0\pm  0.1$ & $  7.9\pm  0.1$ & $0.167\pm0.001$ & $1.39\pm0.01$ \\
 & & 2 & $103.6\pm  0.1$ & $  9.8\pm  0.2$ & $0.175\pm0.001$ & $1.47\pm0.02$ \\
 & & Total & \nodata & $ 53.5\pm  0.1$ & $0.342\pm0.001$ & $1.43\pm0.01$ \\
\hline
HE2217 & 1.627857 & 1 & \nodata & $ 24.6\pm  0.0$ & $0.628\pm0.001$ & $1.41\pm0.00$ \\
\hline
PKS0237 & 1.637117 & 1 & $-140.7\pm  0.1$ & $  9.1\pm  0.1$ & $0.158\pm0.001$ & $1.42\pm0.01$ \\
 & & 2 & $ 14.3\pm  0.1$ & $ 24.1\pm  0.2$ & $0.354\pm0.001$ & $1.55\pm0.01$ \\
 & & Total & \nodata & $ 75.1\pm  0.1$ & $0.512\pm0.001$ & $1.51\pm0.01$ \\
\hline
PKS0237 & 1.657433 & 1 & \nodata & $ 28.1\pm  0.1$ & $0.683\pm0.001$ & $1.38\pm0.00$ \\
\hline
PKS0237 & 1.672334 & 1 & \nodata & $ 46.5\pm  0.3$ & $1.283\pm0.001$ & $1.20\pm0.00$ \\
\hline
HE2217 & 1.692150 & 1 & \nodata & $ 83.7\pm  0.1$ & $1.693\pm0.001$ & $1.37\pm0.00$ \\
\hline
HE0940 & 1.789119 & 1 & \nodata & $ 54.6\pm  0.1$ & $1.121\pm0.001$ & $1.40\pm0.00$ \\
\hline
PKS2126 & 2.022556 & 1 & \nodata & $ 28.6\pm  0.3$ & $0.702\pm0.002$ & $1.06\pm0.00$ \\
\hline
Q0002 & 2.167886 & 1 & \nodata & $ 22.2\pm  0.1$ & $0.353\pm0.001$ & $1.50\pm0.01$ \\
\hline
HE0001 & 2.184439 & 1 & $-84.5\pm  0.2$ & $ 48.9\pm  0.1$ & $0.540\pm0.002$ & $1.65\pm0.01$ \\
 & & 2 & $ 84.8\pm  0.2$ & $ 14.8\pm  0.3$ & $0.148\pm0.001$ & $1.61\pm0.02$ \\
 & & 3 & $246.9\pm  0.1$ & $ 16.3\pm  0.1$ & $0.247\pm0.001$ & $1.48\pm0.01$ \\
 & & Total & \nodata & $152.7\pm  0.2$ & $0.935\pm0.002$ & $1.60\pm0.01$ \\
\hline
Q0002 & 2.301944 & 1 & \nodata & $ 55.4\pm  0.3$ & $1.635\pm0.001$ & $1.13\pm0.00$ \\
\hline
Q0002 & 2.464123 & 1 & \nodata & $ 15.7\pm  0.1$ & $0.371\pm0.001$ & $1.40\pm0.01$ \\
\hline
Q0109 & 1.349585 & 1 & \nodata & $ 76.2\pm  0.1$ & $1.978\pm0.001$ & $1.26\pm0.00$ \\
\hline
Q0453 & 0.908513 & 1 & \nodata & $ 30.2\pm  0.1$ & $0.855\pm0.001$ & $1.26\pm0.00$ \\
\hline
Q0453 & 1.149861 & 1 & $ 24.7\pm  0.7$ & $136.0\pm  0.5$ & $4.444\pm0.002$ & $1.11\pm0.00$ \\
 & & 2 & $405.3\pm  0.7$ & $  8.4\pm  0.7$ & $0.016\pm0.001$ & $1.34\pm0.12$ \\
 & & Total & \nodata & $136.4\pm  0.5$ & $4.460\pm0.002$ & $1.11\pm0.00$ \\
\hline
Q0453 & 1.629994 & 1 & \nodata & $ 40.1\pm  0.1$ & $0.303\pm0.001$ & $1.68\pm0.01$ \\
\hline
Q0453 & 2.304569 & 1 & \nodata & $ 23.9\pm  0.1$ & $0.464\pm0.002$ & $1.55\pm0.01$ \\
\hline
\enddata
\label{tab:tab2}
\end{deluxetable}

\begin{deluxetable}{lccccccc}
\tablewidth{0pc}
\tablenum{3}
\tabletypesize{\tiny}
\tablecolumns{8}
\tablecaption{Subfeature Equivalent Widths for Target Transitions}
\tablehead
{
 &
 &
\colhead{Mg I} &
 &
 &
\colhead{Fe II} &
 &
 \\
\cline{3-3}\cline{4-8}\colhead{System} &
\colhead{$(v^+, v^-)$} &
\colhead{$W_{\rm r}(2853)$} &
\colhead{$W_{\rm r}(2344)$} &
\colhead{$W_{\rm r}(2374)$} &
\colhead{$W_{\rm r}(2383)$} &
\colhead{$W_{\rm r}(2587)$} &
\colhead{$W_{\rm r}(2600)$}
}
\startdata
\cutinhead{$B2311-373\qquad z_{\rm abs} = 0.339862$}
1 & (-66.1,69.9) & $0.9548\pm0.0044$ &  &  &  & $0.4715\pm0.0074$ & $0.7306\pm0.0063$\\
\cutinhead{$Q0122\qquad z_{\rm abs} = 0.443791$}
1 & (-48.1,93.0) & $0.0469\pm0.0041$ & $0.0123\pm0.0039$ & $<0.0069$ & $0.0491\pm0.0045$ & $0.0628\pm0.0026$ & $0.1640\pm0.0038$\\
\cutinhead{$HE0151\qquad z_{\rm abs} = 0.663069$}
1 & (-288.6,-195.1) & $0.0023\pm0.0006$ & $0.2527\pm0.0013$ & $0.1667\pm0.0015$ & $0.0068\pm0.0012$ & $0.0535\pm0.0013$ & $0.4808\pm0.0009$\\
2 & (-27.5,69.2) & $0.0196\pm0.0010$ & $0.0789\pm0.0015$ & $0.0287\pm0.0015$ & $0.1298\pm0.0015$ & $0.0984\pm0.0015$ & $0.1249\pm0.0012$\\
\cutinhead{$HE1122\qquad z_{\rm abs} = 0.682246$}
1 & (-194.5,-118.0) & $0.0038\pm0.0005$ & $0.0124\pm0.0005$ & $0.0357\pm0.0005$ & $0.0392\pm0.0005$ & $0.0140\pm0.0006$ & $0.0298\pm0.0005$\\
2 & (-95.7,146.5) & $0.1343\pm0.0012$ & $1.1387\pm0.0007$ & $0.4259\pm0.0009$ & $1.2393\pm0.0006$ & $0.8137\pm0.0009$ & $1.2773\pm0.0008$\\
\cutinhead{$Q0453\qquad z_{\rm abs} = 0.726110$}
1 & (-106.3,98.7) & $0.4586\pm0.0026$ & $0.8266\pm0.0025$ & $0.5212\pm0.0024$ & $1.1622\pm0.0019$ & $0.7110\pm0.0012$ & $1.0525\pm0.0012$\\
\cutinhead{$Q0329\qquad z_{\rm abs} = 0.762783$}
1 & (-61.5,163.5) & $0.0555\pm0.0035$ & $0.4660\pm0.0031$ & $0.0207\pm0.0017$ & $0.1546\pm0.0029$ & $0.1050\pm0.0053$ & $0.1665\pm0.0048$\\
\cutinhead{$Q0002\qquad z_{\rm abs} = 0.836643$}
1 & (-326.9,297.7) & $1.5856\pm0.0022$ & $3.2925\pm0.0026$ & $1.4011\pm0.0032$ & $3.0639\pm0.0025$ & $1.8329\pm0.0026$ & $2.9964\pm0.0025$\\
\cutinhead{$Q0122\qquad z_{\rm abs} = 0.859777$}
1 & (-287.7,-195.5) & $<0.0062$ & $<0.0038$ & $<0.0038$ & $0.0185\pm0.0025$ & $<0.0038$ & $0.0143\pm0.0026$\\
2 & (-22.5,35.1) & $0.0255\pm0.0022$ & $0.1008\pm0.0017$ & $0.0081\pm0.0013$ & $0.0582\pm0.0019$ & $0.0208\pm0.0016$ & $0.0518\pm0.0015$\\
3 & (141.7,182.1) & $0.0060\pm0.0025$ & $<0.0037$ & $<0.0043$ & $0.0037\pm0.0013$ & $<0.0037$ & $0.0039\pm0.0015$\\
\cutinhead{$HE1341\qquad z_{\rm abs} = 0.872808$}
1 & (-81.9,41.2) & $0.0783\pm0.0062$ & $0.1094\pm0.0071$ & $0.0301\pm0.0060$ & $0.2176\pm0.0093$ & $0.1082\pm0.0070$ & $0.2341\pm0.0061$\\
\cutinhead{$HE2217\qquad z_{\rm abs} = 0.942415$}
1 & (-340.9,-216.7) & $0.0128\pm0.0005$ & $0.0092\pm0.0009$ & $0.0057\pm0.0008$ & $0.0135\pm0.0007$ & $0.0084\pm0.0006$ & $0.0102\pm0.0004$\\
2 & (-87.0,86.9) & $0.0157\pm0.0006$ & $0.0330\pm0.0011$ & $0.0094\pm0.0009$ & $0.0844\pm0.0008$ & $0.0286\pm0.0007$ & $0.0763\pm0.0005$\\
\cutinhead{$HE0001\qquad z_{\rm abs} = 0.949148$}
1 & (-96.5,104.3) & $0.0203\pm0.0014$ & $0.0125\pm0.0015$ & $<0.0024$ & $0.0530\pm0.0026$ & $0.0383\pm0.0011$ & $0.0406\pm0.0011$\\
\cutinhead{$Q0122\qquad z_{\rm abs} = 1.243967$}
1 & (-36.9,70.6) & $0.1021\pm0.0025$ & $0.1771\pm0.0027$ & $0.1124\pm0.0028$ & $0.2268\pm0.0023$ & \nodata & \nodata\\
2 & (77.8,111.2) & $<0.0054$ & $<0.0049$ & $<0.0049$ & $<0.0052$ & \nodata & \nodata\\
\cutinhead{$HE1341\qquad z_{\rm abs} = 1.276756$}
1 & (-92.4,70.1) & $0.2389\pm0.0077$ & $0.4092\pm0.0065$ & $0.2289\pm0.0071$ & $0.6554\pm0.0060$ & $0.3725\pm0.0069$ & $0.6575\pm0.0056$\\
2 & (112.5,166.6) & $0.0657\pm0.0045$ & $0.1209\pm0.0039$ & $0.0639\pm0.0040$ & $0.1535\pm0.0040$ & $0.1497\pm0.0047$ & $0.1614\pm0.0034$\\
3 & (225.5,265.5) & $<0.0102$ & $<0.0101$ & $<0.0095$ & $<0.0099$ & $<0.0101$ & $<0.0090$\\
4 & (298.5,329.1) & $<0.0103$ & $0.0310\pm0.0036$ & $<0.0097$ & $0.0153\pm0.0036$ & $<0.0099$ & $<0.0092$\\
\cutinhead{$PKS0237\qquad z_{\rm abs} = 1.365055$}
1 & (-125.8,184.7) & $0.2690\pm0.0015$ & $0.4994\pm0.0012$ & $0.2063\pm0.0013$ & $0.8855\pm0.0010$ & $0.4105\pm0.0011$ & $0.8702\pm0.0011$\\
\cutinhead{$Q0329\qquad z_{\rm abs} = 1.438025$}
1 & (-102.3,-60.5) & $0.0069\pm0.0015$ & $<0.0032$ & \nodata & \nodata & $0.0090\pm0.0014$ & $0.0076\pm0.0015$\\
2 & (-49.5,115.4) & $0.0537\pm0.0026$ & $0.0284\pm0.0025$ & \nodata & \nodata & $0.0315\pm0.0024$ & $0.0500\pm0.0029$\\
\cutinhead{$Q0002\qquad z_{\rm abs} = 1.541850$}
1 & (-152.2,31.3) & $0.0523\pm0.0011$ & $0.0319\pm0.0011$ & $0.0153\pm0.0011$ & $0.0711\pm0.0010$ & $0.0238\pm0.0008$ & $0.0659\pm0.0011$\\
\cutinhead{$HE0001\qquad z_{\rm abs} = 1.585523$}
1 & (-25.2,23.5) & $0.0087\pm0.0011$ & $0.0029\pm0.0005$ & $0.0061\pm0.0005$ & $0.0085\pm0.0005$ & $<0.0017$ & $0.0096\pm0.0008$\\
2 & (56.7,133.4) & $0.0124\pm0.0012$ & $0.0155\pm0.0006$ & $0.0057\pm0.0007$ & $0.0351\pm0.0006$ & $0.0126\pm0.0006$ & $0.0333\pm0.0010$\\
\cutinhead{$HE2217\qquad z_{\rm abs} = 1.627857$}
1 & (-80.1,62.7) & $0.0372\pm0.0008$ & $0.0208\pm0.0005$ & $0.0048\pm0.0004$ & $0.0571\pm0.0005$ & $0.0236\pm0.0007$ & $0.0444\pm0.0008$\\
\cutinhead{$PKS0237\qquad z_{\rm abs} = 1.637117$}
1 & (-162.9,-97.9) & $0.0120\pm0.0007$ & $0.0151\pm0.0006$ & $0.0033\pm0.0005$ & $0.0477\pm0.0007$ & $0.0093\pm0.0006$ & $0.0212\pm0.0006$\\
2 & (-32.8,109.5) & $0.0314\pm0.0010$ & $0.0436\pm0.0008$ & $0.0635\pm0.0009$ & $0.1399\pm0.0009$ & $0.0317\pm0.0009$ & $0.1026\pm0.0010$\\
\cutinhead{$PKS0237\qquad z_{\rm abs} = 1.657433$}
1 & (-85.4,100.1) & $0.0366\pm0.0015$ & $0.0345\pm0.0008$ & $0.0227\pm0.0009$ & $0.1079\pm0.0011$ & $0.3328\pm0.0012$ & $0.2173\pm0.0013$\\
\cutinhead{$PKS0237\qquad z_{\rm abs} = 1.672334$}
1 & (-188.0,94.8) & $0.9832\pm0.0015$ & $0.3946\pm0.0011$ & $0.2626\pm0.0013$ & $0.5483\pm0.0013$ & $0.5514\pm0.0016$ & $0.5566\pm0.0012$\\
\cutinhead{$HE2217\qquad z_{\rm abs} = 1.692149$}
1 & (-204.9,145.5) & $0.2560\pm0.0013$ & $0.1609\pm0.0008$ & $0.0475\pm0.0007$ & $0.3368\pm0.0007$ & $0.1307\pm0.0011$ & $0.3520\pm0.0012$\\
\cutinhead{$HE0940\qquad z_{\rm abs} = 1.789119$}
1 & (-109.9,211.1) & $0.1695\pm0.0017$ & $0.1239\pm0.0013$ & $0.0439\pm0.0015$ & $0.2875\pm0.0015$ & $0.1701\pm0.0015$ & $0.5414\pm0.0014$\\
\cutinhead{$PKS2126\qquad z_{\rm abs} = 2.022556$}
1 & (-82.7,77.0) &  & $0.1386\pm0.0010$ & $0.0758\pm0.0012$ & $0.3377\pm0.0012$ & $0.1062\pm0.0014$ & $0.2836\pm0.0016$\\
\cutinhead{$Q0002\qquad z_{\rm abs} = 2.167886$}
1 & (-26.2,77.0) & $0.0284\pm0.0018$ & $0.0345\pm0.0008$ & $0.0135\pm0.0007$ & $0.0764\pm0.0007$ & $0.0611\pm0.0011$ & $0.1102\pm0.0010$\\
\cutinhead{$HE0001\qquad z_{\rm abs} = 2.184439$}
1 & (-151.9,34.9) & $0.8641\pm0.0012$ & $0.0214\pm0.0015$ & $0.0080\pm0.0011$ & $0.0552\pm0.0015$ & $0.1222\pm0.0019$ & $0.1852\pm0.0019$\\
2 & (36.6,144.3) & $0.0765\pm0.0010$ & $0.0098\pm0.0007$ & $0.0124\pm0.0011$ & $0.0357\pm0.0011$ & $0.0527\pm0.0015$ & $0.0709\pm0.0016$\\
3 & (191.5,282.4) & $0.0295\pm0.0009$ & $0.0424\pm0.0011$ & $0.0163\pm0.0010$ & $0.0837\pm0.0013$ & $0.0451\pm0.0013$ & $0.1053\pm0.0016$\\
\cutinhead{$Q0002\qquad z_{\rm abs} = 2.301944$}
1 & (-139.5,113.7) & $0.3138\pm0.0026$ & $0.4345\pm0.0011$ & $0.1938\pm0.0011$ & $0.6565\pm0.0013$ & \nodata & \nodata\\
\cutinhead{$Q0002\qquad z_{\rm abs} = 2.464123$}
1 & (-38.9,49.3) & $0.0102\pm0.0021$ & $0.0034\pm0.0005$ & $0.0182\pm0.0007$ & $0.0353\pm0.0007$ & $0.0097\pm0.0011$ & $0.0625\pm0.0012$\\
\cutinhead{$HE1341\qquad z_{\rm abs} = 1.915397$}
1 & (-37.7,32.1) & $0.1519\pm0.0059$ & $0.0595\pm0.0044$ & $0.0164\pm0.0036$ & $0.1119\pm0.0034$ & $0.0382\pm0.0038$ & $0.1133\pm0.0045$\\
\cutinhead{$Q0109\qquad z_{\rm abs} = 1.349585$}
1 & (-208.4,200.0) & $0.2977\pm0.0021$ & $0.4220\pm0.0017$ & $0.1423\pm0.0019$ & $0.8387\pm0.0016$ & $0.3486\pm0.0018$ & $0.7907\pm0.0017$\\
\cutinhead{$Q0453\qquad z_{\rm abs} = 0.908513$}
1 & (-127.7,80.1) & $0.1184\pm0.0015$ & $0.2078\pm0.0016$ & $0.1193\pm0.0017$ & $0.3733\pm0.0017$ & $0.1869\pm0.0025$ & $0.3744\pm0.0018$\\
\cutinhead{$Q0453\qquad z_{\rm abs} = 1.149861$}
1 & (-250.2,373.1) & $1.5264\pm0.0026$ & $2.7782\pm0.0018$ & $1.7779\pm0.0018$ & $3.5627\pm0.0013$ & $2.6907\pm0.0016$ & $3.6532\pm0.0013$\\
2 & (383.1,430.4) & $<0.0019$ & $0.0034\pm0.0006$ & $0.0011\pm0.0004$ & $0.0098\pm0.0007$ & $0.0017\pm0.0005$ & $0.0063\pm0.0005$\\
\cutinhead{$Q0453\qquad z_{\rm abs} = 1.629994$}
1 & (-64.8,114.6) & $0.0114\pm0.0012$ & $0.0165\pm0.0013$ & $0.0083\pm0.0010$ & $0.0155\pm0.0013$ & $0.0121\pm0.0012$ & $0.0072\pm0.0008$\\
\cutinhead{$Q0453\qquad z_{\rm abs} = 2.304569$}
1 & (-54.8,88.0) & $0.5751\pm0.0017$ & $0.0533\pm0.0009$ & $0.0376\pm0.0010$ & $0.1556\pm0.0011$ & \nodata & \nodata\\
\enddata
\label{tab:tab3}
\end{deluxetable}

\begin{deluxetable}{lccccc}
\tablenum{4}
\tabletypesize{\footnotesize}
\tablewidth{0pc}
\tablecolumns{6}
\tablecaption{Sample Membership}
\tablehead
{
\colhead{ } &
\colhead{Sample A} &
\colhead{Sample B} &
\colhead{Sample C} &
\colhead{Sample D} &
\colhead{Sample E} \\
\colhead{ } &
\colhead{$W_{r} \geq 0.3$~{\AA}} &
\colhead{$0.3 \leq W_{r} < 0.6$~{\AA}} & 
\colhead{$0.6 \leq  W_{r} < 1.0$~{\AA}} &
\colhead{$W_{r} \geq 0.6$~{\AA}} &
\colhead{$W_{r} \geq 1.0$~{\AA}} 
 }
\startdata
$N$ & $33$ & $15$ & $7$ & $18$ & $11$ \\
$\left< W_{r}(2796) \right>$, {\AA} & 1.03 & .42 & .77 & 1.58 & 2.10\\
$\left< z \right>$ & 1.38 & 1.42 & 1.36 & 1.35 & 1.35\\
$\left< DR \right>$ & 1.33 & 1.45 & 1.31 & 1.23 & 1.18\\
\cutinhead{Membership Chart}
$B2311-373~~~0.339862$ & X &   & X & X &   \\
$Q0122-380~~~0.443791$ & X & X &   &   &   \\
$HE0151-4326~0.663069$ & X & X &   &   &   \\
$HE1122-1648~0.682246$ & X &   &   & X & X \\
$Q0453-423~~~0.726110$ & X &   &   & X & X \\
$Q0329-385~~~0.762783$ & X &   & X & X &   \\
$Q0002-422~~~0.836643$ & X &   &   & X & X \\
$Q0122-380~~~0.859777$ & X & X &   &   &   \\
$HE1341-1020~0.872808$ & X & X &   &   &   \\
$Q0453-423~~~0.908513$ & X &   & X & X &   \\
$HE2217-2818~0.942415$ & X & X &   &   &   \\
$HE0001-2340~0.949148$ & X & X &   &   &   \\
$Q0453-423~~~1.149861$ & X &   &   & X & X \\
$Q0122-380~~~1.243967$ & X & X &   &   &   \\
$HE1341-1020~1.276756$ & X &   &   & X & X \\
$Q0109-3518~~1.349585$ & X &   &   & X & X \\
$PKS0237-23~~1.365055$ & X &   &   & X & X \\
$Q0329-385~~~1.438025$ & X & X &   &   &   \\
$Q0002-422~~~1.541850$ & X & X &   &   &   \\
$HE0001-2340~1.586204$ & X & X &   &   &   \\
$HE2217-2818~1.627861$ & X &   & X & X &   \\
$Q0453-423~~~1.629994$ & X & X &   &   &   \\
$PKS0237-23~~1.637118$ & X & X &   &   &   \\
$PKS0237-23~~1.657433$ & X &   & X & X &   \\
$PKS0237-23~~1.672334$ & X &   &   & X & X \\
$HE2217-2818~1.692149$ & X &   &   & X & X \\
$HE0940-1050~1.789119$ & X &   &   & X & X \\
$PKS2126-158~2.022556$ & X &   & X & X &   \\
$Q0002-422~~~2.167886$ & X & X &   &   &   \\
$HE0001-234~~2.184439$ & X &   & X & X &   \\
$Q0002-422~~~2.301944$ & X &   &   & X & X \\
$Q0453-423~~~2.304569$ & X & X &   &   &   \\
$Q0002-422~~~2.464123$ & X & X &   &   &   \\
\enddata 
\label{tab:tab4}
\end{deluxetable}

\begin{deluxetable}{lrccc}
\tablewidth{0pc}
\tablenum{5}
\tablecolumns{5}
\tablecaption{Subsystem AOD Column Densities}
\tablehead
{
\colhead{System} &
\colhead{$(v^+, v^-)$} &
\colhead{Mg II} &
\colhead{Mg I} &
\colhead{Fe II}
}
\startdata
\cutinhead{$B2311-373\qquad z_{\rm abs} = 0.339862$}
1 & (-66,70) & $>14.07$ & \nodata & $>14.41$\\
\cutinhead{$Q0122\qquad z_{\rm abs} = 0.443791$}
1 & (-48,93) & $ 13.263\pm  0.006$ & $ 11.65\pm  0.04$ & $ 14.42\pm  0.06$\\
\cutinhead{$HE0151\qquad z_{\rm abs} = 0.663069$}
1 & (-289,-195) & $ 12.04\pm  0.02$ & $ 10.3\pm  0.1$ & $ 11.82\pm  0.08$\\
2 & (-28,69) & $ 13.282\pm  0.001$ & $ 11.23\pm  0.02$ & $ 13.085\pm  0.006$\\
\cutinhead{$HE1122\qquad z_{\rm abs} = 0.682246$}
1 & (-195,-118) & $ 12.739\pm  0.004$ & $ 12.81\pm  0.05$ & $ 12.546\pm  0.008$\\
2 & (-96,147) & $>14.46$ & $ 12.068\pm  0.004$ & $ 14.598\pm  0.001$\\
\cutinhead{$Q0453\qquad z_{\rm abs} = 0.726110$}
1 & (-106,99) & $>14.30$ & $ 12.696\pm  0.002$ & $ 14.581\pm  0.001$\\
\cutinhead{$Q0329\qquad z_{\rm abs} = 0.762783$}
1 & (-62,164) & $ 13.534\pm  0.009$ & $ 11.95\pm  0.03$ & $ 13.17\pm  0.01$\\
\cutinhead{$Q0002\qquad z_{\rm abs} = 0.836643$}
1 & (-327,298) & $>14.82$ & $ 13.229\pm  0.001$ & $ 14.885\pm  0.001$\\
\cutinhead{$Q0122\qquad z_{\rm abs} = 0.859777$}
1 & (-288,-196) & $ 12.78\pm  0.01$ & $<10.47$ & $ 12.3\pm  0.1$\\
2 & (-23,35) & $ 13.04\pm  0.01$ & $ 11.63\pm  0.04$ & $ 12.75\pm  0.02$\\
3 & (142,182) & $ 11.81\pm  0.05$ & $ 10.7\pm  0.2$ & $ 11.6\pm  0.3$\\
\cutinhead{$HE1341\qquad z_{\rm abs} = 0.872808$}
1 & (-82,41) & $>13.72$ & $ 12.68\pm  0.03$ & $ 13.41\pm  0.03$\\
\cutinhead{$Q0453\qquad z_{\rm abs} = 0.908513$}
1 & (-128,80) & $ 13.875\pm  0.001$ & $ 12.011\pm  0.005$ & $ 13.639\pm  0.003$\\
\cutinhead{$HE2217\qquad z_{\rm abs} = 0.942415$}
1 & (-341,-217) & $ 12.560\pm  0.002$ & $ 11.04\pm  0.02$ & $ 12.25\pm  0.03$\\
2 & (-87,86) & $ 13.241\pm  0.001$ & $ 11.20\pm  0.02$ & $ 12.804\pm  0.005$\\
\cutinhead{$HE0001\qquad z_{\rm abs} = 0.949148$}
1 & (-97,104) & $ 13.057\pm  0.002$ & $ 11.25\pm  0.03$ & $ 12.88\pm  0.02$\\
\cutinhead{$Q0453\qquad z_{\rm abs} = 1.149861$}
1 & (-250,373) & $>14.85$ & $ 13.235\pm  0.001$ & $ 15.395\pm  0.001$\\
2 & (383,430) & $ 11.66\pm  0.03$ & $<9.81$ & $ 11.82\pm  0.05$\\
\cutinhead{$Q0122\qquad z_{\rm abs} = 1.243967$}
1 & (-37,71) & $>13.71$ & $ 12.07\pm  0.01$ & $ 14.11\pm  0.01$\\
2 & (78,111) & $ 11.68\pm  0.05$ & $10.1759$ & $11.0974$\\
\cutinhead{$HE1341\qquad z_{\rm abs} = 1.276756$}
1 & (-92,70) & $>14.19$ & $ 14.85\pm  0.01$ & $ 14.20\pm  0.01$\\
2 & (113,167) & $>13.47$ & $ 11.79\pm  0.03$ & $ 13.75\pm  0.02$\\
3 & (226,266) & $ 11.93\pm  0.07$ & $<10.50$ & $<11.41$\\
4 & (299,329) & $ 11.83\pm  0.08$ & $<10.44$ & $<11.38$\\
\cutinhead{$Q0109\qquad z_{\rm abs} = 1.349585$}
1 & (-208,200) & $>14.21$ & $ 12.422\pm  0.003$ & $ 13.992\pm  0.001$\\
\cutinhead{$PKS0237\qquad z_{\rm abs} = 1.365055$}
1 & (-126,185) & $>14.41$ & $ 12.381\pm  0.002$ & $ 14.087\pm  0.001$\\
\cutinhead{$Q0329\qquad z_{\rm abs} = 1.438025$}
1 & (-102,-61) & $ 11.95\pm  0.02$ & $ 11.64\pm  0.09$ & $ 11.8\pm  0.1$\\
2 & (-50,115) & $ 13.078\pm  0.005$ & \nodata & $ 12.69\pm  0.03$\\
\cutinhead{$Q0002\qquad z_{\rm abs} = 1.541850$}
1 & (-152,31) & $ 14.311\pm  0.001$ & $ 11.629\pm  0.009$ & $ 12.800\pm  0.009$\\
\cutinhead{$HE0001\qquad z_{\rm abs} = 1.585523$}
1 & (-25,24) & $ 15.174\pm  0.003$ & $ 11.16\pm  0.06$ & $ 12.08\pm  0.04$\\
2 & (57,133) & $ 12.884\pm  0.003$ & $ 11.03\pm  0.04$ & $ 12.48\pm  0.01$\\
\cutinhead{$HE2217\qquad z_{\rm abs} = 1.627857$}
1 & (-80,63) & $ 14.189\pm  0.001$ & $ 11.491\pm  0.009$ & $ 12.632\pm  0.007$\\
\cutinhead{$Q0453\qquad z_{\rm abs} = 1.629994$}
1 & (-65,115) & $ 12.962\pm  0.002$ & $ 11.13\pm  0.05$ & $ 12.36\pm  0.04$\\
\cutinhead{$PKS0237\qquad z_{\rm abs} = 1.637117$}
1 & (-163,-98) & $ 12.839\pm  0.003$ & $ 10.99\pm  0.03$ & $ 12.35\pm  0.02$\\
2 & (-33,110) & $ 13.114\pm  0.002$ & $ 11.41\pm  0.01$ & $ 12.947\pm  0.006$\\
\cutinhead{$PKS0237\qquad z_{\rm abs} = 1.657433$}
1 & (-85,100) & $ 13.625\pm  0.001$ & $ 11.49\pm  0.02$ & $ 12.913\pm  0.005$\\
\cutinhead{$PKS0237\qquad z_{\rm abs} = 1.672334$}
1 & (-188,95) & $>14.21$ & \nodata & $ 14.572\pm  0.002$\\
\cutinhead{$HE2217\qquad z_{\rm abs} = 1.692150$}
1 & (-205,146) & $ 14.015\pm  0.001$ & \nodata & $ 13.480\pm  0.002$\\
\cutinhead{$HE0940\qquad z_{\rm abs} = 1.789119$}
1 & (-110,211) & $ 13.844\pm  0.001$ & \nodata & $ 13.423\pm  0.003$\\
\cutinhead{$PKS2126\qquad z_{\rm abs} = 2.022556$}
1 & (-83,77) & $>13.91$ & \nodata & $ 13.624\pm  0.004$\\
\cutinhead{$Q0002\qquad z_{\rm abs} = 2.167886$}
1 & (-26,77) & $ 13.169\pm  0.002$ & $ 11.40\pm  0.03$ & $ 12.889\pm  0.005$\\
\cutinhead{$HE0001\qquad z_{\rm abs} = 2.184439$}
1 & (-152,35) & $ 13.312\pm  0.002$ & \nodata & $ 12.71\pm  0.02$\\
2 & (37,144) & $ 12.742\pm  0.004$ & \nodata & $ 12.46\pm  0.02$\\
3 & (192,282) & $ 13.088\pm  0.002$ & \nodata & $ 12.950\pm  0.009$\\
\cutinhead{$Q0002\qquad z_{\rm abs} = 2.301944$}
1 & (-140,114) & $>14.33$ & \nodata & $ 14.160\pm  0.001$\\
\cutinhead{$Q0453\qquad z_{\rm abs} = 2.304569$}
1 & (-55,88) & $ 13.277\pm  0.002$ & \nodata & $ 13.119\pm  0.005$\\
\cutinhead{$Q0002\qquad z_{\rm abs} = 2.464123$}
1 & (-39,49) & $ 13.220\pm  0.002$ & \nodata & \nodata\\

\enddata
\label{tab:tab5}
\end{deluxetable}

\clearpage

\clearpage

\clearpage

\clearpage

\clearpage

\clearpage

\clearpage

\clearpage

\clearpage

\clearpage

\clearpage

\clearpage

\clearpage

\clearpage

\clearpage

\clearpage

\clearpage

\clearpage

\clearpage

\clearpage

\clearpage

\clearpage

\clearpage

\clearpage

\clearpage

\clearpage

\clearpage

\clearpage

\clearpage

\clearpage

\clearpage

\clearpage

\clearpage

\end{document}